\documentclass[fleqn,usenatbib]{mnras}

\usepackage{newtxtext,newtxmath}
\usepackage[T1]{fontenc}
\usepackage{ae,aecompl}

\usepackage{graphicx}	
\usepackage{amsmath}	
\usepackage{amssymb}	
\usepackage{siunitx}
\usepackage{booktabs}
\usepackage{multirow}
\usepackage{color, colortbl}
\usepackage{xcolor}
\usepackage{eso-pic}

\DeclareMathOperator*{\argmax}{argmax}
\let\vec\mathbf
\DeclareMathOperator*{\argmin}{arg\,min}
\newcommand{\ra}[1]{\renewcommand{\arraystretch}{#1}}

\definecolor{Gray}{gray}{0.9}

\AddToShipoutPictureBG*{%
  \AtPageUpperLeft{%
    \hspace{0.75\paperwidth}%
    \raisebox{-3.5\baselineskip}{%
      \makebox[0pt][l]{\textnormal{DES 2018-0377}}
}}}%

\AddToShipoutPictureBG*{%
  \AtPageUpperLeft{%
    \hspace{0.75\paperwidth}%
    \raisebox{-4.5\baselineskip}{%
      \makebox[0pt][l]{\textnormal{FERMILAB-PUB-19-011-AE}}
}}}%

\title[Phenotypic redshifts with self-organizing maps]{Phenotypic redshifts with self-organizing maps:\\ A novel method to characterize redshift distributions of source galaxies for weak lensing}

\author[DES Collaboration]{
\parbox{\textwidth}{
\Large
R.~Buchs$^{1,2,3,4}$,\thanks{E-mail: romainbuchs@hotmail.com}
C.~Davis$^{1,2,5}$,\thanks{E-mail: chris@descarteslabs.com}
D.~Gruen$^{6,2,1}$,\thanks{Einstein Fellow}
J.~DeRose$^{6,2}$,
A.~Alarcon$^{7,8}$,
G.~M.~Bernstein$^{9}$,
C.~S{\'a}nchez$^{9}$,
J.~Myles$^{6}$,
A.~Roodman$^{2,1}$,
S.~Allen$^{6}$,
A.~Amon$^{2}$,
A.~Choi$^{10}$,
D.~C.~Masters$^{11,12}$,
R.~Miquel$^{13,14}$,
M.~A.~Troxel$^{15}$,
R.~H.~Wechsler$^{6,2,1}$,
T.~M.~C.~Abbott$^{16}$,
J.~Annis$^{17}$,
S.~Avila$^{18}$,
K.~Bechtol$^{19,20}$,
S.~L.~Bridle$^{21}$,
D.~Brooks$^{22}$,
E.~Buckley-Geer$^{17}$,
D.~L.~Burke$^{2,1}$,
A.~Carnero~Rosell$^{23,24}$,
M.~Carrasco~Kind$^{25,26}$,
J.~Carretero$^{14}$,
F.~J.~Castander$^{7,8}$,
R.~Cawthon$^{20}$,
C.~B.~D'Andrea$^{9}$,
L.~N.~da Costa$^{24,27}$,
J.~De~Vicente$^{23}$,
S.~Desai$^{28}$,
H.~T.~Diehl$^{17}$,
P.~Doel$^{22}$,
A.~Drlica-Wagner$^{17,29}$,
T.~F.~Eifler$^{30,12}$,
A.~E.~Evrard$^{31,32}$,
B.~Flaugher$^{17}$,
P.~Fosalba$^{7,8}$,
J.~Frieman$^{17,29}$,
J.~Garc\'ia-Bellido$^{33}$,
E.~Gaztanaga$^{7,8}$,
R.~A.~Gruendl$^{25,26}$,
J.~Gschwend$^{24,27}$,
G.~Gutierrez$^{17}$,
W.~G.~Hartley$^{22,34}$,
D.~L.~Hollowood$^{35}$,
K.~Honscheid$^{10,36}$,
D.~J.~James$^{37}$,
K.~Kuehn$^{38}$,
N.~Kuropatkin$^{17}$,
M.~Lima$^{39,24}$,
H.~Lin$^{17}$,
M.~A.~G.~Maia$^{24,27}$,
M.~March$^{9}$,
J.~L.~Marshall$^{40}$,
P.~Melchior$^{41}$,
F.~Menanteau$^{25,26}$,
R.~L.~C.~Ogando$^{24,27}$,
A.~A.~Plazas$^{41}$,
E.~S.~Rykoff$^{2,1}$,
E.~Sanchez$^{23}$,
V.~Scarpine$^{17}$,
S.~Serrano$^{7,8}$,
I.~Sevilla-Noarbe$^{23}$,
M.~Smith$^{42}$,
M.~Soares-Santos$^{43}$,
F.~Sobreira$^{44,24}$,
E.~Suchyta$^{45}$,
M.~E.~C.~Swanson$^{26}$,
G.~Tarle$^{32}$,
D.~Thomas$^{18}$,
V.~Vikram$^{46}$
\begin{center} (DES Collaboration) \end{center}
}
\vspace{0.4cm}
\\
\parbox{\textwidth}{
(Affiliations are listed at the end of the paper)
\vspace{0.8cm}
}
}

\date{Accepted XXX. Received YYY; in original form ZZZ}

\pubyear{2019}

\begin{document}
\label{firstpage}
\pagerange{\pageref{firstpage}--\pageref{lastpage}}
\maketitle

\begin{abstract}
Wide-field imaging surveys such as the Dark Energy Survey (DES) rely on coarse measurements of spectral energy distributions in a few filters to estimate the redshift distribution of source galaxies. In this regime, sample variance, shot noise, and selection effects limit the attainable accuracy of redshift calibration and thus of cosmological constraints. We present a new method to combine wide-field, few-filter measurements with
catalogs from deep fields with additional filters and sufficiently low photometric noise to break degeneracies in photometric redshifts.
The multi-band deep field is used as an intermediary between wide-field observations and accurate redshifts, greatly reducing sample variance, shot noise, and selection effects. Our implementation of the method uses self-organizing maps to group galaxies into \textit{phenotypes} based on their observed fluxes, and is tested using a mock DES catalog created from $N$-body simulations. 
It yields a typical uncertainty on the mean redshift in each of five tomographic bins for an idealized simulation of the DES Year 3 weak-lensing tomographic analysis of $\sigma_{\Delta z} = 0.007$, which is a 60\% improvement compared to the Year 1 analysis. Although the implementation of the method is tailored to DES, its formalism can be applied to other large photometric surveys with a similar observing strategy.
\end{abstract}

\begin{keywords}
dark energy -- galaxies: distances and redshifts -- gravitational lensing: weak
\end{keywords}



\section{Introduction}
Solving the mysteries surrounding the nature of the cosmic acceleration requires measuring the growth of structure with exquisite precision and accuracy. To this end, one of the most promising probes is weak gravitational lensing. In weak lensing, light from distant source galaxies is deflected by the large-scale structure of the Universe, affecting their
apparent shapes by gravitational shear (see e.g.\ \citealt{Bartelmann2001}, \citealt{Kilbinger2015}, or \citealt{Mandelbaum2018} for a review on the subject). The amplitude of the shear depends on the distribution of the matter causing the lensing, and the distance ratios of source and lens galaxies.  The physical interpretation of the signal is thus sensitive to systematic errors in redshift estimates of source galaxies \citep{Ma2006,Huterer2006}. For precision cosmology from weak lensing probes, an accurate measurement of galaxy shapes must be coupled with a robust characterization of the redshift distribution of source galaxies.

In imaging surveys, redshift must be inferred from the electromagnetic spectral energy distribution (SED) of distant galaxies, integrated over a number of filter bands.
Ongoing broadband imaging surveys, such as the Dark Energy Survey \citep[DES;][]{DarkEnergySurveyCollaboration2005}, the Hyper Suprime-Cam Subaru Strategic Program \citep[HSC-SPP;][]{Aihara2018} and the Kilo Degree Survey \citep[KiDS;][]{deJong2013}, as well as upcoming ones such as the Large Synoptic Survey Telescope \citep[LSST;][]{Ivezic2008}, the Euclid survey \citep{Laureijs2011}, and the Wide-Field InfraRed Survey Telescope \citep[WFIRST;][]{Spergel2013}, rely on measurements of flux in a small number of bands (three to six) to determine redshifts of source galaxies. 

\begin{figure*}
\centering
\includegraphics[width=0.75\linewidth]{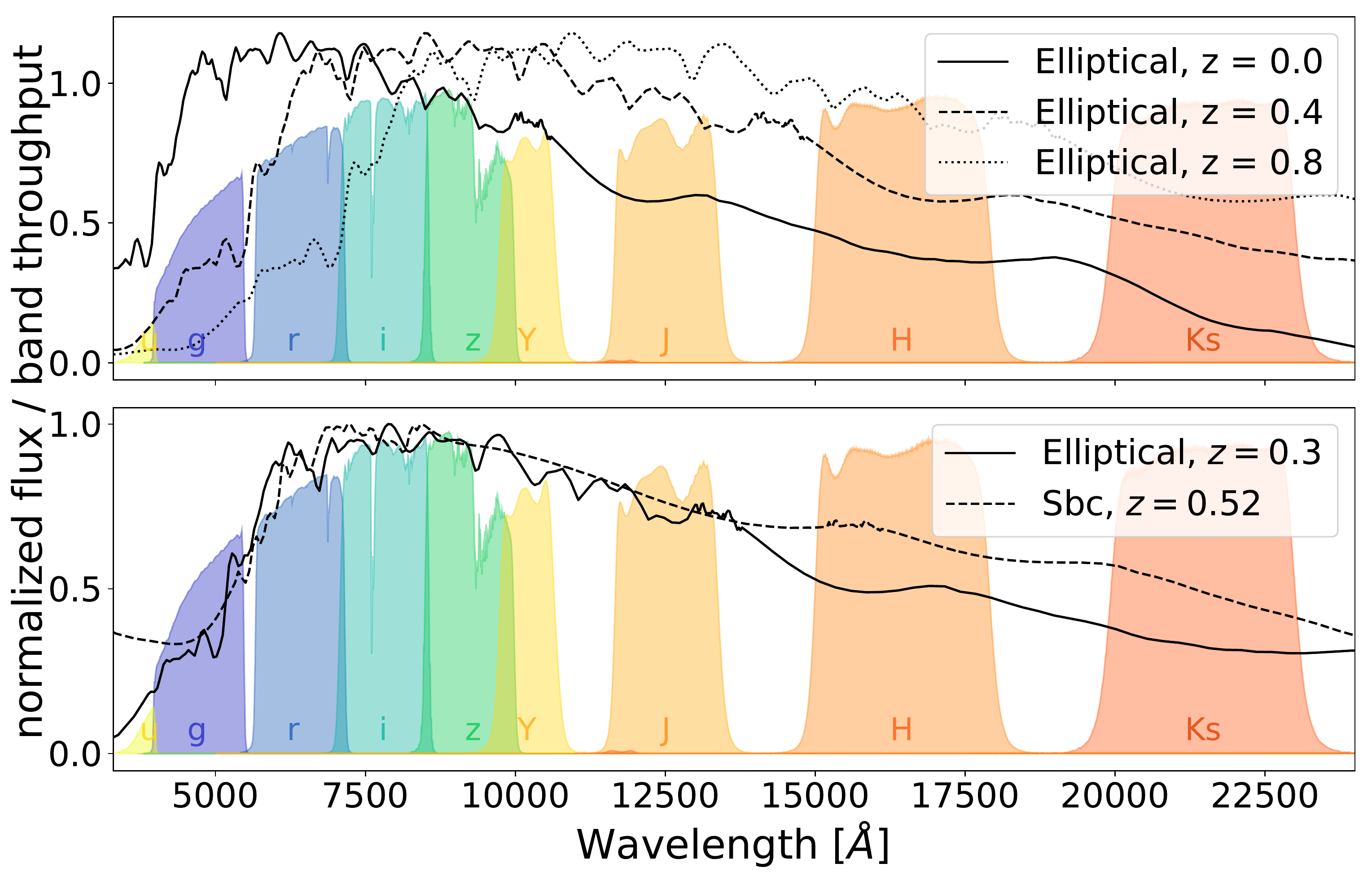}
\caption{Illustration of how redshift can be estimated from broadband images, yet not always unambiguously. \textit{Top}: The same template of an elliptical galaxy is redshifted at $z=0.4$ and $z=0.8$. These objects exhibit clearly different colors. \textit{Bottom}: Templates of an elliptical galaxy and a Sbc galaxy at different redshifts are plotted. In the optical (e.g.\ from \textit{griz} information), those two objects are indistinguishable: type and redshift are degenerate. Adding $u$ and near-infrared bands -- especially the \textit{H} and \textit{Ks} bands -- differentiates them. Colored areas show relative throughput of DES \textit{ugriz} and VISTA \textit{YJHKs} bands. Galaxy templates are taken from \citet{Benitez2004}.\label{fig:SED}}
\end{figure*}

The coarse measurement of a galaxy's redshifted SED often does not uniquely determine its redshift and type: two different rest-frame SEDs at two different redshifts can be indistinguishable, as illustrated in \autoref{fig:SED} (lower panel). This type/redshift degeneracy is the fundamental cause of uncertainty in redshift calibration, i.e.\ in the constraint of the mean and shape of the redshift distribution of an ensemble of galaxies, across methods. It can bias template-fitting methods \citep[e.g.][]{Benitez2000,Ilbert2006}, even with a Bayesian treatment of sufficiently flexible template sets, because the choice of priors determines the mix of estimated type/redshift combinations at fixed ambiguous broad-band fluxes. It can bias empirical methods based on machine learning \citep[e.g.][]{Collister2004, CarrascoKind2013, DEVicente2016} or direct calibration from spectroscopic samples, because present spectroscopic samples are subject to selection effects at fixed broad-band observables \citep{Bonnett2016, Gruen2017}. These can be both explicit (i.e.~because spectroscopic targets were selected by properties not observed in a wide-field survey) or implicit (i.e.~because success of spectroscopic redshift determination depends on type/redshift). Type/redshift degeneracy contributes to the dominant systematic uncertainty in redshifts derived from cross-correlations \citep{Schneider2006,Newman2008,Menard2013,Schmidt2013,Hildebrandt2017, Samuroff2017, Davis2017, Davis2018}, the evolution of clustering bias with redshift \citep{Gatti2018}. The latter is due in part to the evolution of the mix of galaxy types as a function of redshift. Because of type/redshift degeneracy, such an evolution is present in any sample that can be selected from broad-band photometry. Finally, type/redshift degeneracy is the fundamental reason for sample variance in redshift calibration from fields like COSMOS \citep{Lima2008, Amon2018, Hoyle2018}: a criterion based on few broad-band colors selects a mix of galaxy types/redshifts that depends on the large-scale structure present in a small calibration field.

A substantial improvement in redshift calibration therefore requires that the type/redshift degeneracy in wide-field surveys be broken more effectively. While the collection of large, representative samples of faint galaxy spectra remains unfeasible, recent studies indicate that broad-band photometry that covers the full range of optical and near-infrared wavelengths substantially improves the accuracy of redshift calibration \citep{Masters2017,Hildebrandt2018}. This is again illustrated by the lower panel of \autoref{fig:SED}: the additional bands can break type/redshift degeneracies that are present in, e.g., \textit{(g)riz} information. Over large areas and to the depth of upcoming surveys, only few-band photometry is readily available, primarily due to a lack of efficient near-infrared survey facilities (but cf.~KiDS+VIKING, \citealt{Wright2018}). However, in the next decade, imaging surveys such as the Euclid survey \citep{Laureijs2011} and WFIRST \citep{Spergel2013} have been designed to address this lack of near-infrared data.

In this paper, we develop a method that leverages photometric data in additional filters (and with sufficiently low photometric noise), available over a limited area of a survey, to break degeneracies and thus overcome the key limitations of redshift distribution characterization from few-band data. Optical surveys commonly observe some regions more often than the wide-field, and these can be chosen to overlap with auxiliary near-infrared data and spectroscopy. The galaxies observed in these `deep fields' can be grouped into a sufficiently fine-grained set of \textit{phenotypes} based on their observed many-band fluxes \citep{Sanchez2018}. The average density with which galaxies from each phenotype appear in the sky can be measured more accurately from the deep fields than from a smaller spectroscopic sample.

The redshift distribution of a deep-field phenotype can be estimated from a sub-sample for which both the multi-band fluxes and accurate redshifts are available. This could be the result of a future targeted spectroscopic campaign \citep[cf.][]{Masters2015} or, as long as spectroscopic redshifts do not cover all phenotypes, a photometric campaign with high-quality and broad wavelength coverage, to be used with a template fitting method. Members of a phenotype have very similar multi-band colors, giving typically a compact redshift distribution. This substantially reduces redshift biases that might arise from non-representative or incomplete spectroscopic follow-up, sample variance, or from variation of clustering bias with redshift. The larger volume and depth of the deep fields allow the estimation of the density of galaxies from each phenotype in the sky with a lower sample variance, lower shot noise, and higher completeness than would be possible from redshift samples alone. Knowing this density and the multi-band properties of a phenotype and applying the distribution of measurement noise in the wide survey, we can determine the probability that an observation in the wide field originates from a given phenotype. That is, we learn how to statistically break the type/redshift degeneracy at given broad-band flux from a larger sample of galaxies than is possible to obtain accurate redshifts for. In this scheme, the multi-band deep measurements thus mediate an indirect mapping between wide-field measurements and accurate redshifts. 

For the purpose of developing the method in this paper, we will assume that the subset of galaxies with known redshifts (1) is representative at any position in multi-band deep field color space, and (2) has their redshifts characterized accurately. While progress is being made towards achieving this with spectroscopy \citep[e.g.][]{Masters2019}, or many-band photometric redshifts, which show promising performance \citep[at least for a large subset of the source galaxies measured in DES;][]{Laigle2016,Eriksen2019}, substantial work remains to be done on validating this assumption in practical applications of our scheme, and extending its validity to the fainter galaxy samples required by future lensing surveys.

The paper is organized as follows. 
In \autoref{sec:formalism}, we develop the formalism of the method which is tested on a mock galaxy catalog presented in \autoref{sec:Galaxy_catalogs}.
The implementation of the method with self-organizing maps is presented in \autoref{sec:implementation} and the fiducial choices of features and hyperparameters are described in \autoref{sec:fiducial_SOM}. The performance of the method with unlimited samples is assessed in \autoref{sec:capabilities}. We then apply the method to a simulated DES catalog in order to forecast its performance on ongoing and future surveys. The DES Year 3 (Y3), i.e.\ the analysis of the data taken in the first 3 years of DES, targets an uncertainty in the mean redshift of each source tomographic bin $\sigma_{\Delta z}\sim0.01$, which is unmatched for wide-field galaxy samples with comparable data, and a main motivation for this work. The sources of uncertainty and their impacts on a DES Y3-like calibration are characterized in \autoref{sec:sources_uncertainty}. The impact of the DES Y3 weak lensing analysis choices on redshift calibration are assessed in \autoref{sec:refinement}. We describe the redshift uncertainty on a DES Y3-like analysis in \autoref{sec:z_uncertainty_Y3} and explore possible improvements of the calibration in \autoref{sec:possible_improvements}. Finally, we conclude in \autoref{sec:conclusion}. A reader less interested in technical aspects may wish to focus on \autoref{sec:formalism} and \autoref{sec:sources_uncertainty}.

We define three terms used in this paper that have varying uses in the cosmological literature. By \emph{sample variance} we mean a statistical uncertainty introduced by the limited volume of a survey. By \emph{shot noise} we mean a statistical uncertainty introduced by the limited number of objects in a sample. And the term \emph{bias} is used for a mean offset of an estimated quantity from a true one that remains after averaging over many (hypothetical) random realizations of a survey.

\section{Formalism}
\label{sec:formalism}
In this section we develop a method based on galaxy phenotypes to estimate redshift distributions in tomographic bins. The method is applicable to any photometric survey with a similar observation strategy to DES.

Assume two kinds of photometric measurements are obtained over the survey: \textit{wide} data (e.g.~flux or colors), available for every galaxy in the survey, and \textit{deep} data, available only for a subset of galaxies. The dimensionality of the deep data is higher by having flux measurements in more bands of the electromagnetic spectrum. We shall denote the wide data by $\vec{\hat{x}}$ with errors $\vec{\hat{\Sigma}}$. We will call the deep data $\vec{x}$. We will assume noiseless deep data, but confirm that obtainable levels of deep field noise do not affect our conclusions. The deep sample is considered complete in the sense that any galaxy included in the wide data would be observed if its location was within a deep field.

A third sample contains galaxies with confidently known redshifts, $z$. The \textit{redshift} data can be obtained using many-band photometric observations or spectroscopy. In this work, we will assume the redshift sample to be a representative subset of the deep data, with perfect redshift information. This is a fair assumption when the redshift sample is populated with high-quality photometric redshifts (e.g.\ from COSMOS or multi-medium/narrow-band surveys), and the deep sample is complete to sufficiently faint magnitudes. It is, however, a strong assumption for spectroscopic redshifts when matched to a photometric sample with only few observed bands \citep{Gruen2017}. As one increases the wavelength coverage, the assumption of representativeness becomes less problematic: in our scheme, it is only required to hold at each position in deep multi-color space, and thus only for subsets of galaxies with close to uniform type and well-constrained redshift. This is confirmed by the observation that, at a given position in seven-color optical-NIR data space, the dependence of mean redshift on magnitude is small. Once in a discretized cell of the Euclid/WFIRST color space, \citet{Masters2019} quantify the dependence on galaxy brightness as $\sim 0.0029$ change in $\Delta z/(1 + \bar{z})$ per magnitude. Thus despite a selection effect in the spectroscopic survey to only observe a brighter subset of the galaxies at given deep multi-band color, the inferred redshift distribution would still be close to unbiased and representative of the full sample. 

In order to estimate the conditional probability $p(z|\vec{\hat{x}})$, the deep sample can be used as an intermediary between wide-field photometry and redshift.
Redshifts inferred directly from wide measurements with only a small number of bands can be degenerate when distinct galaxy types/redshifts yield the same observables. This is the ultimate reason behind sample variance and selection biases in redshift calibration: the same observed wide-field data can correspond to different distributions of redshift depending on the line of sight or additional selections, e.g.\ based on the success of a spectroscopic redshift determination. Sample variance, shot noise, and selection effects may thus cause the mix of types/redshifts in a redshift sample at given $\vec{\hat{x}}$ to deviate from the mean of a complete sample collected over a larger area. 

The type/redshift degeneracy is mitigated for a deep sample in which supplementary bands and more precise photometry reduce the type mixing at a given point in multi-color space. A tighter relation can in this case be found between redshifts and deep observables. At the same time, the small sample of galaxies with known accurate redshifts can be reweighted to match the density of deep field galaxies in this multi-color space. Because the position in this multi-color space is highly indicative of type and redshift, and because larger, complete samples of deep photometric galaxies can be collected, this reweighting evens out the type/redshift mix of the redshift sample at given wide-field flux \citep{Lima2008}. As a result, sample variance and selection effects present in the redshift sample are reduced. By statistically relating the deep to the wide data that would be observed for the same galaxy, the deep sample enables estimating wide galaxy redshifts with reduced susceptibility to sample variance and selection effects. 

The deep and wide data sets do not necessarily represent the same population of galaxies. Not all galaxies seen in the deep field are detected in the wide field, and for a particular science case, not all the galaxies detected in the wide data are used. Only the ones satisfying some selection, $\hat{s}$, are taken into account. The wide data, $\vec{\hat{x}}$, and its errors, $\vec{\hat{\Sigma}}$, are correlated with other properties that may enter in the selection, $\hat{s}$, such as ellipticity or size. These properties are linked to colors in the deep observations, $\vec{x}$, that are unobserved in the wide data. For example, morphology correlates with galaxy color \citep[e.g.][]{Larson1980, Strateva2001}. Assume one sample, $\hat{s}_0$, selects elliptical galaxies and another, $\hat{s}_1$, selects spiral galaxies. Those two samples will have a different distribution of $\vec{x}$ given $\vec{\hat{x}}$. Therefore, the mapping of $\vec{\hat{x}}$ to $z$ will be different for different science samples, $\hat{s}$. In our case $\hat{s}$ is the selection of observed DES galaxies that end up in our weak lensing source catalog.

A photometric redshift estimator for an individual galaxy is given by
\begin{equation}
\label{p_z_x}
p(z|\vec{\hat{x}},\vec{\hat{\Sigma}},\hat{s}) = \int \vec{dx} \;  p(z|\vec{x}) \; p(\vec{x}|\vec{\hat{x}},\vec{\hat{\Sigma}}, \hat{s}),
\end{equation}
where we marginalized over the deep measurements, $\vec{x}$. In \autoref{p_z_x}, we assume that $p(z|\vec{x},\vec{\hat{x}}, \vec{\hat{\Sigma}}, \hat{s})=p(z|\vec{x})$. The validity of this assumption in our scheme is tested in \autoref{sec:choice_hyperparam}. For an ensemble of $N$ galaxies, an estimator \citep[see][for a discussion]{Malz2018}
of the redshift distribution, ${d}N/\mathrm{d}z\propto p(z)$, is given by
\begin{equation}
\label{p_z_s}
p(z|\hat{s}) = \frac{1}{N} \sum_{i=1}^N \int \vec{dx}\; p(z|\vec{x}) \; p(\vec{x}|\vec{\hat{x}}_i,\vec{\hat{\Sigma}}_i, \hat{s}),
\end{equation}
where the contribution of each galaxy is summed. The conditional probability distributions $p(z|\vec{x})$ and $p(\vec{x}|\vec{\hat{x}},\vec{\hat{\Sigma}}, \hat{s})$ must be learned but this might be infeasible because the variables $\vec{x}$ and $\vec{\hat{x}}$ are continuous and multidimensional. In DES for example, $\vec{\hat{x}}$ has 4 dimensions and $\vec{x}$ can have up to 9 dimensions. To overcome the problem of a complicated mapping from a 4-dimensional space to a 9-dimensional space, we discretize those spaces. We can consider that $\vec{\hat{x}}$ and $\vec{x}$ are observable characteristics of an underlying variable that defines a galaxy's SED and redshift. We can also consider that galaxies with similar $\vec{x}$ will have similar redshift. It is therefore reasonable to cluster the deep data, $\vec{x}$, in discretized deep cells, $c$, and the wide data, $\vec{\hat{x}}$, in discretized wide cells, $\hat{c}$. Each deep cell, $c$, represents a galaxy phenotype. For our purpose, all galaxies in the same cells have the same observables, and some underlying variables -- the galaxy's `genes' -- determine these observables. We therefore call this method \textit{phenotypic} redshifts (hereafter pheno-$z$).

Using a variety of clustering methods, a galaxy with wide measurement, $\vec{\hat{x}}$, and error, $\vec{\hat{\Sigma}}$, can be assigned to a unique cell $\hat{c}$. Similarly, a galaxy with deep measurement, $\vec{x}$, can be assigned to a unique cell $c$. Those two assignments need not necessarily be produced using the same method. This reduces, for each galaxy, the continuous multi-dimensional vectors $\vec{x}$ and $\vec{\hat{x}}$ (with their errors) to two integers. In this scheme, given a selection of galaxies, $\hat{s}$, each wide cell, $\hat{c}$, has a redshift distribution:
\begin{equation}
p(z|\hat{c},\hat{s}) = \sum_{c} p(z|c) \; p(c|\hat{c}, \hat{s}),
\end{equation}
which is analogous to \autoref{p_z_x}, and where we have assumed $p(z | c, \hat{c}, \hat{s}) = p(z | c)$. For an ensemble of galaxies with selection $\hat{s}$, its redshift distribution is given by 
\begin{equation}
\label{p_z_hats}
p(z|\hat{s}) = \sum_{c, \hat{c}} p(z|c) \; p(c|\hat{c}, \hat{s}) \; p(\hat{c} | \hat{s}),
\end{equation}
where $p(\hat{c} | \hat{s})$ is the fractional assignment of galaxies to cell $\hat{c}$. This is the discretized version of \autoref{p_z_s}. The quantity $p(c|\hat{c}, \hat{s})$ will be called the transfer function and is specific to a science sample, $\hat{s}$. In our scheme, it is estimated from galaxies for which both deep and (possibly simulated) wide observations are available (hereafter the \textit{overlap} sample). 

It is useful to consider the sources of bias, sample variance and shot noise inherent to this scheme. We note that the transfer function, $p(c|\hat{c}, \hat{s})$, is proportional to $p(\hat{c}, \hat{s}|c)p(c)$ per Bayes' theorem. Sample variance and shot noise in the estimated redshift distribution can thus be caused by the limited volume and count of the deep galaxy sample, introducing noise in $p(c)$, or by the limited overlap sample, introducing noise in $p(\hat{c}, \hat{s}|c)$. If it were the case that $c$ uniquely determines the redshift, there would be no variance in $p(z|c)$ as long as a redshift is known for at least one galaxy from each deep cell $c$. For a large enough sample of bands in the deep data, $p(z|c)$ is indeed much narrower than $p(z|\hat{c},\hat{s})$. Sample variance and shot noise due to limited redshift sample \citep[as estimated in][]{Bordoloi2010,Gruen2017} can therefore be reduced in this scheme.

Biases could be introduced by the discretization. \autoref{p_z_hats} is an approximation to \autoref{p_z_s} that breaks when the redshift distribution varies within the confines of a $c$ cell in a way that is correlated with $\hat c$ or $\hat s$. One of the purposes of this paper is to test the validity of this approximation (see \autoref{sec:choice_hyperparam} and \autoref{sec:capabilities}).

\subsection{Source bin definition}
\label{sec:bin}
To perform tomographic analysis of lensing signals \citep[see e.g.][]{Hu1999}, galaxies must be placed into redshift bins. The pheno-$z$ method, developed in \autoref{sec:formalism}, to estimate redshift distributions is independent of the bin assignment method. The simple algorithm presented in this section is aimed to assign galaxies to one bin uniquely, with little overlap of bins, such that each contains roughly the same number of galaxies.

To achieve the binning, two samples are used: the redshift sample for which we have deep measurements and the overlap sample for which we have deep and wide measurements. The redshift sample is assigned to cells $c$ and the mean redshift, $\bar{z}_c$, in each of those cells is computed. Each galaxy in the overlap sample is assigned to cells $c$ and $\hat{c}$. The fractional occupation of those cells $f_{c}=p(c|\hat{s})$ and $f_{\hat{c}}=p(\hat{c}|\hat{s})$ are such that $\sum_c f_{c} = \sum_{\hat{c}} f_{\hat{c}} = 1$. All galaxies in the redshift sample are used whereas only the galaxies respecting the selection criteria enter the overlap sample.

We wish to assign galaxies to $N_{\mathrm{bin}}$ tomographic bins ($N_{\mathrm{bin}}=5$ in this work). The first step consists of assigning cells $c$ to tomographic bins $B$. The cells $\hat{c}$ are then assigned to tomographic bins $\hat{B}$ using the transfer function. The procedure is the following:
\begin{enumerate}
\item Cells $c$ are sorted by their mean redshift, $\bar{z}_c$, in ascending order. Cells are assigned to bin $B$ until $\sum_{c\in B}f_c \geq \frac{1}{N_{\mathrm{bin}}}$, where $B=1,...,N_{\mathrm{bin}}$. We discuss the impact of cells lacking redshift information in \autoref{sec:empty_deep_som_cells}.
\item Each cell $\hat{c}$ is assigned to a bin $\hat{B}$ by finding which bin $B$ it has the highest probability of belonging to through $p(c|\hat{c},\hat{s})$:
\begin{equation}
\hat{B} = \argmax_B p(B|\hat{c},\hat{s}) = \argmax_B \sum_{c\in B} p(c|\hat{c}, \hat{s}).
\end{equation}
\item Individual galaxies are assigned to bin $\hat{B}$ based on their wide cell assignment, $\hat{c}$.
\end{enumerate}

Once the bins are computed, the final quantity of interest, the redshift distribution in bin $\hat{B}$ can be inferred:
\begin{equation}
\label{n_i_z}
p(z|\hat{B},\hat{s}) = \sum_{\hat{c}\in \hat{B}}\sum_c p(z|c) \; p(c|\hat{c}, \hat{s}) \; p(\hat{c} | \hat{s}).
\end{equation}
Throughout this work, we will use \autoref{n_i_z} to estimate redshift distributions in tomographic bins.

\section{Simulated DES galaxy catalogs}
\label{sec:Galaxy_catalogs}

In this work we use simulated galaxy catalogs designed to mimic observational data collected with the Dark Energy Camera \citep[DECam;][]{Honscheid2008,Flaugher2015}. DECam is a 570 megapixel camera with a 3 deg$^2$ field-of-view, installed at the prime focus of the Blanco 4-m telescope at the Cerro Tololo Inter-American Observatory (CTIO) in northern Chile. In addition, we mimic data by surveys conducted with the Visible and Infrared Survey Telescope for Astronomy \citep[VISTA;][]{Emerson2004}, a 4-m telescope located at ESO's Cerro Paranal Observatory in Chile and mounted with a near infrared camera, VIRCAM (VISTA InfraRed CAMera), which has a 1.65 degree diameter field-of-view containing 67 megapixels. Both the underlying real and simulated galaxy samples are described below.

\subsection{The Dark Energy Survey}
The DES is an ongoing ground-based wide-area optical imaging survey which is designed to probe the causes of cosmic acceleration through four independent probes: Type Ia supernovae, baryon acoustic oscillations, weak gravitational lensing, and galaxy clusters. After six years of operations (2013-2019), the survey has imaged about one-eighth of the total sky. DES has conducted two distinct multi-band imaging surveys with DECam: a $\sim 5000\; \text{deg}^2$ wide-area survey in the \textit{grizY} bands\footnote{While there are DES $Y$ band flux measurements available, due to their lower depth we will not use it.} and a $\sim 27\; \text{deg}^2$ deep supernova survey observed in the \textit{griz} bands. The deep supernova survey overlaps with the VISTA \textit{YJHKs} bands measurements, and we have obtained $u$ band imaging of these fields using DECam.

\subsubsection{DES Year 3 samples}
\label{sec:DES_Y3_samples}
The pheno-$z$ method requires four samples to estimate redshift distributions in tomographic bins using \autoref{n_i_z}. The following datasets will be used in the DES Y3 analysis:

\begin{enumerate}
\item \textit{Deep sample}: In DES, \textit{ugriz} deep photometry is obtained in 10 supernova fields ($\sim27\;\mathrm{deg}^2$), as well as in the COSMOS field ($\sim2\;\mathrm{deg}^2$).  Some of those fields overlap with deep VISTA measurements in the \textit{YJHKs} bands from the UltraVista survey \citep{McCracken2012} for COSMOS and from the VISTA Deep Extragalactic Observations \citep[VIDEO;][]{Jarvis2013} survey for the supernova fields. \autoref{tab:vista_bands} summarizes the overlap between the DES deep photometry and the UltraVISTA (COSMOS) and VIDEO fields.  The VISTA \textit{Y} band is available in three of the four fields where \textit{JHKs} bands are available. Including the \textit{Y} band reduces the total available area from 9.93 to $7.99\;\mathrm{deg}^2$. We examine the trade-off between area and \textit{Y} band in \autoref{sec:Yband_area}. In the especially deep supernova fields C3 and X3, DECam \textit{griz} is at an equivalent depth of at least $200\times90s$, while the regular depth supernova field E2 has at least $80\times90s$ exposures, compared to a final wide-field DES exposure time of $10\times90$s.
\begin{table}
\caption{Overlap between DES deep \textit{ugriz} measurements and VISTA \textit{(Y)JHKs} measurements. E2, X3 and C3 refer to the DES supernova fields and $\mathcal{C}$ to the COSMOS field. There are VISTA $Y$ band measurements in COSMOS, E2 and X3 fields but not in the C3 field. The last column shows the reduced deep field area when using \textit{Y} band. \label{tab:vista_bands}}
\ra{1.2}
\centering
\resizebox{\linewidth}{!}{%
\begin{tabular}{@{}cccccS[table-format=2.2]@{\hspace{0.7\tabcolsep}}S[table-format=2.2]@{}}
\toprule
  & \multirow{2}{*}[-3pt]{\textbf{E2}}  & \multirow{2}{*}[-3pt]{\textbf{X3}}  & \multirow{2}{*}[-3pt]{\textbf{C3}}& \multirow{2}{*}[-3pt]{$\boldsymbol{\mathcal{C}}$} & \multicolumn{2}{c}{\textbf{Total}} \\ \cmidrule(lr){6-7} 
 & & & & & \multicolumn{1}{c}{\textit{JHKs}}  & \multicolumn{1}{c}{\textit{YJHKs}}   \\  \cmidrule(lr){2-5} \cmidrule(lr){6-6} \cmidrule(lr){7-7} 
Overlap $[\mathrm{deg}^2]$                      & 3.32                    & 3.29                         & 1.94      & 1.38              &   9.93   &   7.99         \\ \bottomrule \addlinespace[2pt]
\end{tabular}
}
\end{table}

\item \textit{Redshift sample}: The galaxies in the COSMOS field can be assigned a redshift either by using many-band photo-$z$s \citep{Laigle2016} which are available for all galaxies or by using spectroscopic redshifts available for a subset of galaxies as long as this subset is representative of the color space spanned by the deep sample and is a fair sample of $z$ within any given cell.\footnote{See section 5.3 of \citet{Masters2015} for references of spectroscopic data available and \citet{Masters2017} for the Complete Calibration of the Color-Redshift Relation (C3R2) survey, which aims at increasing the representativeness of the spectroscopic data available.} The use of many-band photo-$z$ is advantageous as it avoids selection effects commonly present in spectroscopic samples. The COSMOS catalog provides photometry in 30 different UV, visible and IR bands. For each galaxy, the probability density function (PDF) $p^{\mathrm{C30}}(z)$ of its redshift given its photometry is computed using the LePhare template-fitting code \citep{Arnouts1999,Ilbert2006}. The typical width of $p^{\mathrm{C30}}(z)$ for DES sources is $\approx 0.01(1+z)$ and the typical catastrophic failure rate is 1\%. The available overlap between DES deep and UltraVista for which many-band photo-$z$s are available is $1.38\;\mathrm{deg}^2$ and contains $\sim 135,000$ galaxies.

\item \textit{Overlap sample}: This sample comprises objects for which deep and (possibly synthetic) wide measurements are available. In practice, we will use \textsc{Balrog} \citep{Suchyta2016}, a software package that paints synthetic galaxies into observed images in order to render wide measurements and assess selection effects. The deep sample galaxies are painted several times over the whole DES footprint to produce a number of realizations of each deep field galaxy under different observing conditions and noise realizations. The shape measurement pipeline is also run on those fake galaxies yielding only objects that would end up in the shape catalog after its cuts on e.g.~observed size and signal-to-noise ratio. This method produces a sample of galaxies with deep and wide measurements with the same selection as the real source galaxies used in the weak lensing analysis.
\item \textit{Wide sample}: All galaxies that are selected for the shape catalog are included in the wide sample. These are the galaxies for which we infer the redshift distributions.
\end{enumerate}
\subsection{Buzzard mock galaxy catalog}
\label{sec:buzzard}
We use the \textsc{Buzzard-v1.6} simulation, a mock DES catalog created from a set of dark-matter-only simulations (a detailed description of the simulation and the catalog construction can be found in \citealt{MacCrann2018,DeRose2019}, Wechsler et al.~in preparation). \textsc{Buzzard-v1.6} is constructed from a set of 3 $N$-body simulations run using \textsc{l-gadget2}, a version of \textsc{gadget2} \citep{Springel2005} modified for memory efficiency. The simulation box sizes ranged from 1 to 4 $h^{-1}$Gpc. Light cones from each box were constructed on the fly.

Galaxies are added to the simulations using the Adding Density Dependent GAlaxies to Light-cone Simulations algorithm (\textsc{addgals}; \citealt{DeRose2019}, Wechsler et al.~in preparation). This algorithm pastes galaxies onto dark matter particles in an $N$-body simulation by matching galaxy luminosities with local dark matter densities. This method does not use dark matter host haloes, which are commonly unresolved for the galaxies and simulations used here. SEDs from a training set of spectroscopic data from SDSS DR7 \citep{Cooper2011} are assigned to the simulated galaxies to match the color-environment relation. These SEDs are integrated in the DES pass bands to generate \textit{ugriz} magnitudes and in the VISTA pass bands to generate \textit{YJHKs} magnitudes (see \autoref{fig:SED}). Galaxy sizes and ellipticities are drawn from distributions fit to SuprimeCam \textit{i} band data \citep{Miyazaki2002}. Galaxies are added to the simulation to the projected apparent magnitude limit of the final DES dataset out to redshift $z = 2$. 

The use of SDSS spectra means that the SEDs assigned in Buzzard are limited to bright or low redshift galaxies. In contrast to template fitting methods, the resulting lack of SED evolution with redshift is not a major concern for testing our scheme: there is no assumption made that the same underlying SED exists at different redshifts to produce different but related phenotypes. Changes in galaxy SEDs with redshift could, however, introduce a different degree of type-redshift degeneracy as seen in the mock catalogs, which is a caveat in transferring our findings to real data and should be tested by comparing e.g.~the scatter of redshift within deep SOM cells between mock and data. Note also that as the rest-frame UV part of the SEDs is not recorded by the SDSS spectra, the spectroscopic data must be extrapolated to produce the optical colors at $z\gtrapprox1.5$. The lack of informative colors above this redshift motivates the redshift cut in the samples described below. This may lead to an underestimate of the uncertainty in high-redshift tails. Only a small fraction of observable galaxies in those parts of wide-field color-magnitude space that provide sufficiently constrained redshift distributions for lensing use in DES Y1 (the analysis of the data taken in the first year of DES) and Y3 lie at $z>1.5$ (cf.~e.g.~\citealt{Hoyle2018}), but this will change in deeper future data sets. In addition, our error estimates assume that the overall population density and signal-to-noise distribution of Buzzard galaxies as a function of redshift mimics the data, which is only approximately true.

\subsubsection{Mock samples in \textsc{Buzzard}}

Using the \textsc{Buzzard} simulated catalog, we construct the 4 samples described in \autoref{sec:DES_Y3_samples} to test and refine our method. In the simulations, galaxies are assigned a true redshift $z_{\mathrm{true}}$ and a true flux in each band. Observed fluxes are derived for each galaxy depending on its position on the sky. The error model used is tailored to match DES wide-field observations. In the following, in order for the simulations to mimic the real data, we will use the simulated true and observed information as deep and wide information, respectively.

We use the true redshift for the redshift sample and to compare our inferred redshift distributions to the true ones. We reiterate that this assumption is likely valid for the brighter subset of existing spectroscopic and many-band photometric redshift samples only \citep{Laigle2016,Eriksen2019,Masters2019}, and must be validated when applying any empirical redshift calibration scheme in practice. The simulated true fluxes without errors are used as the deep measurements. This is justified by the significantly longer integrated exposure time of the deep fields relative to the wide survey. We have validated that flux errors at least five times smaller than those that define the limiting magnitudes in Appendix \ref{app:softB}, applied to the simulated deep field catalogs, do not appreciably affect our redshift calibration. We are actively studying the interplay of deep field flux errors and deep SOM calibration on Y3 data (Myles et al.~in preparation). When the measurements are considered noiseless, $\vec{\Sigma}$ is taken to be the identity matrix when training the self-organizing map (SOM; \autoref{sec:SOM}) or when assigning galaxies to the SOM. To mimic the deep and redshift samples, two cuts are performed on the \textsc{Buzzard} catalog. Only galaxies with true redshift $z_{\mathrm{true}}<1.5$ and true magnitude in \textit{i} band $m_{\mathrm{true},\,i}<24.5$ are kept in both samples. The hard boundary at  $z_{\mathrm{true}}=1.5$ has its drawbacks (namely, in the reliability of the error estimate of the highest redshift bin) but it ensures that the colors are sufficiently correct which is not the case at high $z_{\mathrm{true}}$ in the simulations. The redshift sample is expected to be representative of the deep sample at any point in deep multi-color space, as would be the case for COSMOS multi-band redshifts.

\begin{figure}
\centering
\includegraphics[width=\linewidth]{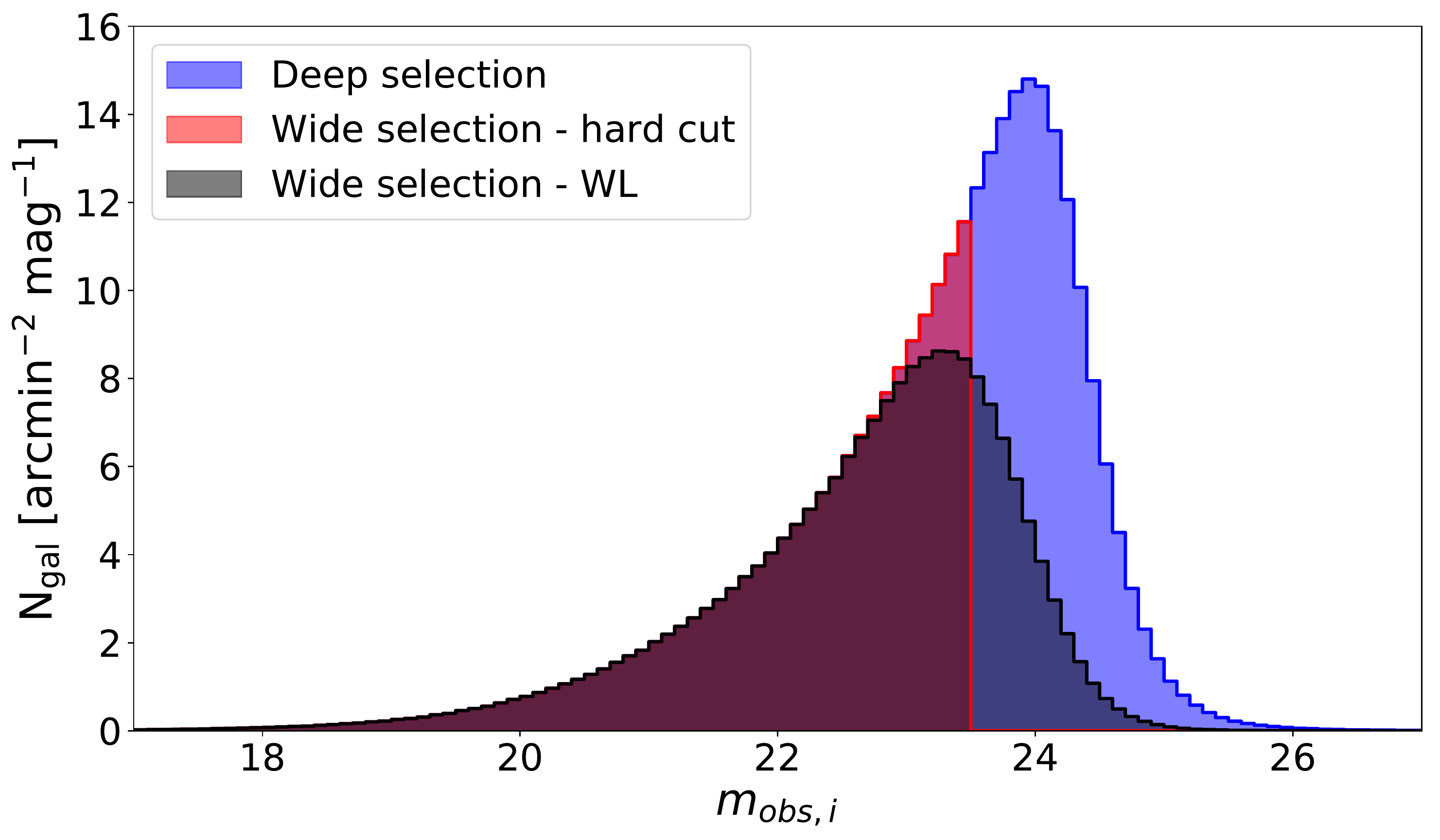}
\caption{Distribution of the observed \textit{i} band magnitude of three selections, $\hat{s}$, applied to all galaxies in a \textsc{Buzzard} tile ($\sim53\;\mathrm{deg}^2$). The deep selection (blue) is $z_{\mathrm{true}}<1.5$ and $m_{\mathrm{true},\,i}<24.5$. The hard cut selection (red) is the deep selection plus $m_{\mathrm{obs},\,i}<23.5$. The weak lensing (WL) selection (black) uses a more complex criterion based on the size of the galaxy and the limiting magnitude of the DES Y3 survey (see \autoref{sec:wl-selection}).\label{fig:samples_mag_i}}
\end{figure}

For the wide sample, we want galaxies whose properties are similar to the ones of galaxies we would use in the DES Y3 cosmology analysis. These galaxies are a subset of all DES Y3 observed galaxies in the wide survey. This subset is the result of both easy-to-mock selections (galaxies with observed magnitude in some band lower than a threshold) and difficult-to-mock selections (cut galaxies which would fail in the shape measurement algorithm). We use the simple selection criterion of observed magnitude in \textit{i} band $m_{\mathrm{obs},\,i}<23.5$ to create the wide sample. A more refined selection, which depends on a galaxy's size and the limiting magnitude in \textit{r} band of the survey at its position on the sky, is used at the end of this work to check the robustness of our uncertainties estimates (see \autoref{sec:wl-selection}). The distributions of the observed \textit{i} band magnitude of those three selections applied to all galaxies in a \textsc{Buzzard} tile ($\sim53\;\mathrm{deg}^2$) is shown in \autoref{fig:samples_mag_i}.

To obtain the overlap sample, we apply the same selection criterion as the one used for the wide sample. To mimic the \textsc{Balrog} algorithm, we can take the galaxies in the deep sample, randomly select positions over the full Y3 footprint and run the error model at their position to obtain a noisy version of the galaxy. This allows us to have multiple wide realizations of the same galaxy. Only galaxies respecting the wide selection criterion are then selected. The galaxies in the deep sample can be spliced a certain number of times giving an overlap sample of variable size.

In summary, only two selections are performed: a deep and a wide. The first is applied to the redshift and deep samples, the second to the overlap and wide samples. 

\section{Implementation}
\label{sec:implementation}
As stated in \autoref{sec:formalism}, a wide variety of clustering methods can be used to achieve the assignment of wide and deep data to cells $\hat{c}$ and $c$, respectively. In this work, we use self-organizing maps to obtain both. This choice is motivated by the visual representation offered by this method which helps interpretation and debugging. Also, recent works have shown the capabilities of this algorithm in dealing with photo-$z$s \citep[see e.g.][]{CarrascoKind2014,Masters2015}.

\subsection{Self-organizing maps}
\label{sec:SOM}
A self-organizing map (SOM) or Kohonen map is a type of artificial neural network that produces a discretized and low-dimensional representation of the input space. Since its introduction by \citet{Kohonen1982}, this algorithm has found a large range of scientific applications (see e.g.\ \citeauthor{Kohonen2001} \citeyear{Kohonen2001}). In astronomy, SOMs have been used in different classification problems: galaxy morphologies \citep{Naim1997}, gamma-ray bursts \citep{Rajaniemi2002} or astronomical light curves \citep{Brett2004}. More recently, this method has been used to compute photo-$z$s: single photo-$z$ estimator \citep{Geach2012,Way2012} and the full redshift PDF \citep{CarrascoKind2014}. It has also been used to characterize the color-redshift relation to determine relevant spectroscopic targets \citep{Masters2015} necessary to meet the photo-$z$ precision requirements for weak lensing cosmology for the Euclid survey \citep{Laureijs2011}. This work resulted in the Complete Calibration of the Color-Redshift Relation \citep[C3R2;][]{Masters2017} survey, which targets missing regions of color space.

The SOM algorithm is an unsupervised method (the output variable, in our case the redshift, is not used in training) which produces a direct mapping from the input space to a lower dimensional grid. The training phase is a competitive process whereby cells of the map (more commonly called neurons or nodes) compete to most closely resemble each galaxy of the training data, until the best match is assigned as that galaxy's phenotype. The SOM is a type of non-linear principal component analysis which preserves separation, i.e.\ distances in input space are reflected in the map.

Consider a training sample of $n$ galaxies. For each galaxy we can build an $m$-dimensional input vector $\vec{x} \in \mathbb{R}^m$ made of measured galaxy attributes such as magnitudes, colors or size (but not the redshift). A SOM is a set of $C$ cells arranged in an $l$-dimensional grid that has a given topology. Here we consider two-dimensional square maps with periodic boundary conditions (the map resembles a torus). Each cell is associated to a weight vector $\vec{\omega_k} \in \mathbb{R}^m$, where $k=1,...,C$, that lives in the same space as the input vectors.

The iterative training process starts by initializing the weight vectors either randomly or by sampling from the input data. At each step of the algorithm, a random galaxy from the training sample is presented to the map. The cells whose weight vector is the closest to the galaxy's vector is the Best Matching Unit (BMU). To define closeness, we use the $\chi^2$ distance:
\begin{equation}
\label{chi_2_dist}
d^2(\vec{x}, \vec{\omega_k}) =  (\vec{x}-\vec{\omega_{k}})^\top\vec{\Sigma^{-1}}(\vec{x}-\vec{\omega_{k}}),
\end{equation}
where $\vec{\Sigma}$ is the covariance matrix for the training vector, $\vec{x}$. The cell minimizing this distance is the BMU and is denoted by the subscript $b$:
\begin{equation}
c_b = \argmin_k d(\vec{x}, \vec{\omega_k}).
\end{equation}
To preserve the topology of the input space, not only the BMU is identified as being similar to the training galaxy but also its neighborhood. Therefore, these cells are all modified to more closely resemble the training galaxy. To update the weights, the following relation is used for all weights $\vec{\omega_k}$: 
\begin{equation}
\vec{\omega_k}(t+1) = \vec{\omega_k}(t) + a(t)\;H_{b,k}(t)\;[\vec{x}(t)-\vec{\omega_k}(t)],
\end{equation}
where $t$ represents the current time step in the training. The learning rate function, $a(t)$, encodes the responsiveness of the map to new data. It is a monotonically decreasing function of the time step: the map gets gradually less sensitive to new training vectors. The learning rate function in terms of the total number of training steps, $t_{\mathrm{max}}$, has the following form:
\begin{equation}
a(t) = a_0^{t/t_{\mathrm{max}}},
\end{equation}
where $a_0\in[0,1]$. The size of the BMU's neighborhood affected by the new training vector also decreases as a function of time steps. This is encoded in the neighborhood function $H_{b,k}(t)$ which is parametrized as a Gaussian kernel centered on the BMU:
\begin{equation}
H_{b,k}(t) = \exp[-D^2_{b,k}/\sigma^2(t)].
\end{equation}
The distance between the BMU, $c_b$, and any cell on the map, $c_k$, is the Euclidean distance on the $l$-dimensional map:
\begin{equation}
D^2_{b,k} = \sum_{i=1}^l (c_{b,i}-c_{k,i})^2,
\end{equation}
where we account for periodic boundary conditions. The width of the Gaussian kernel is parametrized as
\begin{equation}
\sigma(t) = \sigma_s^{1-t/t_{\mathrm{max}}}.
\end{equation}
Its starting value $\sigma_s$ should be large enough such that most of the map is initially affected. As the training progresses the width shrinks until only the BMU and its closest neighbors are significantly affected by new data.

\subsection{Assignment of galaxies to cells}
After the training has converged, we use the $\chi^2$ distance introduced in \autoref{chi_2_dist} to assign galaxies to a cell.
Given its input vector, $\vec{x}$, and its covariance matrix, $\vec{\Sigma}$, a galaxy has a probability of belonging to cell $c$ given by
\begin{equation}
\label{assignment_som}
-2\ln p(c|\vec{x}, \vec{\Sigma}) =  (\vec{x}-\vec{\omega_{c}})^\top\vec{\Sigma^{-1}}(\vec{x}-\vec{\omega_{c}}) + \mathrm{const.},
\end{equation}
where $\omega_{c}$ is the weight vector of cell $c$. In this paper, the deep measurements are considered noiseless, while the wide measurements have noise. When the measurement is considered noiseless, the identity is used for the covariance matrix, $\vec{\Sigma}$. When the measurement has noise, we compute the full inverse covariance matrix in \autoref{assignment_som}. We present how to calculate the inverse covariance matrix in Appendix \ref{app:covariance}. For computational efficiency and tractability, we would like to keep a single integer for each galaxy instead of a vector in $\mathbb{R}^C$ where C is the number of cells. To this end, we keep only the cell which maximizes the above probability.

\subsection{Scheme implementation}

\begin{figure*}
\centering
\includegraphics[width=\linewidth]{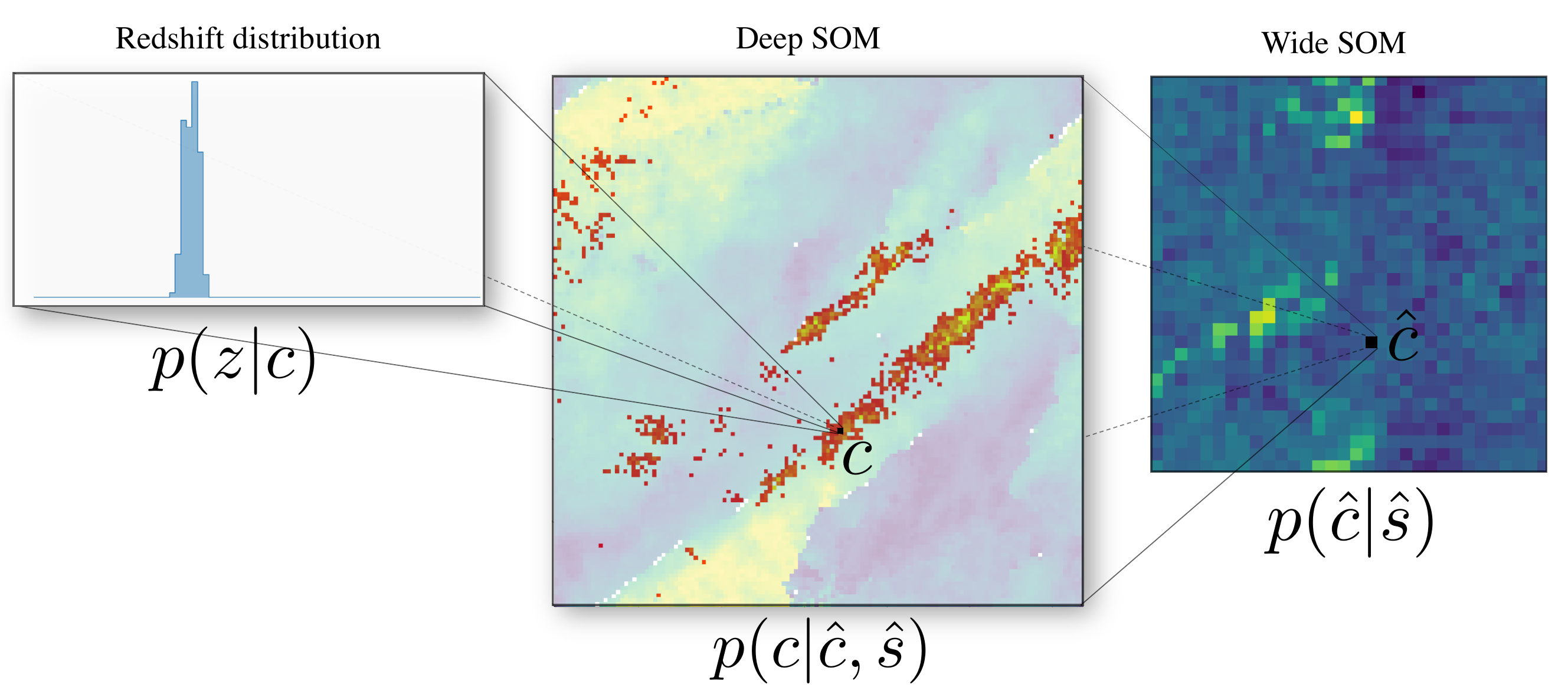}
\caption{The pheno-$z$ scheme using self-organizing maps (SOMs). This illustration depicts the estimation of redshift distributions using \autoref{n_i_z}. The Wide and Deep SOMs are trained using the wide and deep samples, respectively. The term $p(z|c)$ is computed using the assignment of the redshift sample to the Deep SOM. For each cell $c$, a normalized redshift histogram is computed using the galaxies assigned to the cell. The transfer function, $p(c|\hat{c},\hat{s})$, is obtained by assigning the overlap sample to both the Wide and Deep SOMs. For all galaxies assigned to a cell $\hat{c}$, the probability of belonging to any cell $c$ can be computed. The last piece of our scheme, $p(\hat{c}|\hat{s})$, is the fractional occupation of the wide sample in the Wide SOM.}
\label{fig:2som_scheme}
\end{figure*}

The computation of the redshift distributions with the pheno-$z$ scheme (see \autoref{n_i_z}) is depicted in \autoref{fig:2som_scheme}. To compute $p(z|c)$, a `Deep SOM' is trained using all galaxies in the deep sample. A redshift distribution can be computed for each SOM cell $c$ by assigning the redshift sample to the Deep SOM. A second SOM, the `Wide SOM', is trained on the wide sample. Assignment of the galaxies in this sample to the Wide SOM yields $p(\hat{c}|\hat{s})$. The transfer function, $p(c|\hat{c}, \hat{s})$, is computed by assigning the galaxies in the overlap sample to the Deep and the Wide SOMs. The tomographic bins are obtained using the procedure described in \autoref{sec:bin} using the assignment of the redshift and overlap samples to the Deep SOM as well as the transfer function, $p(c|\hat{c}, \hat{s})$.

In our scheme, three probability distributions must be obtained to compute \autoref{n_i_z}. We need to know the probability that a galaxy ends up in wide cell, $\hat{c}$, if it passes selection $\hat{s}$.
This is obtained as the fractional occupation of $\hat{c}$ by the sample of interest. We can make use of our single cell assignment to compute it: 
\begin{equation}
p(\hat{c}|\hat{s})=\frac{1}{n_{\hat{s}}} \sum_{i\in \hat{s}} \delta_{\hat{c}, \hat{c}_i},
\end{equation}
where $n_{\hat{s}}$ is the number of galaxies in the sample of interest, $\delta$ is the Kronecker delta and $\hat{c}_i$ is a number representing the cell that maximizes the probability given in \autoref{assignment_som} for the \textit{i}th object in the sample.

The second necessary piece of our scheme is the transfer function, $p(c|\hat{c},\hat{s})$, which characterizes the mapping between wide and deep measurements. Using the definition of conditional probability, we can define it as 
\begin{equation}
\label{transfer_matrix_bayes}
p(c|\hat{c},\hat{s}) = \frac{p(c,\hat{c}|\hat{s})}{p(\hat{c}|\hat{s})}.
\end{equation}
This transfer function is the fractional occupation of deep cell, $c$, given wide cell, $\hat{c}$. Galaxies of the overlap sample are assigned to cells $c$ and $\hat{c}$ based on their deep and wide information, respectively. 
To compute \autoref{transfer_matrix_bayes}, we count the number of instances of the unique combination $(c,\hat{c})$ and divide it by the number of instances of $\hat{c}$. The transfer function becomes:
\begin{equation}
\label{transfer_matrix_assign_unique}
p(c|\hat{c},\hat{s}) = \frac{\sum_{i\in\hat{s}}\delta_{c,c_i}\delta_{\hat{c},\hat{c_i}}}{\sum_{i\in\hat{s}}\delta_{c,c_i}},
\end{equation}
where $c_i$, $\hat{c_i}$ are the best matching deep and wide cells, respectively, of the \textit{i}th galaxy in the overlap sample which has selection $\hat{s}$. This overlap sample can be obtained using either actual galaxies which are measured in both the deep and the wide survey or artificial wide-field measurement of the deep sample as discussed in \autoref{sec:DES_Y3_samples}.

The last piece of our scheme is the redshift distribution $p(z|c)$ of deep cell $c$. We use the assignment of the redshift sample to the deep cells $c$. For each cell $c$, we compute $p(z|c)$ as a normalized redshift histogram with bin spacing $\Delta z = 0.02$. This resolution is sufficient since we only use combinations of these histograms, corresponding to wide-field bins with relatively wide redshift distributions, for our metrics.
 
\label{sec:implementation_Y3}

\section{Fiducial SOMs}
\label{sec:fiducial_SOM}
We must choose the features (\autoref{sec:choice_features}) used to train the SOMs as well as their hyperparameters (\autoref{sec:choice_hyperparam}), i.e.\ parameters whose values are not learned during the training process. Intuition guides the search for the best parameters but empirical evidence settles the final choices. The choice of the number of cells for both SOMs is specific to the samples available in DES Y3.

\subsection{Choice of features}
\label{sec:choice_features}
A SOM needs input vectors, $\vec{x}$, on which it is trained. The available data consist of flux measurements, $f_x$, in a set of electromagnetic bands ($x = u, g, r, ...$). Using those raw fluxes would not be optimal as the value in each band is highly correlated with the overall luminosity of the galaxy, which spans several orders of magnitude. This would result in an overweighting of the brightest or the faintest galaxies. A common choice to overcome this problem is to use magnitudes:
\begin{equation}
m_x = m_{0,x} - 2.5 \log_{10} f_x,
\end{equation}
for the zeropoint $m_{0,x}$ in $x$ band. The drawback of this method is that some faint galaxies will have a zero -- or even negative -- measured flux in some bands. Those measurements are undefined in the magnitude system. Removing those faint objects is unacceptable as it would introduce an additional selection that may bias cosmological analyses. Preserving the information about a galaxy's SED contained in the non-measurement of some band is not easily achievable using the magnitude system. 

We instead adopt an inverse hyperbolic sine transformation of flux known as `luptitude' after \citeauthor{Lupton1999} (\citeyear{Lupton1999}):
\begin{equation}
\label{luptitutude}
\mu_x=\mu_0 - a\sinh^{-1}\left(\frac{f_x}{2b}\right).
\end{equation}
The zeropoint is $\mu_0=m_0-2.5\log b$, $a=2.5 \log e$ and $b$ is a softening parameter that sets the scale at which
luptitudes transition between logarithmic and linear behavior. For bright galaxies (large $f_x$), luptitudes behave like magnitudes whereas for faint galaxies (small $f_x$), they behave like fluxes. Zero or negative fluxes are well defined with this parametrization which allows us to avoid throwing away any galaxy. Luptitudes properly manage both the bright and faint ends of the luminosity function. Our analysis is robust to the choice of softening parameter $b$ which is discussed in Appendix \ref{app:softB}.

Luptitudes could be used as entries of the input vector, $\vec{x}$, but for a sufficient set of measured colors we expect most of the information regarding redshift to lie in the shape of the SED. The flux of a galaxy in this case is only a weak prior on its redshift. We find that in practice the addition of total flux (or magnitude) to the deep SOM does not improve the performance of the algorithm. Ratios of fluxes (or equivalently difference of magnitudes) appear to encode the most relevant information to discriminate redshifts and types. Similarly to color, which is a difference of magnitudes, we can define `lupticolor' which is a difference of luptitudes. For high signal-to-noise ratios, a lupticolor is equivalent to the ratio of fluxes. For noisy detections, it becomes the preferable flux difference.

Our tests show that adding a luptitude to the input vector of the Deep SOM slightly decreases the ability of the method to estimate the redshift distributions whereas for the Wide SOM, it improves it. The difference lies in the number of bands available. The deep input vector has 8 lupticolors which are enough to characterize the redshift of the galaxy. For the wide input vector with only three lupticolors, the luptitude adds information and helps break degeneracies at low redshift.

The input vector of the Deep SOM is chosen to be a list of lupticolors with respect to the luptitude in \textit{i} band:
\begin{equation*}
\vec{x} = (\mu_{x_1}{-}\mu_i, ..., \mu_{x_8}{-}\mu_i),
\end{equation*}
where the bands $x_1$ to $x_8$ are $ugrzYJHK$. This choice will be referred to as a lupticolor Deep SOM. For the input vector of the Wide SOM, we also use lupticolors with respect to the luptitude in \textit{i} band, and we add the luptitude in \textit{i} band:
\begin{equation*}
\vec{\hat{x}} = (\mu_i, \mu_g{-}\mu_i, \mu_r{-}\mu_i, \mu_z{-}\mu_i).
\end{equation*}
In the case of the wide field, where only few colors are measures, we find empirically that addition of the luptitude improves the performance of the scheme. This choice will be referred to as a `lupticolor-luptitude' Wide SOM.

\subsection{Choice of hyperparameters}
\label{sec:choice_hyperparam}
As presented in \autoref{sec:SOM}, the SOM has various hyperparameters. Apart from one key parameter, the number of cells in the SOM, both the Wide and the Deep SOMs share the same hyperparameters. 

The topology of the two-dimensional grid (square, rectangular, spherical or hexagonal), the boundary conditions (periodic or not) as well as the number of cells must be decided. \citeauthor{CarrascoKind2014} (\citeyear{CarrascoKind2014}) showed that spherical or rectangular grids with periodic boundary conditions performed better. The drawback of the spherical topology is that the number of cells cannot be easily tuned. This leads us to choose the square grid with periodic boundary conditions. 

Our pheno-$z$ scheme assumes that $p(z|c,\hat{c}, \hat{s})=p(z|c)$, i.e.\ once the assignment to a Deep SOM cell, $c$, is known, a galaxy's noisy photometry, embodied by its assignment to the Wide SOM cell, $\hat{c}$, does not add information. This is only true if the cell $c$ represents a negligible volume in the \textit{griz} color space compared to the photometric uncertainty. This assumption requires a sufficient number of Deep SOM cells. A second assumption of our method is that we have a $p(z|c)$ for each Deep SOM cell, $c$, which is only true if we have a sufficient number of galaxies with redshifts to sample the distribution in each cell. While for a narrow distribution $p(z|c)$ a small number of galaxies suffices, this still limits the number of Deep SOM cells. Those two competing effects lead us to set the Deep SOM to a 128 by 128 grid (16,384 cells). This setup reduces the number of empty cells for a COSMOS-like redshift sample ($\sim135,000$ galaxies) while producing rather sharp phenotypes, i.e.\ the volume of each cell in color space is small.

The number of cells of the Wide SOM is dictated by the photometric uncertainty in the wide measurements. By scanning over the number of Wide SOM cells, we found that a 32 by 32 grid offers a sufficient amount of cells to describe the possible phenotypes observed in the wide survey, and that larger numbers of cells did not significantly change the calibration. 

The pheno-$z$ method is robust to other available hyperparameters. The learning rate, $a_0$, which governs how much each step in the training process affects the map, has a negligible impact unless we choose extreme values (0.01, 0.99). It is set to $a_0=0.5$. The initial width of the neighborhood function, $\sigma_s$, is set to the full size of the SOM. This allows the first training vectors to affect the whole map. The maximum number of training steps, $t_{\mathrm{max}}$, must be large enough such that the SOM converges. By scanning over $t_{\mathrm{max}}$, we found that two million steps are sufficient.

We also looked at three-dimensional SOMs and found that an extra dimension for the same number of cells had a negligible impact on the results.

\subsection{Validation of the fiducial SOMs}
Our fiducial pheno-$z$ scheme uses a 128 by 128 lupticolor Deep SOM and a 32 by 32 lupticolor-luptitude Wide SOM. In Appendix \ref{app:val_features_size}, we present the assessment of this choice on our redshift calibration procedure. Using other feature combinations for the Deep and Wide SOMs results in a similar calibration. The feature selection does not matter much but our choice has the conceptual advantages described in \autoref{sec:choice_features}. While using a limited redshift sample, increasing the number of cells of the Deep SOM leads to a higher calibration error whereas increasing the number of cells of the Wide SOM does not affect it.

\section{Pheno-$z$ scheme performance with unlimited samples}
\label{sec:capabilities}
In this section, we use our pheno-$z$ scheme with the fiducial SOMs presented in the preceding section to test its capabilities. We show that, with large enough redshift, deep, and overlap samples, the choices made in the methodology allow a redshift calibration without relevant biases. The effect of limited samples is evaluated separately in \autoref{sec:sources_uncertainty}.

The most relevant metrics to assess performance for weak lensing purposes \citep{Bonnett2016,Hoyle2018} are the differences in the mean and the width of the true redshift distribution and the one estimated with the pheno-$z$ scheme, in each tomographic bin:
\begin{subequations}
\label{metrics}
\begin{align}
\label{metric_dz}
\Delta \langle z \rangle &= \langle z_{\mathrm{true}}\rangle-\langle z_{\mathrm{pheno}}\rangle, \; \text{and} \\ \label{metric_dsigmas}
\Delta \sigma(z) &= \sigma(z_{\mathrm{true}})-\sigma(z_{\mathrm{pheno}}).
\end{align}
\end{subequations}
These metrics are the calibration error of the method. Averaging them over many (hypothetical) random realizations of a survey gives the bias of the method. In the Y1 analysis, the detailed shape of the redshift distributions had little impact. Switching the redshift distribution shape directly estimated from resampled COSMOS objects \citep{Hoyle2018} to the one estimated using Bayesian Photometric Redshifts \citep[\textsc{BPZ;}][]{Benitez2000}, a template fitting method, had little impact on the cosmological inference from cosmic shear as long as the mean redshift of the distributions agreed within uncertainties \citep{Troxel2018}. This is consistent with the finding of \citet{Bonnett2016} for the DES Science Verification analysis. For future, statistically improved, lensing measurements, this simplification may however become invalid. We therefore focus our attention on the first metric for tuning and validating the method, but aim to be able to characterize the biases in general (i.e.\ in terms of possible realizations of the redshift distributions).

The bias is determined under `perfect' conditions that are defined by the following requirements: the redshift sample is identical to the deep sample; the overlap sample is identical to the wide sample; and both are large. The galaxies of all samples are randomly sampled from the full DES Y3 footprint. We use our usual selection $m_{\mathrm{obs},i}<23.5$ for the wide/overlap sample. A hundred iterations of this best case scenario are run where the redshift/deep sample is made of $10^6$ galaxies and the wide/overlap sample is made of $2\cdot10^6$ galaxies. \autoref{tab:bias_best_case} presents the means of the metrics defined in \autoref{metrics} for this best case scenario for our fiducial lupticolor 128 by 128 Deep SOM coupled to a lupticolor-luptitude 32 by 32 Wide SOM. For comparison the same test is performed with a lupticolor 256 by 256 Deep SOM. As expected, from the reduction of biases related to discretization, increasing the number of cells in the Deep SOM results in a lower bias. This means that there are more than 16,384 possible phenotypes (as the 128 by 128 Deep SOM has 16,384 cells). The first two bins are the most affected: increasing the number of cells by a factor of four reduces the bias, $\langle\Delta \langle z \rangle\rangle$, by a factor of two. Note that this increase in resolution is only possible with the idealized, large redshift/deep sample used in this test. If our available redshift sample were larger, we would use a larger SOM.

\begin{table}
\caption{Bias of the pheno-$z$ method in the best case scenario of a large redshift sample. Shown are the biases on mean redshift of tomographic bins, $\langle\Delta \langle z \rangle\rangle$, and width of redshift distribution of a bin, $\langle\Delta \sigma( z )\rangle$. The  fiducial 128 by 128 Deep SOM is compared to a 256 by 256 Deep SOM. For such a large redshift sample, the increased number of Deep SOM cells is beneficial. Note that the standard deviation of both metrics in this best case scenario is an order of magnitude smaller than their means. The mean of the true redshift distribution in each bin, $\langle z_{\mathrm{true}} \rangle$, is also shown.}
\ra{1.2}
\centering
\resizebox{\linewidth}{!}{%
\begin{tabular}{cl*{5}{S[table-format=1.4, round-mode=places,round-precision=4, group-digits=false]}}
\toprule
\multicolumn{1}{l}{\textbf{metric}} & \textbf{size} & \multicolumn{1}{c}{Bin 1} & \multicolumn{1}{c}{Bin 2} & \multicolumn{1}{c}{Bin 3} & \multicolumn{1}{c}{Bin 4} & \multicolumn{1}{c}{Bin 5} \\
\multicolumn{2}{c}{\textit{$\langle z_{\mathrm{true}} \rangle$}} & {\textit{0.34}}  & {\textit{0.48}} & {\textit{0.68}} & {\textit{0.87}} & {\textit{1.07}}\\
 \cmidrule(lr){1-2} \cmidrule(lr){3-7} 
\multirow{2}{*}{$\langle\Delta \langle z \rangle\rangle$}                  
& 128  & -0.005028 & -0.002351 &	0.000106 &	0.002450 &	0.002449   \\
& 256             & -0.002560 &	-0.001041 &	0.000265 &	0.001785 &	0.001969 \\[3pt]
\multirow{2}{*}{$\langle\Delta \sigma( z )\rangle$}                  
& 128 & -0.003871 &	-0.002664 &	-0.002899 &	-0.001804 &	-0.001426\\
& 256  & -0.002338 & -0.001499 & -0.001495 & -0.000861 & -0.000305 \\
\bottomrule                
\end{tabular}
}
\label{tab:bias_best_case}
\end{table}

\section{Sources of uncertainty due to limited samples}
\label{sec:sources_uncertainty}

Deep multi-band observations and, more so, observations that accurately determine galaxy redshifts with spectroscopy or otherwise, require substantial telescope resources. As a result, in practice, deep and redshift samples are limited in galaxy count and area. In this section, we determine the impact of these limited samples on redshift calibration using our scheme. 

Limited samples can impact redshift calibration both as a statistical error -- i.e.\ depending on the field or sample of galaxies chosen for deep and redshift observations -- or as a systematic error -- i.e.\ as a bias due to the limited resolution by which galaxies sample color space (see also \autoref{sec:capabilities}). We use the metrics presented in \autoref{metrics} to assess this limitation:  $\langle \Delta \langle z \rangle \rangle$ over many realizations of our samples assesses the systematic error in mean redshift, whereas $\sigma(\Delta \langle z \rangle)$ is a statistical error due to variance in the samples used.

To this end, we use the \textsc{Buzzard} simulated catalogs to assess systematic and statistical errors. Each source of uncertainty, i.e.\ the effect of limiting each sample, is separately probed in the subsections below. At each iteration of a test, only the sample of interest is modified and the other fixed samples are sampled randomly over the full DES footprint and with a sufficient number to avoid a sample variance or shot noise contribution. In \autoref{sec:strangecv}, we discuss the perhaps counter-intuitive finding that the statistical error in a realistic use case is limited by the size of the \emph{deep} sample, not the redshift sample. 

\subsection{Limited redshift sample}
\label{sec:lim_spec_sample}
The redshift sample used to estimate $p(z|c)$ is limited in two ways. First, it contains a finite number of galaxies; second, the galaxies it contains come from a small field on the sky: COSMOS. This implies that the scatter of the redshift calibration error, $\sigma(\Delta \langle z \rangle)$, has contributions from shot noise and sample variance, respectively. 

\subsubsection{Shot noise}
\label{sec:sample_variance}
One can assess the effect of shot noise in the redshift sample by computing the redshift distribution of a sample of galaxies many times using our pheno-$z$ method. At each iteration we randomly select a fixed number of galaxies for our redshift sample. In this test, we do not want to include sample variance, thus the redshift sample is also composed of galaxies randomly selected over the full DES footprint. 

\begin{figure*}
\centering
\includegraphics[width=\linewidth]{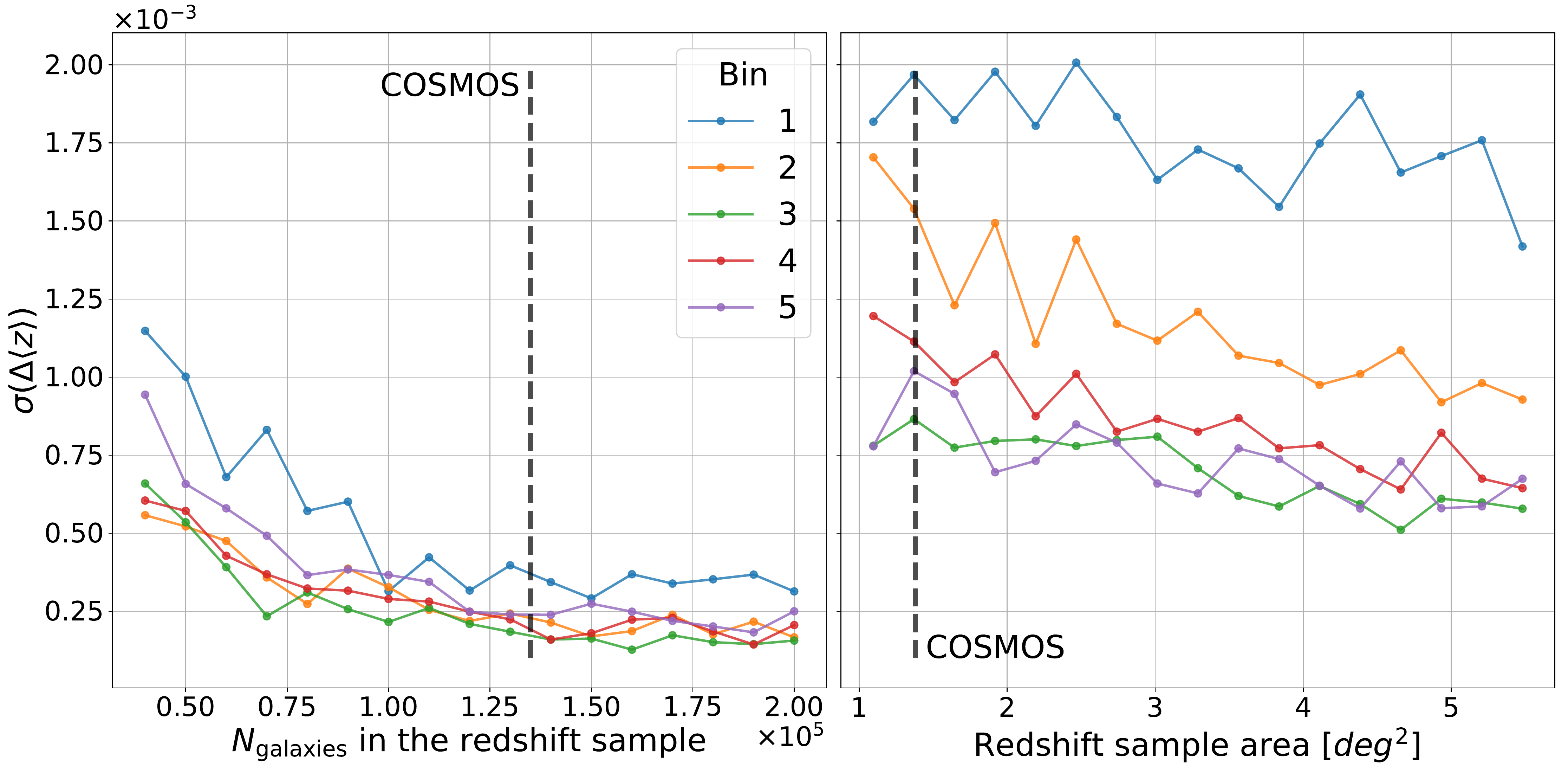}
\caption{Shot noise due to limited sample size (\textit{left panel}) compared to the sample variance due to limited sample area (\textit{right panel}) in the redshift sample. The gray dashed line highlights the number of galaxies in the COSMOS field and its size. The standard deviation of the difference in mean redshift between the true and estimated distribution over 100 iterations is plotted. \textit{Left panel:} Effect of shot noise as a function of the number of galaxies in the redshift sample. The galaxies used to compute $p(z|c)$ are sampled from the whole sky to avoid any sample variance contribution. Above $\sim10^5$ galaxies, increasing the number of objects does not yield a significant improvement. \textit{Right panel}: Effect of sample variance as a function of redshift field area sampled. One hundred thousand galaxies are sampled over different contiguous areas. The calibration of redshift distribution with the pheno-$z$ scheme is not limited by the number of galaxies in COSMOS but by their common location on the same line of sight.
\label{fig:SV_CV_spec_sample}}
\end{figure*}

The left panel of \autoref{fig:SV_CV_spec_sample} shows the shot noise as a function of the number of galaxies in the redshift sample. If the number of galaxies is too low, there is a significant scatter in $\Delta\langle z \rangle$, but with more than $\sim10^5$ galaxies, the scatter reaches a plateau at the $2\text{--}4 \cdot 10^{-4}$ level which is well below our requirements ($\sigma_{\Delta z} \sim 0.01$). Note that the first and last bins exhibit more shot noise probably due to the hard boundaries at $z=0$ and $z=1.5$.

\subsubsection{Sample variance}

This effect stems from the fact that the selection of galaxies depends on the environment: as the matter field is not homogeneous on small scales, different lines of sight have different distributions of galaxies.

\begin{figure*}
\centering
\includegraphics[width=0.7\linewidth]{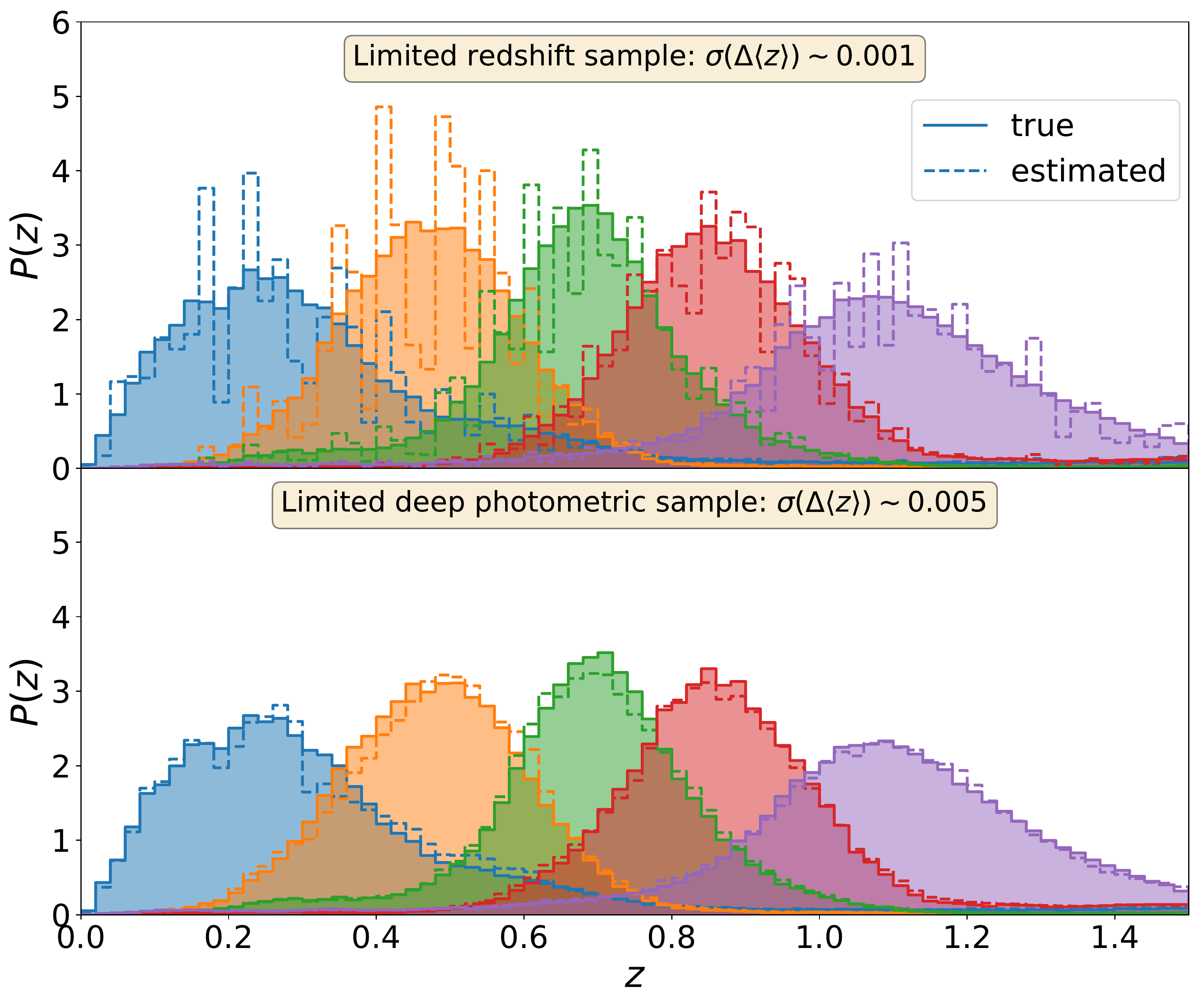}
\caption{\textit{Top panel:} Impact of limited redshift sample in the redshift distribution calibration. The redshift sample is made of 135,000 galaxies sampled from a 1.38 $\mathrm{deg}^2$ field. The spikes in the estimated distributions are due to the particular redshift distribution of this small area. \textit{Bottom panel:} Impact of limited deep sample in the redshift distribution calibration. The deep field is made of three fields of 3.32, 3.29 and 1.38 $\mathrm{deg}^2$ respectively. Each deep field galaxy is painted 10 times at random positions over the whole DES footprint to yield an overlap sample of $\sim 4.6\cdot10^6$ galaxies. Although the redshift distributions do not exhibit the spikes visible in the upper panel, the scatter of the calibration error, $\Delta\langle z \rangle$, is three to five times larger, meaning that the sample variance in the deep sample dominates over the one in the redshift sample.
\label{fig:CV_spec_overlap}}
\end{figure*}

With a redshift sample that is small on the sky, subsets of galaxies contained in this sample have the same environment which influences their overall properties (notably redshift and colors). This sample variance was a major limitation in the DES Y1 redshift calibration. To test the effect of sample variance we can repeat our calibration method many times with a redshift sample coming from a different part of the sky at each iteration. The top plot in \autoref{fig:CV_spec_overlap} shows the result of one iteration with a redshift sample made of 135,000 galaxies sampled from a 1.38 $\mathrm{deg}^2$ field. The estimated distribution has many spikes. Those are caused by the incomplete population of galaxies in the sample: galaxies have similar redshifts and colors. Many Deep SOM cells that should have broader redshift distributions end up being peaked due to the presence of a galaxy cluster in the redshift field. When the redshift sample is limited to a small field on the sky, the $p(z|c)$ is strongly structured by sample variance. 

We test the effect of sample variance as a function of the redshift field area available. To avoid shot noise effects, we sample the same number of galaxies for redshift fields of different sizes. The sample variance is measured as the standard deviation of the difference between the mean of the true redshift distribution and the mean of the pheno-$z$ estimation. As expected, it decreases as the area increases. This effect is shown in the right panel of \autoref{fig:SV_CV_spec_sample} for a fixed number of galaxies of $10^5$. The first tomographic bin has a higher level of sample variance because of higher density fluctuations due to the smaller volume at low redshift.

For the DES Y3 calibration, we expect that the redshift sample will contain about 135,000 galaxies in a 1.38 $\mathrm{deg}^2$ field from COSMOS. The expected sample variance from such a field is quoted in \autoref{tab:vistaYimpact} for two different sets of VISTA bands used. Using the \textit{Y} band, we expect uncertainties of the order $\sigma(\Delta\langle z \rangle)\sim0.001$ from sample variance alone. Relative to \autoref{sec:sample_variance}, we find that for COSMOS, this effect dominates by a factor of five, compared to shot noise. For comparison, DES Y1 redshift calibration \citep{Hoyle2018} achieved a typical
$\sigma(\Delta\langle z \rangle) \sim 0.02$, with sample variance (labeled `COSMOS footprint sampling' in their Table 2) contributing $\sim0.007$ in quadrature to the uncertainty. Despite using an identical sample of galaxies as \citet{Hoyle2018}, our pheno-$z$ method reveals a net reduction of the sample variance in the COSMOS redshift information, owing to augmentation of the estimate of multi-color density of galaxies with a larger, purely photometric, deep sample. The main source of sample variance is the limited size of the deep sample (\autoref{sec:lim_deep_sample}) which, however, can be more easily extended than the redshift sample.

\begin{table*}
\caption{Sources of bias and uncertainty of redshift calibration with the pheno-$z$ scheme. $\Delta\langle z \rangle$ is the difference between the means of the true and estimated redshift distribution. The mean (i.e.\ bias) and standard deviation of this metric over 100 iterations are shown, with the last column (in bold) showing the root-mean-square of the latter over the tomographic bins. To isolate the effect of limited redshift and limited deep samples, only one sample is modified in each iteration. All other samples are fixed, sufficiently large, and sampled from the whole DES footprint. The upper two and lower two lines show the impact of using VISTA \textit{Y} band in our pheno-$z$ scheme. Using it reduces the area of deep field available but improves deep color information. For the limited redshift sample, 135,000 galaxies are sampled from the 1.38 $\mathrm{deg}^2$ field.  \label{tab:vistaYimpact}}
\ra{1.2}
\centering
\resizebox{\linewidth}{!}{%
\begin{tabular}{ccc*{10}{S[table-format=1.4, round-mode=places,round-precision=4, group-digits=false]}c}
\toprule
\multicolumn{3}{c}{\textbf{Test}}            & \multicolumn{5}{c}{\textbf{$\boldsymbol{\langle \Delta \langle z \rangle\rangle}$ in bin}}                                                  & \multicolumn{5}{c}{\textbf{$\boldsymbol{\sigma(\Delta\langle z \rangle)}$ in bin}}& \multirow{2}{*}[-3pt]{$\boldsymbol{\sigma_{\mathrm{RMS}}(\Delta\langle z \rangle)}$}                                                           \\ \cmidrule(lr){1-3} \cmidrule(lr){4-8}\cmidrule(lr){9-13}
\textit{VISTA bands} & \textit{Sample} & \textit{Size} [$\mathrm{deg}^2$]  &  \multicolumn{1}{c}{\textit{1}} & \multicolumn{1}{c}{\textit{2}} & \multicolumn{1}{c}{\textit{3}} & \multicolumn{1}{c}{\textit{4}} & \multicolumn{1}{c}{5} & \multicolumn{1}{c}{\textit{1}} & \multicolumn{1}{c}{\textit{2}} & \multicolumn{1}{c}{\textit{3}} & \multicolumn{1}{c}{\textit{4}} & \multicolumn{1}{c}{\textit{5}} &\\ \midrule
\multirow{2}{*}{YJHKs}        & Redshift & 1.38     & -0.005054 & -0.002376 & -0.000604 & 0.002091 & 0.002235 & 0.001608 & 0.001362 & 0.000765 & 0.001029 & 0.000779 & \bfseries 0.0012\\
&Deep       & 7.99         & -0.006199 & -0.004922 & -0.000573 & 0.002009 & 0.004667 & 0.007691 & 0.004191 & 0.003050 & 0.002604 & 0.003606 & \bfseries 0.0046 \\[5pt]
\multirow{2}{*}{JHKs}  & Redshift & 1.38          & -0.005402 & -0.002729 & -0.000091 & 0.002773 & 0.004754 & 0.001689 & 0.001528 & 0.000671 & 0.000757 & 0.000984 & \bfseries 0.0012\\
&Deep       & 9.93         & -0.005032 & -0.004904 & -0.000157 & 0.001728 & 0.007832 & 0.006476 & 0.003705 & 0.002654 & 0.002687 & 0.003180 & \bfseries 0.0040\\ \bottomrule 
\end{tabular}
}
\end{table*}

\subsection{Limited overlap sample}

We estimate the overlap sample by drawing galaxies from the deep fields (i.e.\ the overlap between deep DES \textit{ugriz} and VISTA \textit{YJHKs} or \textit{JHKs}; see \autoref{sec:DES_Y3_samples}) over the full DES footprint with the \textsc{Balrog} algorithm. In this section we test what size of the overlap sample is required.

We assume the deep sample is artificially drawn at random locations over the footprint, with $N$ realizations of each galaxy over the full footprint. $N$ must be sufficient to provide enough deep-wide tuples to populate the transfer function, $p(c|\hat{c},\hat{s})$, and avoid noise introduced by unevenly sampling observing conditions.

Our investigation shows that increasing $N$ from 5 to 50 has no impact on the mean and standard deviation of the calibration error, $\Delta \langle z \rangle$. We thus use 10 realizations at different random positions (i.e.\ with different noise realizations) of each deep field galaxy. This corresponds to 1--2\% ratio of galaxy count in the overlap to wide sample for DES Y3.

\subsection{Limited deep sample}
\label{sec:lim_deep_sample}
The overlap sample used to compute the transfer function, $p(c|\hat{c},\hat{s})$, is limited by the deep sample. Indeed, \textsc{Balrog} takes as input the galaxies measured in the deep survey, which spans only a limited area (see \autoref{sec:DES_Y3_samples}). We first look at the sample variance in the overlap sample due to the limited area of the deep sample. Secondly, we look at the trade-offs between the number of VISTA bands used and the area available.

\subsubsection{Sample variance}
As the area is bigger than the one of the redshift sample, we might expect less sample variance coming from the overlap sample. Unfortunately, that is not the case. The transfer function is sensitive to changes in $p(c, \hat{c})$ due to sample variance. Although the reconstructed redshift distributions, shown on the bottom plot in \autoref{fig:CV_spec_overlap}, do not exhibit the spikes produced by the limited redshift sample shown on the top plot, the scatter of the calibration error, $\Delta\langle z \rangle$, is three to five times larger, as reported in \autoref{tab:vistaYimpact}. The sample variance of the deep sample dominates over the one of the redshift sample.  We are learning a noisy realization of the distribution of multi-band deep colors given a wide-field flux measurement, and so are incorrectly learning the distribution of SEDs given our selection and observed galaxy colors.

\subsubsection{Number of bands vs.~deep area}
\label{sec:Yband_area}
As described in \autoref{sec:DES_Y3_samples} and \autoref{tab:vista_bands}, depending on which VISTA bands are used the available area in the deep sample will be different. Either we use \textit{YJHKs} and have 7.99 $\mathrm{deg}^2$ of deep fields in three places or we drop the \textit{Y} band and have 9.93 $\mathrm{deg}^2$ in four fields. Those two possibilities are tested empirically.

We repeat the tests performed on the limited redshift sample (\autoref{sec:lim_spec_sample}) and on the limited deep sample (\autoref{sec:lim_deep_sample}) without the \textit{Y} band and with the increased area. The results, shown in \autoref{tab:vistaYimpact}, show two opposite trends. The bias, $\langle\Delta \langle z \rangle\rangle$, is significantly larger without the \textit{Y} band for the last bin and almost unchanged for the other bins. At large redshift, the \textit{Y} band provides valuable information necessary to estimate correctly the redshift distribution. The bias is not sensitive to the area used but to the number of bands available. On the contrary, the variance of the calibration error is affected by the size of the deep field. Without the \textit{Y} band, the standard deviation of the calibration error, $\sigma(\Delta \langle z \rangle)$, is smaller by about 15\% because this option provides a larger deep field area. 

The two effects -- a bigger deep field area and one less band -- have opposite impact of about the same amplitude. A reduction in bias in the high redshift bin is particularly beneficial and thus may favor including $Y$.

\subsection{Impact of empty Deep SOM cells}
\label{sec:empty_deep_som_cells}
When computing the redshift distributions using \autoref{n_i_z}, the $p(c_e|\hat{c},\hat{s})$ of empty cells, $c_e$, is set to zero. To check that this does not introduce a bias, we compute the `true' redshift distributions of the empty cells by assigning a sample of $5\cdot10^5$ galaxies to the Deep SOM. A redshift distribution is obtained for the initially empty cells and used in our $p(z|\hat{B}, \hat{s})$ computation. In \autoref{tab:bestcase_interpolation}, we compare the resulting bias in the two cases: with empty cells ignored and with empty cells filled with a large number of galaxies to be as close to the `true' redshift distribution as we can get. This latter method is equivalent to a `perfect' interpolation to the empty cells. We therefore conclude that ignoring empty cells does not introduce a relevant bias. In practice, since larger numbers of cells could be empty in the case of sparse redshift samples, and since spectroscopic samples (rather than complete redshifts over a field) may suffer selection biases, the impact of cells without redshift information should be checked.

\begin{table}
\caption{Setting the $p(c_e)$ of empty Deep SOM cells, $c_e$, to zero does not introduce a bias. The bias, $\langle\Delta \langle z \rangle\rangle$, and standard deviation of the calibration error, $\sigma(\Delta \langle z \rangle)$, in five tomographic bins is computed over 162 iterations of our pheno-$z$ scheme with a different 1.38 $\mathrm{deg}^2$ redshift sample at each iteration. The redshift distribution of empty Deep SOM cells, $p(z|c_e)$, is either set to zero or filled with the redshifts of a large sample of galaxies.}
\ra{1.2}
\centering
\resizebox{\linewidth}{!}{%
\begin{tabular}{cl*{5}{S[table-format=1.4, round-mode=places,round-precision=4, group-digits=false]}}
\toprule
\multicolumn{1}{l}{\textbf{metric}} & $\boldsymbol{p(z|c_e)}$ & \multicolumn{1}{c}{\textit{Bin 1}} & \multicolumn{1}{c}{\textit{Bin 2}} & \multicolumn{1}{c}{\textit{Bin 3}} & \multicolumn{1}{c}{\textit{Bin 4}} & \multicolumn{1}{c}{\textit{Bin 5}} \\ \cmidrule(lr){1-2} \cmidrule(lr){3-7} 
\multirow{2}{*}{$\langle\Delta \langle z \rangle\rangle$}                  & Set to 0           & -0.008013                          & -0.003804                          & 0.000385                           & 0.001305                           & 0.003925                           \\
                                    & Filled             & -0.007823                          & -0.003766                          & 0.000722                           & 0.001698                           & 0.004292                           \\[3pt]
\multirow{2}{*}{$\sigma(\Delta \langle z \rangle)$}                  & Set to 0           & 0.001727                           & 0.001416                           & 0.000790                           & 0.001095                           & 0.000845                           \\
                                    & Filled             & 0.001415                           & 0.001167                           & 0.000532                           & 0.000775                           & 0.000622         \\ \bottomrule                
\end{tabular}
}

\label{tab:bestcase_interpolation}
\end{table}

Some of the cells ($\sim50$) remain empty even when the very large sample is assigned to the Deep SOM. Those cells are often located where there is a sharp color and redshift gradient. This results from the SOM training: both sides of the boundary evolve differently pulling the cells to empty regions of color space. These cells are not a problem in our scheme as they never enter any computation.

\subsection{Discussion of statistical error budget}
\label{sec:strangecv}

The comparison of \autoref{sec:lim_spec_sample} and \autoref{sec:lim_deep_sample} shows that the limited area of the deep photometric sample is dominating the statistical error budget of redshift calibration for a DES-like setting, by a factor of several, rather than the limited size of a COSMOS-like redshift sample (see \autoref{fig:CV_spec_overlap}).

This finding can be understood from the role of these samples in our scheme. The redshift sample informs the redshift distribution of galaxies at given multi-band color. Because at most multi-band colors this redshift distribution is narrow, there is little room for sample variance -- regardless of their position in the sky, any set of redshift galaxies of the same multi-band color will be very similar in mean redshift. Increasing the number of accurate redshifts, or spreading them over a larger area, reduces this variance further (see \autoref{fig:SV_CV_spec_sample}), but it is already at a tolerable level for a COSMOS-like sample.

The deep sample, while not adding accurate redshift information, constrains the density of galaxies in multi-band color space, i.e.\ the mix of multi-band colors that corresponds to a given few-band color observed in the wide field. Uncertain information about this distribution can be seen as an incorrect prior on the abundance of galaxy templates, causing an inaccurate breaking of the type/redshift degeneracy. 

This finding represents an opportunity: by separating the abundance aspect of sample variance from the redshift sample, it allows us to augment the scarce information on accurate galaxy redshifts with a larger, complete sample for which deep multi-band photometry can be acquired with relatively modest observational effort.

\section{Impact of analysis choices for DES Y3 weak lensing}
\label{sec:refinement}
In this section, we assess the robustness of our method when the quality of the inputs decreases. We first test a more realistic selection, $\hat{s}$, for the wide and overlap samples in \autoref{sec:wl-selection}. The \textsc{Metacalibration} \citep{Huff2017,Sheldon2017} weak lensing analysis requires the use of fluxes measured by the shape measurement algorithm to correct for selection biases. We test the effect of this noisier photometry in \autoref{sec:metacal_flux}. Finally, we test the possibility of dropping the \textit{g} band in the Wide SOM in \autoref{sec:drop_g}. The combined effect of these realistic conditions is discussed in \autoref{sec:z_uncertainty_Y3}.

We compare those variations of the scheme to a `standard' pheno-$z$ scheme which uses DES \textit{ugriz} and VISTA \textit{YJHKs} bands for the Deep SOM and DES \textit{griz} for the Wide SOM, a 1.38 $\mathrm{deg}^2$ redshift sample, a 7.99 $\mathrm{deg}^2$ deep sample and a hard cut $m_{\mathrm{obs},i}<23.5$ as the wide selection. In this standard scheme, 10 realizations of each deep field galaxy at different random positions constitute the overlap sample. The usual metrics for this standard scheme are presented in \autoref{tab:YJHK_comparisons}. 

\begin{table*}
\caption{Comparison of the standard pheno-$z$ scheme to various variations of the scheme. The standard scheme uses DES \textit{ugriz} and VISTA \textit{YJHKs} bands for the Deep SOM and DES \textit{griz} for the Wide SOM. It uses a redshift sample made of 135,000 galaxies sampled from a 1.38 $\mathrm{deg}^2$ field, a 7.99 $\mathrm{deg}^2$ deep sample and a hard cut $m_{\mathrm{obs},i}<23.5$ for the wide selection. In this standard scheme, 10 realizations of each deep field galaxy at different random positions constitute the overlap sample. The mean and standard deviation of the metrics given in \autoref{metrics} over 100 iterations are presented.}
\centering
\begin{tabular}{@{}l*{11}{S[table-format=1.4, round-mode=places,round-precision=4, group-digits=false]}@{}}
\toprule
\multirow{2}{*}[-3pt]{\textbf{Variation}}   & \multicolumn{1}{c}{\textit{Bin 1}}     & \multicolumn{1}{c}{\textit{Bin 2}}   & \multicolumn{1}{c}{\textit{Bin 3}}    & \multicolumn{1}{c}{\textit{Bin 4}}     & \multicolumn{1}{c}{\textit{Bin 5}}     & \multicolumn{1}{c}{\textit{Bin 1}}     & \multicolumn{1}{c}{\textit{Bin 2}}   & \multicolumn{1}{c}{\textit{Bin 3}}    & \multicolumn{1}{c}{\textit{Bin 4}}     & \multicolumn{1}{c}{\textit{Bin 5}}     \\ \cmidrule(lr){2-11} 
 & \multicolumn{5}{c}{\textbf{$\boldsymbol{\langle\Delta \langle z \rangle\rangle}$}}     & \multicolumn{5}{c}{\textbf{$\boldsymbol{\sigma(\Delta \langle z \rangle)}$}}  \\  \cmidrule(lr){1-1} \cmidrule(lr){2-6}\cmidrule(lr){7-11} \addlinespace[2pt]
Standard                           & -0.007282 & -0.004013 & 0.000637  & 0.001587  & 0.005594 & 0.007652 & 0.004196 & 0.002786 & 0.003273 & 0.003691 \\ \addlinespace[2pt]
w/ weak lensing selection$^a$      & -0.005708 & -0.004044 & 0.000311  & 0.002744  & 0.005748 & 0.008316 & 0.004626 & 0.004231 & 0.002967 & 0.004212 \\ \addlinespace[2pt]
w/ \textsc{Metacalibration} fluxes$^b$ & -0.008891 & -0.004413 & 0.000081  & 0.002180  & 0.006886 & 0.007659 & 0.004911 & 0.003906 & 0.003317 & 0.003739 \\ \addlinespace[2pt]
w/ only \textit{riz}$^c$             & -0.007003 & -0.004999 & 0.001929  & 0.000635  & 0.006117 & 0.007059 & 0.005132 & 0.003564 & 0.003626 & 0.003833 \\ \addlinespace[2pt]
w/ decreased softening parameter$^d$ & -0.007430 & -0.005286 & -0.001229 & -0.000533 & 0.004381 & 0.007169 & 0.003961 & 0.003432 & 0.002794 & 0.003253 \\  \addlinespace[5pt]
                & \multicolumn{5}{c}{\textbf{$\boldsymbol{\langle\Delta \sigma(z)\rangle}$}} & \multicolumn{5}{c}{\textbf{$\boldsymbol{\sigma(\Delta \sigma(z))}$}}                              \\ \cmidrule(lr){1-1} \cmidrule(lr){2-6}\cmidrule(lr){7-11} \addlinespace[2pt]
Standard                           & -0.004825 & -0.004463 & -0.005671 & -0.005382 & -0.004385 & 0.003638 & 0.002610 & 0.002915 & 0.002405 & 0.003688 \\ \addlinespace[2pt]
w/ weak lensing selection$^a$      & -0.000912 & -0.004316 & -0.004418 & -0.004202 & -0.003882 & 0.004292 & 0.003533 & 0.003902 & 0.002734 & 0.003516 \\ \addlinespace[2pt]
w/ \textsc{Metacalibration} fluxes$^b$ & -0.004657 & -0.004000 & -0.005294 & -0.004816 & -0.003629 & 0.003928 & 0.003073 & 0.003687 & 0.002956 & 0.003894 \\ \addlinespace[2pt]
w/ only \textit{riz}$^c$             & -0.004292 & -0.002877 & -0.005380 & -0.005109 & -0.003423 & 0.003550 & 0.004250 & 0.003410 & 0.002873 & 0.003458 \\ \addlinespace[2pt]
w/ decreased softening parameter$^d$ & -0.003444 & -0.004135 & -0.005149 & -0.004773 & -0.003886 & 0.003550 & 0.003274 & 0.003017 & 0.002645 & 0.003460 \\ \bottomrule \addlinespace[2pt]
\multicolumn{11}{l}{\footnotesize $^a$The selection is given in \autoref{wl_selection}.} \\
\multicolumn{11}{l}{\footnotesize $^b$Mock \textsc{Metacalibration} fluxes are used for the wide bands (see \autoref{metacal_flux}).} \\
\multicolumn{11}{l}{\footnotesize $^c$The \textit{g} band is not used in the Wide SOM.} \\
\multicolumn{11}{l}{\footnotesize $^d$The limiting magnitudes are increased by one magnitude in each band (see Appendix \ref{app:softB}).} 
\end{tabular}
\label{tab:YJHK_comparisons}
\end{table*}

\subsection{Weak lensing selection $\hat{s}$}
\label{sec:wl-selection}
In the above tests, we used a simple selection for the wide and overlap samples. Only galaxies with $m_{\mathrm{obs},\,i}<23.5$ were selected. Here we run our pheno-$z$ scheme with a more refined selection criterion. The goal is to more accurately mimic the selection effect produced by the shape measurement algorithm. For this purpose, we select only galaxies for which 
\begin{equation}
\begin{aligned}
\label{wl_selection}
m_{\mathrm{obs},\,r} < -2.5\;\log_{10}(0.5)+l_r,\; \mathrm{and}  \\
\sqrt{s^2 + (0.13 \cdot \mathrm{psf}_r)^2} > 0.1625 \cdot \mathrm{psf}_r,
\end{aligned}
\end{equation}
where $m_{\mathrm{obs},\,r}$ is the observed \textit{r} band magnitude of the galaxy, $l_r$ is the limiting magnitude in the \textit{r} band of the survey at the galaxy's position, $s$ is the size of the galaxy and $\text{psf}_r$ is the full width at half maximum of the point spread function in \textit{r} band, both in pixels. The latter is a function of the telescope optics and the astronomical seeing. As many observations of the same line of sight are combined to produce the catalog, the variation is averaged out. We therefore approximate it by $\text{psf}_r = 0.9''$ over the full footprint. The distribution in magnitude of such a sample is shown in \autoref{fig:samples_mag_i}.

The values for the mean and standard deviation of the bias using the weak lensing selection described in \autoref{wl_selection} are presented in \autoref{tab:YJHK_comparisons} (see `w/ weak lensing selection' entry). Using a more refined weak lensing selection than the hard cut at $m_{\mathrm{obs},\,i}<23.5$ used throughout this work does not introduce any bias but slightly increases the variance.

\subsection{\textsc{Metacalibration} fluxes}
\label{sec:metacal_flux}
For our cosmology analysis, we must understand if, when sheared, a galaxy's tomographic bin changes. The shape algorithm -- \textsc{Metacalibration} -- allows us to artificially shear the galaxies and measure their resulting fluxes. The \textsc{Metacalibration} flux measurement is noisier than the usual multi-object fitting \citep[MOF;][their section 6.3]{Drlica-Wagner2018} flux measurement used by DES but the tomographic binning must be performed on \textsc{Metacalibration} fluxes for the reason mentioned above \citep[see also][their section 7.4]{Zuntz2018}. The estimation of the redshift distribution could then be performed using MOF photometry. To achieve this, we would need to introduce a third SOM and compute a transfer function between MOF fluxes and \textsc{Metacalibration} fluxes. We suspect that introducing a third SOM would not improve our calibration. To avoid this complication we can perform the estimation of redshift distributions using \textsc{Metacalibration} fluxes.

The simulated fluxes used throughout this work were tailored to match MOF measurement errors. On average the errors are $\sqrt{2}$ larger for \textsc{Metacalibration} measurements, i.e.\ $\sigma_{\mathrm{MCAL}} = \sqrt{2}\sigma_{\mathrm{MOF}}$. We can build fake \textsc{Metacalibration} fluxes, $f_{\mathrm{MCAL}}$, using our `MOF' fluxes, $f_{\mathrm{MOF}}$:
\begin{equation}
\label{metacal_flux}
f_{\mathrm{MCAL}} = f_{\mathrm{MOF}} + \sqrt{\sigma_{\mathrm{MCAL}}^2-\sigma_{\mathrm{MOF}}^2}\cdot\mathcal{N}(0,1). 
\end{equation}
This results in $f_{\mathrm{MCAL}} = f_{\rm MOF}+\sigma_{\mathrm{MOF}}\cdot\mathcal{N}(0,1)$, where $\mathcal{N}(0,1)$ is a normal distribution with zero mean and standard deviation one. Note that this neglects any systematic differences between \textsc{Metacalibration} and MOF fluxes, which would be compensated by the transfer function derived from deep galaxies with wide-field \textsc{Metacalibration} flux realizations.

We run our pheno-$z$ scheme replacing the wide `MOF' fluxes by the mock \textsc{Metacalibration} fluxes. As can be seen in \autoref{tab:YJHK_comparisons} (entry `w/ \textsc{Metacalibration} fluxes'), this results in a slight increase of the bias and variance in calibration.

\subsection{Dropping the \textit{g} band}
\label{sec:drop_g}
Detailed tests on DES Y1 and Y3 data (Mike Jarvis, private communication) and theoretical considerations \citep{Plazas2012} show that point-spread function (PSF) modeling in DES is most difficult in the \textit{g} band. The expected and observed bias in PSF modeling in \textit{g} band significantly biases shape measurement. At a secondary level, it also biases \textit{g} band photometry. Thus, it may be preferable to run \textsc{Metacalibration} uniquely on the \textit{riz} bands. In this case, the tomographic binning must be performed only on those bands. To simplify the comparisons and separate the different effects, we use only the \textit{riz} bands, but still use `MOF' fluxes. We perform our pheno-$z$ scheme, training the Wide SOM on \textit{riz} bands. The result of dropping the \textit{g} band is shown in \autoref{tab:YJHK_comparisons} (entry `w/ only \textit{riz}'). The observed degradation is similarly slight to the one produced by the use of \textsc{Metacalibration} fluxes. 

\begin{figure}
\centering
\includegraphics[width=\linewidth]{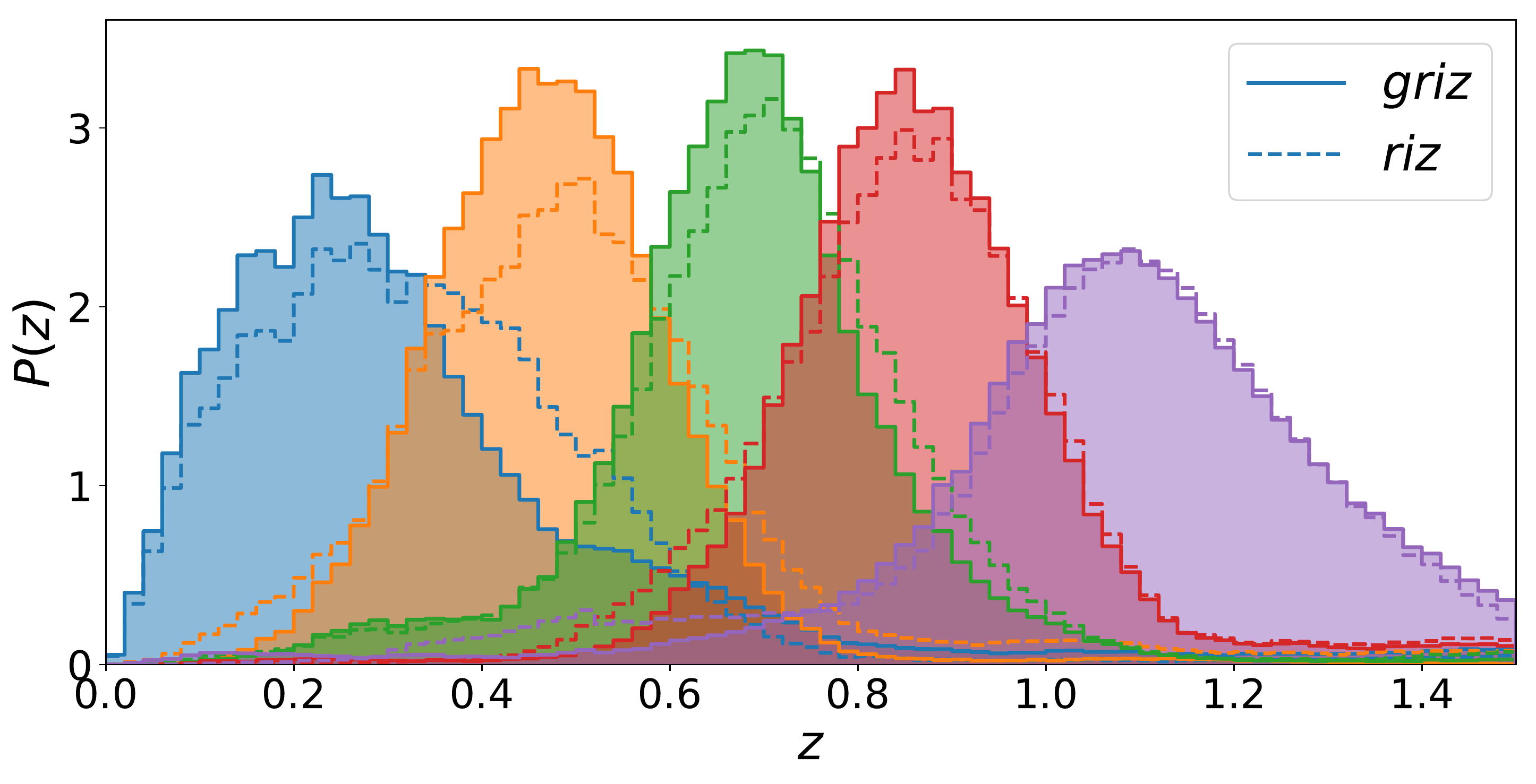}
\caption{Comparison of redshift distributions of bins defined with and without \textit{g} band color in addition to \textit{riz}. Information contained in \textit{g} is particularly useful at low redshift. We show the true redshift distributions. The calibration of mean redshift does not substantially suffer from the loss of \textit{g} band data in the wide field.}
\label{fig:Riz_griz_metacal_mof_comparison}
\end{figure}

The \textit{g} band carries some useful information, especially at low redshift. Indeed, dropping this band results in larger bins. This effect is the strongest for the lowest redshift bin as can be seen in \autoref{fig:Riz_griz_metacal_mof_comparison}, where the true redshift distributions obtained using \textit{griz} or \textit{riz} are compared.

\section{Pheno-$z$ uncertainty for DES Y3}
\label{sec:z_uncertainty_Y3}
We integrate the different variations discussed in \autoref{sec:refinement} to be as close as possible to the actual redshift distribution estimation of DES Y3 weak lensing sources.

\begin{figure}
\centering
\includegraphics[width=\linewidth]{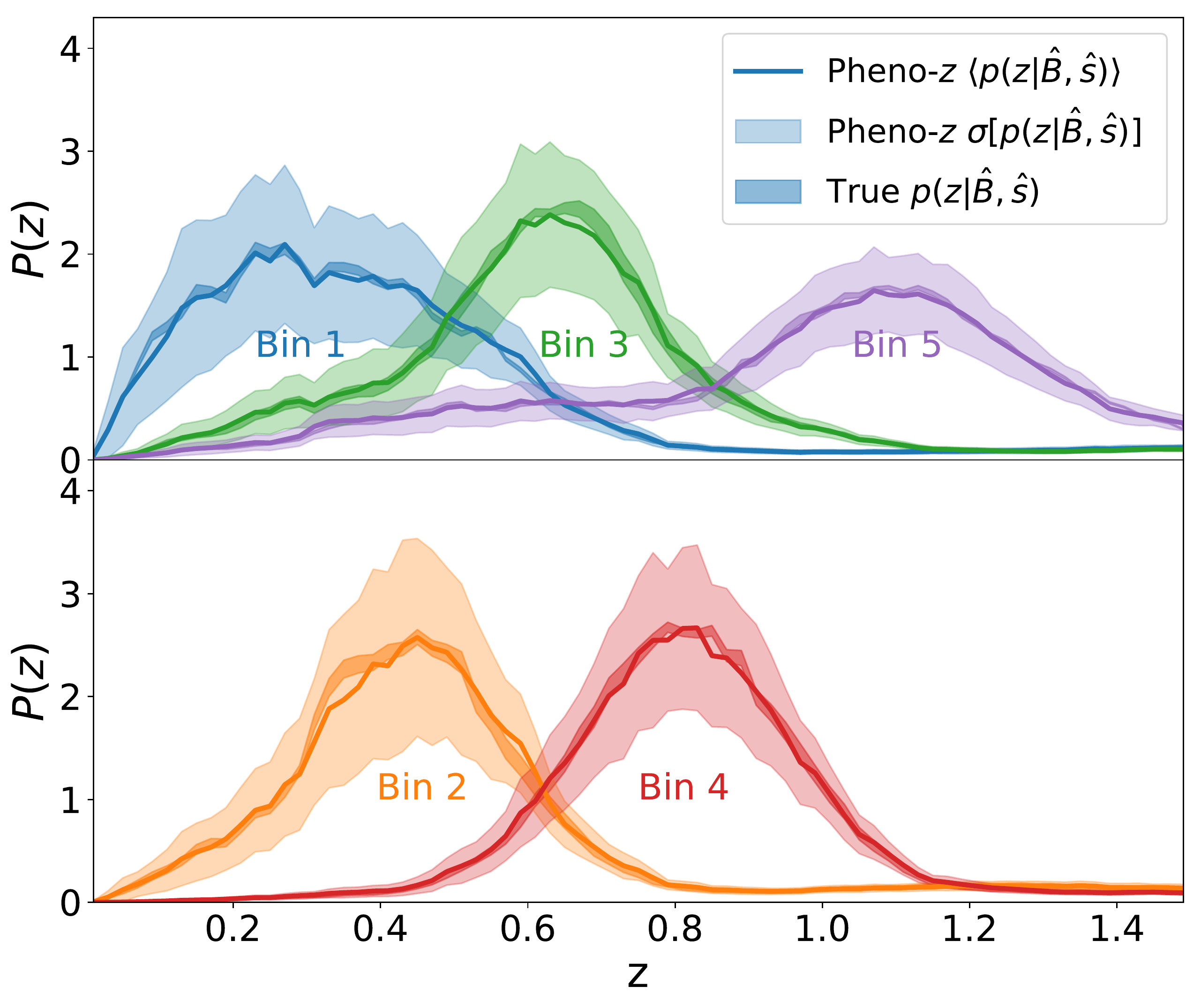}
\caption{Effect of sample variance on the DES Y3 source redshift distributions. 300 realizations of the distributions are computed using the pheno-$z$ scheme. The light shaded regions contain 68\% of these realizations at each redshift. Their means (lines) estimate closely the true redshift distributions (dark shaded regions; mildly different in each realization).}
\label{fig:Pheno-z_DESY3}
\end{figure}

We make use of three deep fields of 3.32, 3.29 and 1.38 $\mathrm{deg}^2$ (see \autoref{tab:vista_bands}), respectively, to train a 128 by 128 Deep SOM.  The input vectors are eight lupticolors relative to the \textit{i} band (using DES \textit{ugriz} and VISTA \textit{YJHKs}). The redshift sample is made of 135,000 galaxies sampled from a 1.38 $\mathrm{deg}^2$ field mimicking COSMOS. Each deep field galaxy is painted 10 times over the full DES footprint to yield the overlap sample used to compute the transfer function. The wide sample is made of randomly selected galaxies over the full DES footprint. The wide and overlap sample selection is performed using the refined weak lensing selection (see \autoref{sec:wl-selection}), and the samples use mock \textsc{Metacalibration} fluxes (see \autoref{sec:metacal_flux}). The 32 by 32 Wide SOM is trained on the wide sample, and does not use the \textit{g} band (see \autoref{sec:drop_g}). Its input vector is $\vec{x} = (\mu_i, \mu_r-\mu_i, \mu_z-\mu_i)$, where $\mu_x$ is the luptitude in $x$ band (see \autoref{luptitutude}). 300 iterations of this pheno-$z$ fiducial scheme are performed with different deep and redshift fields at each iteration. The resulting redshift distributions of the wide sample are presented in \autoref{fig:Pheno-z_DESY3} and the associated metrics in \autoref{tab:final_uncertainty} (entry `DES Y3'). The expectation value of the realizations estimates closely the shape of the true redshift distribution. At each redshift, 68\% of the realizations are comprised in the light shaded area. This broad region is the result of sample variance.
\begin{table*}
\ra{1.2}
\centering
\caption{All effects affecting the calibration of Y3 source redshift distributions are included yielding the expected uncertainty on redshift distributions. The DES Y3 uncertainties are computed on 300 iterations and include the predicted redshift and deep samples size, \textsc{Metacalibration} fluxes, only \textit{riz} bands for the galaxies in the wide sample and the weak lensing selection.  Increasing the deep fields available to a total of 29.88 $\mathrm{deg}^2$ reduces the standard deviation on $\Delta\langle z\rangle$ by 34--41\% and on $\Delta\sigma(z)$ by 35--43\%. Increasing the redshift field area by a factor of four reduces the bias, $\langle\Delta\langle z\rangle\rangle$, by 34--41\% in the first two bins and has a marginal impact on the other bins. The standard deviation of this metric decreases by 17--34\%. From these metrics, total uncertainty is estimated according to \autoref{eqn:total_uncertainty} and shown in \autoref{fig:DES_Y3_uncertainty}.}
\begin{tabular}{@{}l*{10}{S[table-format=1.4, round-mode=places,round-precision=4, group-digits=false]}@{}}
\toprule
\multirow{2}{*}[-3pt]{\textbf{Pheno-$z$ scheme}}   & \multicolumn{1}{c}{\textit{Bin 1}}     & \multicolumn{1}{c}{\textit{Bin 2}}   & \multicolumn{1}{c}{\textit{Bin 3}}    & \multicolumn{1}{c}{\textit{Bin 4}}     & \multicolumn{1}{c}{\textit{Bin 5}}     & \multicolumn{1}{c}{\textit{Bin 1}}     & \multicolumn{1}{c}{\textit{Bin 2}}   & \multicolumn{1}{c}{\textit{Bin 3}}    & \multicolumn{1}{c}{\textit{Bin 4}}     & \multicolumn{1}{c}{\textit{Bin 5}}     \\ \cmidrule(lr){2-11} 
 & \multicolumn{5}{c}{\textbf{$\boldsymbol{\langle\Delta \langle z \rangle\rangle}$}}     & \multicolumn{5}{c}{\textbf{$\boldsymbol{\sigma(\Delta \langle z \rangle)}$}}  \\  \cmidrule(lr){1-1} \cmidrule(lr){2-6}\cmidrule(lr){7-11}
 DES Y3                       & -0.006093 & -0.005594 & 0.000819 & 0.001853 & 0.007972 & 0.007534 & 0.006102 & 0.004734 & 0.003693 & 0.005360 \\
 Bigger deep sample   & -0.006321 & -0.005936 & 0.000967 & 0.001820 & 0.007732 & 0.004428 & 0.003882 & 0.003105 & 0.002240 & 0.003315 \\
Bigger redshift sample      & -0.004013 & -0.003293 & 0.001699 & 0.001686 & 0.007697 & 0.004999 & 0.004376 & 0.003866 & 0.003069 & 0.004386 \\\addlinespace[3pt]
                & \multicolumn{5}{c}{\textbf{$\boldsymbol{\langle\Delta \sigma(z)\rangle}$}} & \multicolumn{5}{c}{\textbf{$\boldsymbol{\sigma(\Delta \sigma(z))}$}}                              \\ \cmidrule(lr){1-1} \cmidrule(lr){2-6}\cmidrule(lr){7-11}
 DES Y3                         & -0.000933 & -0.001647 & -0.003395 & -0.004936 & -0.004096 & 0.004213 & 0.004520 & 0.004160 & 0.002872 & 0.003632 \\
 Bigger deep sample & -0.001330 & -0.001964 & -0.003662 & -0.005403 & -0.004159 & 0.002735 & 0.002716 & 0.002368 & 0.001784 & 0.002309 \\
 Bigger redshift sample      & -0.000728 & -0.000918 & -0.002459 & -0.004340 & -0.003572 & 0.003301 & 0.003371 & 0.002775 & 0.002088 & 0.002782\\ \bottomrule 
\end{tabular}
\label{tab:final_uncertainty}
\end{table*}

\begin{figure}
\centering
\includegraphics[width=\linewidth]{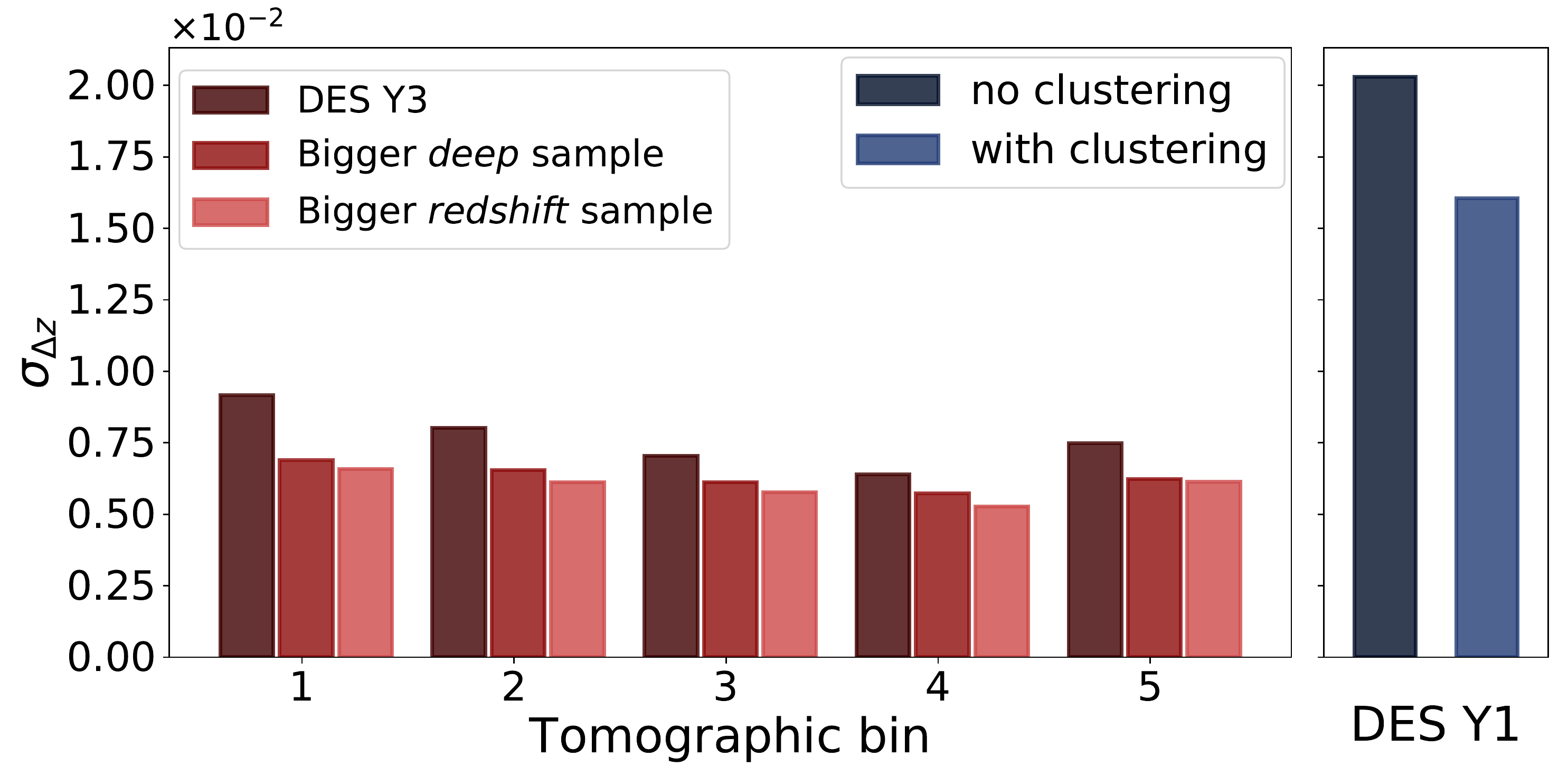}
\caption{Predicted uncertainty in the mean redshift for DES Y3 (\autoref{eqn:total_uncertainty}) compared to the uncertainty of the DES Y1 analysis with and without clustering information \citep{Hoyle2018}. The potential impact on this uncertainty of a bigger deep or redshift sample is also presented (dark red vs. brighter red bars). A caveat to the comparison to DES Y1 is that, unlike the Y1 uncertainty shown here, the pheno-$z$ calibration uncertainty is correlated between bins, however in a way that can be accounted for in the cosmological likelihood.}
\label{fig:DES_Y3_uncertainty}
\end{figure}

To obtain the DES Y3 redshift uncertainty for the \textit{i}th bin, $\sigma_{\Delta z^i}$, we take the root mean square of $\langle\Delta\langle z \rangle\rangle$ and add $\sigma(\Delta\langle z \rangle)$ of the \textit{i}th tomographic bin in quadrature:
\begin{equation}
\sigma_{\Delta z^i}= \sqrt{\frac{1}{N_{\mathrm{bin}}}\sum_{j=1}^{N_{\mathrm{bin}}}\langle \Delta \langle z \rangle \rangle_j^2 +\sigma(\Delta\langle z \rangle)_i^2}.
\label{eqn:total_uncertainty}
\end{equation}
The result is presented in \autoref{fig:DES_Y3_uncertainty} and compared to the DES Y1 results \citep{Hoyle2018}. The pheno-$z$ scheme shows a net improvement by a factor of 2 (55--69\% compared to the Y1 uncertainty without clustering and 43--60\% with clustering; see \citealt{Davis2017}, \citealt{Gatti2018}, and \citealt{Cawthon2018} for details on the clustering redshift method applied to DES Y1).

\begin{figure}
\centering
\includegraphics[width=0.7\linewidth]{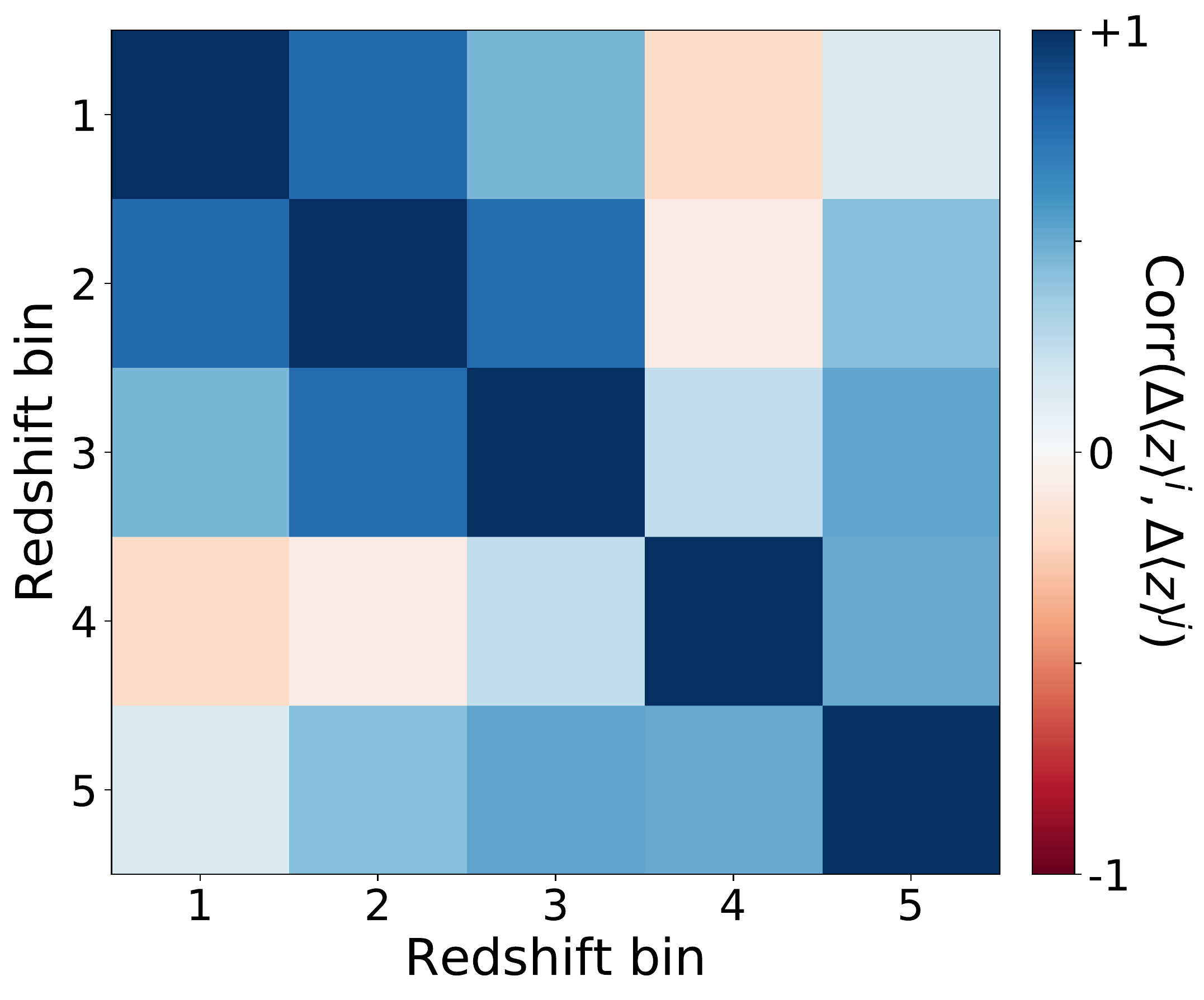}
\caption{Correlation matrix of $\Delta \langle z \rangle$ between redshift bins for the DES Y3 configuration.}
\label{fig:DES_Y3_correlation_matrix}
\end{figure}

The $\Delta \langle z \rangle$ in different bins are correlated which must be accounted for in the inference of cosmological parameters. In the Y1 analysis \citep{Hoyle2018}, the uncertainty on the mean redshift was derived independently for each redshift bin. The off-diagonal elements of the covariance matrix of $\Delta \langle z \rangle$ could not be estimated accurately. \citet{Hoyle2018} showed that increasing the diagonal elements of the covariance matrix by a factor $(1.6)^2$ and zeroing the off-diagonal elements ensured that the uncertainties of any inferred parameters were conservatively estimated for reasonable values of the off-diagonal elements. The DES Y1 values presented in \autoref{fig:DES_Y3_uncertainty} have also been increased by this factor 1.6. We do not include this factor in our DES Y3 estimate, as there we plan to fully marginalize over the correlated redshift distribution uncertainty. Given multiple realizations of the distributions, whose variability due to sample variance is estimated here, we can marginalize over redshift uncertainty fully by directly sampling from these realizations in the cosmological likelihood (Cordero et al.~in preparation). This fully accounts for the correlation between the redshift bins and we expect it to yield reduced -- yet still conservative -- errors on derived quantities, which adds to the improvement in calibration possible with our scheme. The covariance matrix of $\Delta \langle z \rangle$ is presented in \autoref{fig:DES_Y3_correlation_matrix}. As expected, neighboring bins are more correlated and the correlation is higher at low redshift. 

\section{Possible improvements}
\label{sec:possible_improvements}
Our pheno-$z$ scheme applied to the simulated \textsc{Buzzard} catalog allows us to investigate how the calibration could be improved with more data.  As we have previously seen, the bias is limited by the size of our redshift sample whereas the standard deviation of the calibration error is limited by the size of the deep fields. We investigate how both those effects could be mitigated.

\begin{figure}
\centering
\includegraphics[width=\linewidth]{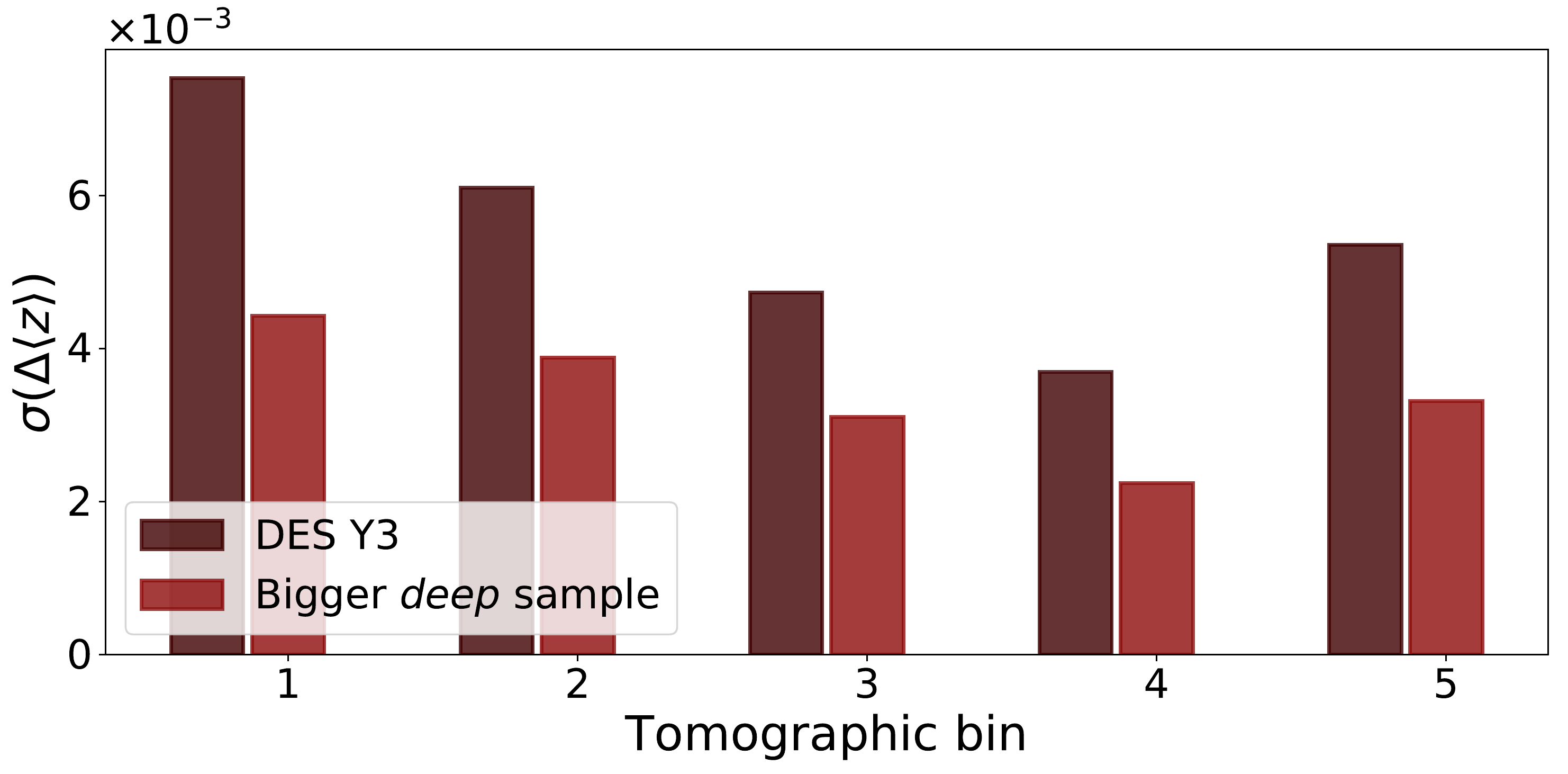}
\caption{Taking VISTA \textit{YJHKs} measurements in the DES supernova fields would increase the deep field area available by $\sim22\;\mathrm{deg}^2$ which would result in a significant decrease of the sample variance.}
\label{fig:DES_Y3_cosmic_variance}
\end{figure}

The major contributor to the cosmic variance in our pheno-$z$ scheme are the deep fields. Increasing their area by taking VISTA \textit{YJHKs} images of the DES supernova fields would allow us to reduce the sample variance. The VISTA Extragalactic Infrared Legacy Survey (VEILS)\footnote{\url{https://www.ast.cam.ac.uk/~mbanerji/VEILS/}} is currently imaging some of the lacking photometry in \textit{J} and \textit{Ks} bands. We estimate that 15, 8, 14 and 8 VISTA pointings in \textit{Y}, \textit{J}, \textit{H} and \textit{Ks} bands, respectively, would be needed to acquire the remaining uncovered DES supernova fields area. Achieving similar depth to the VIDEO survey (see Table 1 of \citealt{Jarvis2013} for planned time per pointing) would require $\sim395$ hours of telescope time. We test this possibility by assuming the availability of five deep fields: one 1.38 $\mathrm{deg}^2$, one 9 $\mathrm{deg}^2$, one 7.5 $\mathrm{deg}^2$ and two 6 $\mathrm{deg}^2$ fields. The mean and standard deviation of the calibration error, $\Delta\langle z \rangle$, over 300 iterations are presented in \autoref{tab:final_uncertainty} (see entry `Bigger deep sample'). As expected the mean of the bias is marginally reduced by the increase of the deep fields area. As shown in \autoref{fig:DES_Y3_cosmic_variance}, the standard deviation decreases by 34--41\%. This significant reduction of the sample variance can also be seen in the standard deviation of $\Delta\sigma(z)$ which decreases by 35--43\%.

The bias of the method is limited by the number of galaxies in the redshift sample. As we have seen in \autoref{sec:capabilities}, increasing the number of cells in the Deep SOM reduces the bias, but those cells must be populated. Therefore we need a large enough sample to populate a bigger SOM. Let us assume that we can increase the number of galaxies for which we have many-band photo-$z$ by a factor of four and use a 256 by 256 Deep SOM. We suppose that we take many-band measurements in three supplementary COSMOS-like (i.e.\ 1.38 $\mathrm{deg}^2$) fields in the DES footprint. In each of these fields we sample 135,000 galaxies. As we have seen in \autoref{sec:lim_spec_sample}, the increase of area should not be contiguous but at different locations on the sky to maximize the sample variance reduction. Furthermore, as the many-bands must include DES \textit{ugriz} and VISTA \textit{YJHKs}, these fields can also be used in the transfer function computation.

The results of increasing the redshift field area by a factor of four is presented in \autoref{tab:final_uncertainty} (entry `Bigger redshift sample'). The effect on $\Delta\langle z\rangle$ and $\Delta\sigma(z)$ is assessed over 300 iterations. The RMS of the calibration error of mean redshift is decreased by 18\%. As the redshift field is also part of the deep fields and is used in the transfer function computation, the standard deviation of $\Delta\langle z\rangle$ also decreases by 17--34\%. 

While there are advantages to the spatial resolution and wavelength coverage of the space-based observations of the COSMOS field, a multi-medium/narrow-band survey like the Advanced Large Homogeneous Area Medium-Band Redshift Astronomical \citep[ALHAMBRA;][]{Moles2008} survey with the appropriate depth might also offer the necessary sample of reliable photometric redshifts. The data could be provided by the ongoing Physics of the Accelerating Universe \citep[PAU;][]{Marti2014, Eriksen2019} survey or by the planned Javalambre Physics of the Accelerating Universe Astrophysical Survey \citep[J-PAS;][]{Benitez2014}. Another option would be to use a sufficiently accurate photometric code on the weight vectors of the Deep SOM. This would yield a redshift PDF for each Deep SOM cell.

The method could be extended by incorporating the information contained in the clustering of sources. A hierarchical Bayesian model can be used to combine the pheno-$z$ method with the information contained in the galaxy clustering against a well-characterized tracer population \citep{Sanchez2018}.

\subsection{Reliability of redshift samples}

One aspect not considered in this work is the reliability of present or obtainable redshift samples. Photometric redshifts based on multi-band fluxes, such as COSMOS \citep{Laigle2016}, are known to suffer from increasing outlier rates towards faint magnitudes. These have recently been found to be a significant concern for the purposes of lensing cosmology, at least under some conditions \citep{Hildebrandt2018}. Likewise, spectroscopic samples can suffer from outliers due to erroneous line identifications or blends. In addition, such a sample may be incomplete at a given position in multi-band color space, albeit superior to a photometric sample with a limited number of bands \citep{Gruen2017}.

Unless complementary information, e.g.\ from clustering, is able to counter these sources of potential bias \citep{Sanchez2018}, we thus emphasize that application of our method requires validating that the redshift samples used are sufficiently reliable. While this is not within the scope of the present work, which is primarily meant to establish the statistical benefits of a phenotypic approach that uses deep field photometry as part of wide field redshift calibration, such tests need to be part of any practical application to data.

\section{Conclusion}
\label{sec:conclusion}

Inferring accurate redshift distributions from coarse measurements of redshifted source photometry is a difficult task. Improving the characterization of redshift distributions requires breaking type/redshift degeneracies. To this end, we propose a novel method -- \textit{phenotypic} redshift -- which uses photometric deep fields, where measurements in more bands are available. The information from multi-band deep fields acts as an intermediary between wide-field photometry and accurate redshifts to produce a better mapping in two ways. Firstly, because the deep fields in surveys like DES are larger than existing samples of galaxies with accurately known redshifts, it provides an improved estimate of the distribution of galaxies in color space. Secondly, the deep many-band photometry better breaks type/redshift degeneracies, thereby improving the color/redshift relation applied in the redshift estimation. Importantly, this reduces sample variance and selection effects due to the sparse sample of galaxies with accurate redshifts and therefore leverages this scarce resource towards a more accurate characterization of the target sample's redshift distribution.

Our implementation of this method uses two self-organizing maps: one to group galaxies into phenotypes based on their observed fluxes in the deep fields, and one to discretize the wide-field flux measurements. By taking actual or simulated observations of the deep fields under wide-field conditions, the transfer of galaxies from cells in one of these maps to the other can be accurately quantified.

Application of the method to simulated galaxy samples allows us to probe the various sources of uncertainty in a coherent manner. 
We tested the method on a mock DES catalog, emulating a calibration of the DES Year 3 weak lensing analysis using DES deep fields with near-infrared auxiliary data, and COSMOS for redshifts. With these samples, the typical uncertainty on the mean redshift in five tomographic bins is $\sigma_{\Delta z} = 0.00$7, which is about a factor of 2 improvement compared to the Year 1 analysis. The method yields realizations of redshift distributions which can be marginalized in the cosmological parameter likelihood, accounting for the correlation between redshift distributions in different tomographic bins. This finding comes with the caveat, shared among all redshift calibration methods that are based on reference samples, that it assumes perfectly accurate redshifts to be known for the COSMOS-like sample used in the calibration.

About half the error is due to systematic biases, $\langle \Delta z \rangle$, in the method. This bias is limited by our ability to populate the deep field SOM on a fine enough grid. If we had more galaxies with accurate redshifts, we could increase the number of phenotypes. The potential gains are rather modest for the effort required: if DES had three additional COSMOS-like fields, the RMS value of the calibration error would be reduced by 18\%. This is a tall order and unlikely to be fulfilled on the timescale of DES. A different solution to the requirement of a high-resolution deep field SOM with redshifts in each cell may be to use a template fitting technique to assign redshift distributions to any cells not covered by spectroscopy.

The error due to sample variance can, on the contrary, be reduced with a somewhat unexpected strategy. We find the sample variance of the method to be dominated by the area covered with deep multi-band photometric observations, rather than the sample of accurate redshifts (\autoref{sec:strangecv}). Designs of future imaging surveys should thus maximize the overlap of their deep fields with complementary photometric surveys. For example, $\sim395$ hours of telescope time would be needed to obtain VISTA \textit{YJHKs} measurements over the rest of the DES supernova fields. This would reduce the sample variance by 34--41\%, and would also be beneficial to redshift calibrations with the overlapping LSST. The Dark Energy Science Collaboration of LSST aims for $\sigma_{\Delta z}=0.002(1+z)$ in their Year 1 analysis and $\sigma_{\Delta z}=0.001(1+z)$ in their Year 10 analysis \citep{LSST2018}, which are challenging requirements. Our tests indicate that while in principle they could be met by a scheme like the one presented here, this would require both an increase in the resolution of the deep field SOM (and thus a larger sample of galaxies with known redshift and accurate multi-band photometry) and a larger volume of purely photometric optical and near-infrared deep fields.

In DES Y1, the information contained in the clustering of sources was used separately to constrain the redshift distribution. The pheno-$z$ method provides a way of combining flux measurements with information contained in the sources' position. A hierarchical Bayesian model allows us to combine the pheno-$z$ method and the information contained in the galaxy clustering against a well-characterized tracer population in a robust way \citep{Sanchez2018}. We intend to apply a variation of this method on DES Y3 data.

Obtaining reliable redshifts to cover the many-color optical/NIR space remains a major observational and modeling challenge. The pheno-$z$ framework can leverage this effort by efficiently using complementary information about the abundance and redshift of observable galaxy types to accurately estimate redshift distributions of ensembles of galaxies selected from photometric data sets.  

\section*{Acknowledgements}

The authors thank Jamie McCullough and numerous members of the DES Collaboration for helpful comments on the manuscript.

This work was supported in part by the U.S. Department of Energy under
contract number DE-AC02-76SF00515. 
Support for DG was provided by NASA through Einstein Postdoctoral
Fellowship grant number PF5-160138 awarded by the Chandra X-ray
Center, which is operated by the Smithsonian Astrophysical Observatory
for NASA under contract NAS8-03060.

Funding for the DES Projects has been provided by the U.S. Department of Energy, the U.S. National Science Foundation, the Ministry of Science and Education of Spain, 
the Science and Technology Facilities Council of the United Kingdom, the Higher Education Funding Council for England, the National Center for Supercomputing 
Applications at the University of Illinois at Urbana-Champaign, the Kavli Institute of Cosmological Physics at the University of Chicago, 
the Center for Cosmology and Astro-Particle Physics at the Ohio State University,
the Mitchell Institute for Fundamental Physics and Astronomy at Texas A\&M University, Financiadora de Estudos e Projetos, 
Funda{\c c}{\~a}o Carlos Chagas Filho de Amparo {\`a} Pesquisa do Estado do Rio de Janeiro, Conselho Nacional de Desenvolvimento Cient{\'i}fico e Tecnol{\'o}gico and 
the Minist{\'e}rio da Ci{\^e}ncia, Tecnologia e Inova{\c c}{\~a}o, the Deutsche Forschungsgemeinschaft and the Collaborating Institutions in the Dark Energy Survey. 

The Collaborating Institutions are Argonne National Laboratory, the University of California at Santa Cruz, the University of Cambridge, Centro de Investigaciones Energ{\'e}ticas, 
Medioambientales y Tecnol{\'o}gicas-Madrid, the University of Chicago, University College London, the DES-Brazil Consortium, the University of Edinburgh, 
the Eidgen{\"o}ssische Technische Hochschule (ETH) Z{\"u}rich, 
Fermi National Accelerator Laboratory, the University of Illinois at Urbana-Champaign, the Institut de Ci{\`e}ncies de l'Espai (IEEC/CSIC), 
the Institut de F{\'i}sica d'Altes Energies, Lawrence Berkeley National Laboratory, the Ludwig-Maximilians Universit{\"a}t M{\"u}nchen and the associated Excellence Cluster Universe, 
the University of Michigan, the National Optical Astronomy Observatory, the University of Nottingham, The Ohio State University, the University of Pennsylvania, the University of Portsmouth, 
SLAC National Accelerator Laboratory, Stanford University, the University of Sussex, Texas A\&M University, and the OzDES Membership Consortium.

Based in part on observations at Cerro Tololo Inter-American Observatory, National Optical Astronomy Observatory, which is operated by the Association of 
Universities for Research in Astronomy (AURA) under a cooperative agreement with the National Science Foundation.

The DES data management system is supported by the National Science Foundation under Grant Numbers AST-1138766 and AST-1536171.
The DES participants from Spanish institutions are partially supported by MINECO under grants AYA2015-71825, ESP2015-66861, FPA2015-68048, SEV-2016-0588, SEV-2016-0597, and MDM-2015-0509, 
some of which include ERDF funds from the European Union. IFAE is partially funded by the CERCA program of the Generalitat de Catalunya.
Research leading to these results has received funding from the European Research
Council under the European Union's Seventh Framework Program (FP7/2007-2013) including ERC grant agreements 240672, 291329, and 306478.
We  acknowledge support from the Australian Research Council Centre of Excellence for All-sky Astrophysics (CAASTRO), through project number CE110001020, and the Brazilian Instituto Nacional de Ci\^encia
e Tecnologia (INCT) e-Universe (CNPq grant 465376/2014-2).

This manuscript has been authored by Fermi Research Alliance, LLC under Contract No. DE-AC02-07CH11359 with the U.S. Department of Energy, Office of Science, Office of High Energy Physics. The United States Government retains and the publisher, by accepting the article for publication, acknowledges that the United States Government retains a non-exclusive, paid-up, irrevocable, world-wide license to publish or reproduce the published form of this manuscript, or allow others to do so, for United States Government purposes.
\begin{center}
\end{center}

This paper has gone through internal review by the DES collaboration.



\bibliography{pheno-z}

\begin{thebibliography}{}
\makeatletter
\relax
\def\mn@urlcharsother{\let\do\@makeother \do\$\do\&\do\#\do\^\do\_\do\%\do\~}
\def\mn@doi{\begingroup\mn@urlcharsother \@ifnextchar [ {\mn@doi@}
  {\mn@doi@[]}}
\def\mn@doi@[#1]#2{\def\@tempa{#1}\ifx\@tempa\@empty \href
  {http://dx.doi.org/#2} {doi:#2}\else \href {http://dx.doi.org/#2} {#1}\fi
  \endgroup}
\def\mn@eprint#1#2{\mn@eprint@#1:#2::\@nil}
\def\mn@eprint@arXiv#1{\href {http://arxiv.org/abs/#1} {{\tt arXiv:#1}}}
\def\mn@eprint@dblp#1{\href {http://dblp.uni-trier.de/rec/bibtex/#1.xml}
  {dblp:#1}}
\def\mn@eprint@#1:#2:#3:#4\@nil{\def\@tempa {#1}\def\@tempb {#2}\def\@tempc
  {#3}\ifx \@tempc \@empty \let \@tempc \@tempb \let \@tempb \@tempa \fi \ifx
  \@tempb \@empty \def\@tempb {arXiv}\fi \@ifundefined
  {mn@eprint@\@tempb}{\@tempb:\@tempc}{\expandafter \expandafter \csname
  mn@eprint@\@tempb\endcsname \expandafter{\@tempc}}}

\bibitem[\protect\citeauthoryear{{Aihara} et~al.,}{{Aihara}
  et~al.}{2018}]{Aihara2018}
{Aihara} H.,  et~al., 2018, \mn@doi [\pasj] {10.1093/pasj/psx066}, \href
  {http://ads.nao.ac.jp/abs/2018PASJ...70S...4A} {70, S4}

\bibitem[\protect\citeauthoryear{{Amon} et~al.,}{{Amon}
  et~al.}{2018}]{Amon2018}
{Amon} A.,  et~al., 2018, \mn@doi [\mnras] {10.1093/mnras/sty859}, \href
  {http://adsabs.harvard.edu/abs/2018MNRAS.477.4285A} {477, 4285}

\bibitem[\protect\citeauthoryear{{Arnouts}, {Cristiani}, {Moscardini},
  {Matarrese}, {Lucchin}, {Fontana}  \& {Giallongo}}{{Arnouts}
  et~al.}{1999}]{Arnouts1999}
{Arnouts} S.,  {Cristiani} S.,  {Moscardini} L.,  {Matarrese} S.,  {Lucchin}
  F.,  {Fontana} A.,   {Giallongo} E.,  1999, \mn@doi [\mnras]
  {10.1046/j.1365-8711.1999.02978.x}, \href
  {https://ui.adsabs.harvard.edu/#abs/1999MNRAS.310..540A} {310, 540}

\bibitem[\protect\citeauthoryear{{Bartelmann} \& {Schneider}}{{Bartelmann} \&
  {Schneider}}{2001}]{Bartelmann2001}
{Bartelmann} M.,  {Schneider} P.,  2001, \mn@doi [\physrep]
  {10.1016/S0370-1573(00)00082-X}, \href
  {http://adsabs.harvard.edu/abs/2001PhR...340..291B} {340, 291}

\bibitem[\protect\citeauthoryear{{Ben{\'\i}tez}}{{Ben{\'\i}tez}}{2000}]{Benitez2000}
{Ben{\'\i}tez} N.,  2000, \mn@doi [\apj] {10.1086/308947}, \href
  {https://ui.adsabs.harvard.edu/#abs/2000ApJ...536..571B} {536, 571}

\bibitem[\protect\citeauthoryear{{Ben{\'\i}tez} et~al.,}{{Ben{\'\i}tez}
  et~al.}{2004}]{Benitez2004}
{Ben{\'\i}tez} N.,  et~al., 2004, \mn@doi [\apjs] {10.1086/380120}, \href
  {https://ui.adsabs.harvard.edu/#abs/2004ApJS..150....1B} {150, 1}

\bibitem[\protect\citeauthoryear{{Benitez} et~al.,}{{Benitez}
  et~al.}{2014}]{Benitez2014}
{Benitez} N.,  et~al., 2014, preprint, \href
  {http://adsabs.harvard.edu/abs/2014arXiv1403.5237B} {} (\mn@eprint {arXiv}
  {1403.5237})

\bibitem[\protect\citeauthoryear{{Bonnett} et~al.,}{{Bonnett}
  et~al.}{2016}]{Bonnett2016}
{Bonnett} C.,  et~al., 2016, \mn@doi [\prd] {10.1103/PhysRevD.94.042005}, \href
  {http://adsabs.harvard.edu/abs/2016PhRvD..94d2005B} {94, 042005}

\bibitem[\protect\citeauthoryear{{Bordoloi}, {Lilly}  \& {Amara}}{{Bordoloi}
  et~al.}{2010}]{Bordoloi2010}
{Bordoloi} R.,  {Lilly} S.~J.,   {Amara} A.,  2010, \mn@doi [\mnras]
  {10.1111/j.1365-2966.2010.16765.x}, \href
  {http://adsabs.harvard.edu/abs/2010MNRAS.406..881B} {406, 881}

\bibitem[\protect\citeauthoryear{{Brett}, {West}  \& {Wheatley}}{{Brett}
  et~al.}{2004}]{Brett2004}
{Brett} D.~R.,  {West} R.~G.,   {Wheatley} P.~J.,  2004, \mn@doi [\mnras]
  {10.1111/j.1365-2966.2004.08093.x}, \href
  {https://ui.adsabs.harvard.edu/#abs/2004MNRAS.353..369B} {353, 369}

\bibitem[\protect\citeauthoryear{{Carrasco Kind} \& {Brunner}}{{Carrasco Kind}
  \& {Brunner}}{2013}]{CarrascoKind2013}
{Carrasco Kind} M.,  {Brunner} R.~J.,  2013, \mn@doi [\mnras]
  {10.1093/mnras/stt574}, \href
  {http://adsabs.harvard.edu/abs/2013MNRAS.432.1483C} {432, 1483}

\bibitem[\protect\citeauthoryear{{Carrasco Kind} \& {Brunner}}{{Carrasco Kind}
  \& {Brunner}}{2014}]{CarrascoKind2014}
{Carrasco Kind} M.,  {Brunner} R.~J.,  2014, \mn@doi [\mnras]
  {10.1093/mnras/stt2456}, \href
  {https://ui.adsabs.harvard.edu/#abs/2014MNRAS.438.3409C} {438, 3409}

\bibitem[\protect\citeauthoryear{{Cawthon} et~al.,}{{Cawthon}
  et~al.}{2018}]{Cawthon2018}
{Cawthon} R.,  et~al., 2018, \mn@doi [\mnras] {10.1093/mnras/sty2424}, \href
  {https://ui.adsabs.harvard.edu/\#abs/2018MNRAS.481.2427C} {481, 2427}

\bibitem[\protect\citeauthoryear{{Collister} \& {Lahav}}{{Collister} \&
  {Lahav}}{2004}]{Collister2004}
{Collister} A.~A.,  {Lahav} O.,  2004, \mn@doi [\pasp] {10.1086/383254}, \href
  {http://adsabs.harvard.edu/abs/2004PASP..116..345C} {116, 345}

\bibitem[\protect\citeauthoryear{{Cooper} et~al.,}{{Cooper}
  et~al.}{2011}]{Cooper2011}
{Cooper} M.~C.,  et~al., 2011, \mn@doi [\apjs] {10.1088/0067-0049/193/1/14},
  \href {https://ui.adsabs.harvard.edu/#abs/2011ApJS..193...14C} {193, 14}

\bibitem[\protect\citeauthoryear{{Dark Energy Survey Collaboration}}{{Dark
  Energy Survey Collaboration}}{2005}]{DarkEnergySurveyCollaboration2005}
{Dark Energy Survey Collaboration} 2005, preprint, \href
  {http://adsabs.harvard.edu/abs/2005astro.ph.10346T} {} (\mn@eprint {arXiv}
  {astro-ph/0510346})

\bibitem[\protect\citeauthoryear{{Davis} et~al.,}{{Davis}
  et~al.}{2017}]{Davis2017}
{Davis} C.,  et~al., 2017, preprint, \href
  {http://adsabs.harvard.edu/abs/2017arXiv171002517D} {} (\mn@eprint {arXiv}
  {1710.02517})

\bibitem[\protect\citeauthoryear{{Davis} et~al.,}{{Davis}
  et~al.}{2018}]{Davis2018}
{Davis} C.,  et~al., 2018, \mn@doi [\mnras] {10.1093/mnras/sty787}, \href
  {http://adsabs.harvard.edu/abs/2018MNRAS.477.2196D} {477, 2196}

\bibitem[\protect\citeauthoryear{{De Vicente}, {S{\'a}nchez}  \&
  {Sevilla-Noarbe}}{{De Vicente} et~al.}{2016}]{DEVicente2016}
{De Vicente} J.,  {S{\'a}nchez} E.,   {Sevilla-Noarbe} I.,  2016, \mn@doi
  [\mnras] {10.1093/mnras/stw857}, \href
  {http://adsabs.harvard.edu/abs/2016MNRAS.459.3078D} {459, 3078}

\bibitem[\protect\citeauthoryear{{DeRose} et~al.,}{{DeRose}
  et~al.}{2019}]{DeRose2019}
{DeRose} J.,  et~al., 2019, preprint, \href
  {https://ui.adsabs.harvard.edu/\#abs/2019arXiv190102401D} {} (\mn@eprint
  {arXiv} {1901.02401})

\bibitem[\protect\citeauthoryear{{Drlica-Wagner} et~al.,}{{Drlica-Wagner}
  et~al.}{2018}]{Drlica-Wagner2018}
{Drlica-Wagner} A.,  et~al., 2018, \mn@doi [\apjs] {10.3847/1538-4365/aab4f5},
  \href {http://adsabs.harvard.edu/abs/2018ApJS..235...33D} {235, 33}

\bibitem[\protect\citeauthoryear{{Emerson}, {Sutherland}, {McPherson}, {Craig},
  {Dalton}  \& {Ward}}{{Emerson} et~al.}{2004}]{Emerson2004}
{Emerson} J.~P.,  {Sutherland} W.~J.,  {McPherson} A.~M.,  {Craig} S.~C.,
  {Dalton} G.~B.,   {Ward} A.~K.,  2004, The Messenger, \href
  {https://ui.adsabs.harvard.edu/#abs/2004Msngr.117...27E} {117, 27}

\bibitem[\protect\citeauthoryear{{Eriksen} et~al.,}{{Eriksen}
  et~al.}{2019}]{Eriksen2019}
{Eriksen} M.,  et~al., 2019, \mn@doi [\mnras] {10.1093/mnras/stz204}, \href
  {https://ui.adsabs.harvard.edu/abs/2019MNRAS.484.4200E} {484, 4200}

\bibitem[\protect\citeauthoryear{{Flaugher} et~al.,}{{Flaugher}
  et~al.}{2015}]{Flaugher2015}
{Flaugher} B.,  et~al., 2015, \mn@doi [\aj] {10.1088/0004-6256/150/5/150},
  \href {https://ui.adsabs.harvard.edu/#abs/2015AJ....150..150F} {150, 150}

\bibitem[\protect\citeauthoryear{{Gatti} et~al.,}{{Gatti}
  et~al.}{2018}]{Gatti2018}
{Gatti} M.,  et~al., 2018, \mn@doi [\mnras] {10.1093/mnras/sty466}, \href
  {http://adsabs.harvard.edu/abs/2018MNRAS.477.1664G} {477, 1664}

\bibitem[\protect\citeauthoryear{{Geach}}{{Geach}}{2012}]{Geach2012}
{Geach} J.~E.,  2012, \mn@doi [\mnras] {10.1111/j.1365-2966.2011.19913.x},
  \href {https://ui.adsabs.harvard.edu/#abs/2012MNRAS.419.2633G} {419, 2633}

\bibitem[\protect\citeauthoryear{{Gruen} \& {Brimioulle}}{{Gruen} \&
  {Brimioulle}}{2017}]{Gruen2017}
{Gruen} D.,  {Brimioulle} F.,  2017, \mn@doi [\mnras] {10.1093/mnras/stx471},
  \href {http://adsabs.harvard.edu/abs/2017MNRAS.468..769G} {468, 769}

\bibitem[\protect\citeauthoryear{{Hildebrandt} et~al.,}{{Hildebrandt}
  et~al.}{2017}]{Hildebrandt2017}
{Hildebrandt} H.,  et~al., 2017, \mn@doi [\mnras] {10.1093/mnras/stw2805},
  \href {http://adsabs.harvard.edu/abs/2017MNRAS.465.1454H} {465, 1454}

\bibitem[\protect\citeauthoryear{{Hildebrandt} et~al.,}{{Hildebrandt}
  et~al.}{2018}]{Hildebrandt2018}
{Hildebrandt} H.,  et~al., 2018, preprint, \href
  {http://adsabs.harvard.edu/abs/2018arXiv181206076H} {} (\mn@eprint {arXiv}
  {1812.06076})

\bibitem[\protect\citeauthoryear{{Honscheid}, {DePoy}  \& {for the DES
  Collaboration}}{{Honscheid} et~al.}{2008}]{Honscheid2008}
{Honscheid} K.,  {DePoy} D.~L.,   {for the DES Collaboration} 2008, preprint,
  \href {http://adsabs.harvard.edu/abs/2008arXiv0810.3600H} {} (\mn@eprint
  {arXiv} {0810.3600})

\bibitem[\protect\citeauthoryear{{Hoyle} et~al.,}{{Hoyle}
  et~al.}{2018}]{Hoyle2018}
{Hoyle} B.,  et~al., 2018, \mn@doi [\mnras] {10.1093/mnras/sty957}, \href
  {https://ui.adsabs.harvard.edu/#abs/2018MNRAS.478..592H} {478, 592}

\bibitem[\protect\citeauthoryear{{Hu}}{{Hu}}{1999}]{Hu1999}
{Hu} W.,  1999, \mn@doi [\apj] {10.1086/312210}, \href
  {https://ui.adsabs.harvard.edu/#abs/1999ApJ...522L..21H} {522, L21}

\bibitem[\protect\citeauthoryear{{Huff} \& {Mandelbaum}}{{Huff} \&
  {Mandelbaum}}{2017}]{Huff2017}
{Huff} E.,  {Mandelbaum} R.,  2017, preprint, \href
  {http://adsabs.harvard.edu/abs/2017arXiv170202600H} {} (\mn@eprint {arXiv}
  {1702.02600})

\bibitem[\protect\citeauthoryear{{Huterer}, {Takada}, {Bernstein}  \&
  {Jain}}{{Huterer} et~al.}{2006}]{Huterer2006}
{Huterer} D.,  {Takada} M.,  {Bernstein} G.,   {Jain} B.,  2006, \mn@doi
  [\mnras] {10.1111/j.1365-2966.2005.09782.x}, \href
  {https://ui.adsabs.harvard.edu/#abs/2006MNRAS.366..101H} {366, 101}

\bibitem[\protect\citeauthoryear{{Ilbert} et~al.,}{{Ilbert}
  et~al.}{2006}]{Ilbert2006}
{Ilbert} O.,  et~al., 2006, \mn@doi [\aap] {10.1051/0004-6361:20065138}, \href
  {https://ui.adsabs.harvard.edu/#abs/2006A&A...457..841I} {457, 841}

\bibitem[\protect\citeauthoryear{{Ivezi{\'c}} et~al.,}{{Ivezi{\'c}}
  et~al.}{2008}]{Ivezic2008}
{Ivezi{\'c}} {\v Z}.,  et~al., 2008, preprint, \href
  {http://adsabs.harvard.edu/abs/2008arXiv0805.2366I} {} (\mn@eprint {arXiv}
  {0805.2366})

\bibitem[\protect\citeauthoryear{{Jarvis} et~al.,}{{Jarvis}
  et~al.}{2013}]{Jarvis2013}
{Jarvis} M.~J.,  et~al., 2013, \mn@doi [\mnras] {10.1093/mnras/sts118}, \href
  {https://ui.adsabs.harvard.edu/#abs/2013MNRAS.428.1281J} {428, 1281}

\bibitem[\protect\citeauthoryear{{Kilbinger}}{{Kilbinger}}{2015}]{Kilbinger2015}
{Kilbinger} M.,  2015, \mn@doi [Reports on Progress in Physics]
  {10.1088/0034-4885/78/8/086901}, \href
  {http://adsabs.harvard.edu/abs/2015RPPh...78h6901K} {78, 086901}

\bibitem[\protect\citeauthoryear{Kohonen}{Kohonen}{1982}]{Kohonen1982}
Kohonen T.,  1982, \mn@doi [Biological Cybernetics] {10.1007/BF00337288}, 43,
  59

\bibitem[\protect\citeauthoryear{Kohonen}{Kohonen}{2001}]{Kohonen2001}
Kohonen T.,  2001, Self-organizing maps, 3rd edn.
Springer series in information sciences, 30, Springer-Verlag, Berlin,
  Heidelberg, \mn@doi{10.1007/978-3-642-56927-2}

\bibitem[\protect\citeauthoryear{{Laigle} et~al.,}{{Laigle}
  et~al.}{2016}]{Laigle2016}
{Laigle} C.,  et~al., 2016, \mn@doi [\apjs] {10.3847/0067-0049/224/2/24}, \href
  {https://ui.adsabs.harvard.edu/#abs/2016ApJS..224...24L} {224, 24}

\bibitem[\protect\citeauthoryear{{Larson}, {Tinsley}  \& {Caldwell}}{{Larson}
  et~al.}{1980}]{Larson1980}
{Larson} R.~B.,  {Tinsley} B.~M.,   {Caldwell} C.~N.,  1980, \mn@doi [\apj]
  {10.1086/157917}, \href
  {https://ui.adsabs.harvard.edu/#abs/1980ApJ...237..692L} {237, 692}

\bibitem[\protect\citeauthoryear{{Laureijs} et~al.,}{{Laureijs}
  et~al.}{2011}]{Laureijs2011}
{Laureijs} R.,  et~al., 2011, preprint, \href
  {http://adsabs.harvard.edu/abs/2011arXiv1110.3193L} {} (\mn@eprint {arXiv}
  {1110.3193})

\bibitem[\protect\citeauthoryear{{Lima}, {Cunha}, {Oyaizu}, {Frieman}, {Lin}
  \& {Sheldon}}{{Lima} et~al.}{2008}]{Lima2008}
{Lima} M.,  {Cunha} C.~E.,  {Oyaizu} H.,  {Frieman} J.,  {Lin} H.,   {Sheldon}
  E.~S.,  2008, \mn@doi [\mnras] {10.1111/j.1365-2966.2008.13510.x}, \href
  {http://adsabs.harvard.edu/abs/2008MNRAS.390..118L} {390, 118}

\bibitem[\protect\citeauthoryear{{Lupton}, {Gunn}  \& {Szalay}}{{Lupton}
  et~al.}{1999}]{Lupton1999}
{Lupton} R.~H.,  {Gunn} J.~E.,   {Szalay} A.~S.,  1999, \mn@doi [\aj]
  {10.1086/301004}, \href
  {https://ui.adsabs.harvard.edu/#abs/1999AJ....118.1406L} {118, 1406}

\bibitem[\protect\citeauthoryear{{Ma}, {Hu}  \& {Huterer}}{{Ma}
  et~al.}{2006}]{Ma2006}
{Ma} Z.,  {Hu} W.,   {Huterer} D.,  2006, \mn@doi [\apj] {10.1086/497068},
  \href {https://ui.adsabs.harvard.edu/#abs/2006ApJ...636...21M} {636, 21}

\bibitem[\protect\citeauthoryear{{MacCrann} et~al.,}{{MacCrann}
  et~al.}{2018}]{MacCrann2018}
{MacCrann} N.,  et~al., 2018, \mn@doi [\mnras] {10.1093/mnras/sty1899}, \href
  {https://ui.adsabs.harvard.edu/abs/2018MNRAS.480.4614M} {480, 4614}

\bibitem[\protect\citeauthoryear{{Malz}, {Marshall}, {DeRose}, {Graham},
  {Schmidt}  \& {Wechsler}}{{Malz} et~al.}{2018}]{Malz2018}
{Malz} A.~I.,  {Marshall} P.~J.,  {DeRose} J.,  {Graham} M.~L.,  {Schmidt}
  S.~J.,   {Wechsler} R.,  2018, \mn@doi [\aj] {10.3847/1538-3881/aac6b5},
  \href {http://adsabs.harvard.edu/abs/2018AJ....156...35M} {156, 35}

\bibitem[\protect\citeauthoryear{{Mandelbaum}}{{Mandelbaum}}{2018}]{Mandelbaum2018}
{Mandelbaum} R.,  2018, \mn@doi [\araa] {10.1146/annurev-astro-081817-051928},
  \href {https://ui.adsabs.harvard.edu/abs/2018ARA&A..56..393M} {56, 393}

\bibitem[\protect\citeauthoryear{{Mart{\'{\i}}}, {Miquel}, {Castander},
  {Gazta{\~n}aga}, {Eriksen}  \& {S{\'a}nchez}}{{Mart{\'{\i}}}
  et~al.}{2014}]{Marti2014}
{Mart{\'{\i}}} P.,  {Miquel} R.,  {Castander} F.~J.,  {Gazta{\~n}aga} E.,
  {Eriksen} M.,   {S{\'a}nchez} C.,  2014, \mn@doi [\mnras]
  {10.1093/mnras/stu801}, \href
  {http://adsabs.harvard.edu/abs/2014MNRAS.442...92M} {442, 92}

\bibitem[\protect\citeauthoryear{{Masters} et~al.,}{{Masters}
  et~al.}{2015}]{Masters2015}
{Masters} D.,  et~al., 2015, \mn@doi [\apj] {10.1088/0004-637X/813/1/53}, \href
  {https://ui.adsabs.harvard.edu/#abs/2015ApJ...813...53M} {813, 53}

\bibitem[\protect\citeauthoryear{{Masters}, {Stern}, {Cohen}, {Capak},
  {Rhodes}, {Castander}  \& {Paltani}}{{Masters} et~al.}{2017}]{Masters2017}
{Masters} D.~C.,  {Stern} D.~K.,  {Cohen} J.~G.,  {Capak} P.~L.,  {Rhodes}
  J.~D.,  {Castander} F.~J.,   {Paltani} S.,  2017, \mn@doi [\apj]
  {10.3847/1538-4357/aa6f08}, \href
  {https://ui.adsabs.harvard.edu/#abs/2017ApJ...841..111M} {841, 111}

\bibitem[\protect\citeauthoryear{{Masters} et~al.,}{{Masters}
  et~al.}{2019}]{Masters2019}
{Masters} D.~C.,  et~al., 2019, preprint, \href
  {https://ui.adsabs.harvard.edu/abs/2019arXiv190406394M} {} (\mn@eprint
  {arXiv} {1904.06394})

\bibitem[\protect\citeauthoryear{{McCracken} et~al.,}{{McCracken}
  et~al.}{2012}]{McCracken2012}
{McCracken} H.~J.,  et~al., 2012, \mn@doi [\aap] {10.1051/0004-6361/201219507},
  \href {https://ui.adsabs.harvard.edu/#abs/2012A&A...544A.156M} {544, A156}

\bibitem[\protect\citeauthoryear{{M{\'e}nard}, {Scranton}, {Schmidt},
  {Morrison}, {Jeong}, {Budavari}  \& {Rahman}}{{M{\'e}nard}
  et~al.}{2013}]{Menard2013}
{M{\'e}nard} B.,  {Scranton} R.,  {Schmidt} S.,  {Morrison} C.,  {Jeong} D.,
  {Budavari} T.,   {Rahman} M.,  2013, preprint, \href
  {http://adsabs.harvard.edu/abs/2013arXiv1303.4722M} {} (\mn@eprint {arXiv}
  {1303.4722})

\bibitem[\protect\citeauthoryear{{Miyazaki} et~al.,}{{Miyazaki}
  et~al.}{2002}]{Miyazaki2002}
{Miyazaki} S.,  et~al., 2002, \mn@doi [\pasj] {10.1093/pasj/54.6.833}, \href
  {https://ui.adsabs.harvard.edu/#abs/2002PASJ...54..833M} {54, 833}

\bibitem[\protect\citeauthoryear{{Moles} et~al.,}{{Moles}
  et~al.}{2008}]{Moles2008}
{Moles} M.,  et~al., 2008, \mn@doi [\aj] {10.1088/0004-6256/136/3/1325}, \href
  {http://adsabs.harvard.edu/abs/2008AJ....136.1325M} {136, 1325}

\bibitem[\protect\citeauthoryear{{Naim}, {Ratnatunga}  \& {Griffiths}}{{Naim}
  et~al.}{1997}]{Naim1997}
{Naim} A.,  {Ratnatunga} K.~U.,   {Griffiths} R.~E.,  1997, \mn@doi [\apjs]
  {10.1086/313022}, \href
  {https://ui.adsabs.harvard.edu/#abs/1997ApJS..111..357N} {111, 357}

\bibitem[\protect\citeauthoryear{{Newman}}{{Newman}}{2008}]{Newman2008}
{Newman} J.~A.,  2008, \mn@doi [\apj] {10.1086/589982}, \href
  {http://adsabs.harvard.edu/abs/2008ApJ...684...88N} {684, 88}

\bibitem[\protect\citeauthoryear{{Plazas} \& {Bernstein}}{{Plazas} \&
  {Bernstein}}{2012}]{Plazas2012}
{Plazas} A.,  {Bernstein} G.,  2012, \mn@doi [\pasp] {10.1086/668294}, \href
  {http://adsabs.harvard.edu/abs/2012PASP..124.1113A} {124, 1113}

\bibitem[\protect\citeauthoryear{{Rajaniemi} \& {M{\"a}h{\"o}nen}}{{Rajaniemi}
  \& {M{\"a}h{\"o}nen}}{2002}]{Rajaniemi2002}
{Rajaniemi} H.~J.,  {M{\"a}h{\"o}nen} P.,  2002, \mn@doi [\apj]
  {10.1086/337959}, \href
  {https://ui.adsabs.harvard.edu/#abs/2002ApJ...566..202R} {566, 202}

\bibitem[\protect\citeauthoryear{{Samuroff}, {Troxel}, {Bridle}, {Zuntz},
  {MacCrann}, {Krause}, {Eifler}  \& {Kirk}}{{Samuroff}
  et~al.}{2017}]{Samuroff2017}
{Samuroff} S.,  {Troxel} M.~A.,  {Bridle} S.~L.,  {Zuntz} J.,  {MacCrann} N.,
  {Krause} E.,  {Eifler} T.,   {Kirk} D.,  2017, \mn@doi [\mnras]
  {10.1093/mnrasl/slw201}, \href
  {http://adsabs.harvard.edu/abs/2017MNRAS.465L..20S} {465, L20}

\bibitem[\protect\citeauthoryear{{S{\'a}nchez} \& {Bernstein}}{{S{\'a}nchez} \&
  {Bernstein}}{2018}]{Sanchez2018}
{S{\'a}nchez} C.,  {Bernstein} G.~M.,  2018, preprint, \href
  {http://adsabs.harvard.edu/abs/2018arXiv180711873S} {} (\mn@eprint {arXiv}
  {1807.11873})

\bibitem[\protect\citeauthoryear{{Schmidt}, {M{\'e}nard}, {Scranton},
  {Morrison}  \& {McBride}}{{Schmidt} et~al.}{2013}]{Schmidt2013}
{Schmidt} S.~J.,  {M{\'e}nard} B.,  {Scranton} R.,  {Morrison} C.,   {McBride}
  C.~K.,  2013, \mn@doi [\mnras] {10.1093/mnras/stt410}, \href
  {http://adsabs.harvard.edu/abs/2013MNRAS.431.3307S} {431, 3307}

\bibitem[\protect\citeauthoryear{{Schneider}, {Knox}, {Zhan}  \&
  {Connolly}}{{Schneider} et~al.}{2006}]{Schneider2006}
{Schneider} M.,  {Knox} L.,  {Zhan} H.,   {Connolly} A.,  2006, \mn@doi [\apj]
  {10.1086/507675}, \href {http://adsabs.harvard.edu/abs/2006ApJ...651...14S}
  {651, 14}

\bibitem[\protect\citeauthoryear{{Sheldon} \& {Huff}}{{Sheldon} \&
  {Huff}}{2017}]{Sheldon2017}
{Sheldon} E.~S.,  {Huff} E.~M.,  2017, \mn@doi [\apj]
  {10.3847/1538-4357/aa704b}, \href
  {http://adsabs.harvard.edu/abs/2017ApJ...841...24S} {841, 24}

\bibitem[\protect\citeauthoryear{{Spergel} et~al.,}{{Spergel}
  et~al.}{2013}]{Spergel2013}
{Spergel} D.,  et~al., 2013, preprint, \href
  {http://adsabs.harvard.edu/abs/2013arXiv1305.5422S} {} (\mn@eprint {arXiv}
  {1305.5422})

\bibitem[\protect\citeauthoryear{{Springel}}{{Springel}}{2005}]{Springel2005}
{Springel} V.,  2005, \mn@doi [\mnras] {10.1111/j.1365-2966.2005.09655.x},
  \href {https://ui.adsabs.harvard.edu/#abs/2005MNRAS.364.1105S} {364, 1105}

\bibitem[\protect\citeauthoryear{{Strateva} et~al.,}{{Strateva}
  et~al.}{2001}]{Strateva2001}
{Strateva} I.,  et~al., 2001, \mn@doi [\aj] {10.1086/323301}, \href
  {https://ui.adsabs.harvard.edu/#abs/2001AJ....122.1861S} {122, 1861}

\bibitem[\protect\citeauthoryear{{Suchyta} et~al.,}{{Suchyta}
  et~al.}{2016}]{Suchyta2016}
{Suchyta} E.,  et~al., 2016, \mn@doi [\mnras] {10.1093/mnras/stv2953}, \href
  {https://ui.adsabs.harvard.edu/#abs/2016MNRAS.457..786S} {457, 786}

\bibitem[\protect\citeauthoryear{{The LSST Dark Energy Science
  Collaboration}}{{The LSST Dark Energy Science
  Collaboration}}{2018}]{LSST2018}
{The LSST Dark Energy Science Collaboration} 2018, preprint, \href
  {http://adsabs.harvard.edu/abs/2018arXiv180901669T} {} (\mn@eprint {arXiv}
  {1809.01669})

\bibitem[\protect\citeauthoryear{{Troxel} et~al.,}{{Troxel}
  et~al.}{2018}]{Troxel2018}
{Troxel} M.~A.,  et~al., 2018, \mn@doi [\prd] {10.1103/PhysRevD.98.043528},
  \href {https://ui.adsabs.harvard.edu/abs/2018PhRvD..98d3528T} {98, 043528}

\bibitem[\protect\citeauthoryear{{Way} \& {Klose}}{{Way} \&
  {Klose}}{2012}]{Way2012}
{Way} M.~J.,  {Klose} C.~D.,  2012, \mn@doi [\pasp] {10.1086/664796}, \href
  {https://ui.adsabs.harvard.edu/#abs/2012PASP..124..274W} {124, 274}

\bibitem[\protect\citeauthoryear{{Wright} et~al.,}{{Wright}
  et~al.}{2018}]{Wright2018}
{Wright} A.~H.,  et~al., 2018, preprint, \href
  {http://adsabs.harvard.edu/abs/2018arXiv181206077W} {} (\mn@eprint {arXiv}
  {1812.06077})

\bibitem[\protect\citeauthoryear{{Zuntz} et~al.,}{{Zuntz}
  et~al.}{2018}]{Zuntz2018}
{Zuntz} J.,  et~al., 2018, \mn@doi [\mnras] {10.1093/mnras/sty2219}, \href
  {http://adsabs.harvard.edu/abs/2018MNRAS.481.1149Z} {481, 1149}

\bibitem[\protect\citeauthoryear{{de Jong}, {Verdoes Kleijn}, {Kuijken}  \&
  {Valentijn}}{{de Jong} et~al.}{2013}]{deJong2013}
{de Jong} J.~T.~A.,  {Verdoes Kleijn} G.~A.,  {Kuijken} K.~H.,   {Valentijn}
  E.~A.,  2013, \mn@doi [Experimental Astronomy] {10.1007/s10686-012-9306-1},
  \href {http://adsabs.harvard.edu/abs/2013ExA....35...25D} {35, 25}

\makeatother
\end{thebibliography}



\appendix
\section{Softening parameter of luptitudes}
\label{app:softB}
The use of luptitudes to build the input vector of our scheme requires the choice of a softening parameter, $b$, which sets the scale at which luptitudes transition between logarithmic and linear behavior (see \autoref{luptitutude}). 

Two requirements guide the choice of this parameter. The differences between luptitudes and magnitudes for high signal-to-noise data as well as the luptitude variance at low flux levels should be minimized. The former is the intrinsic goal of luptitudes while the latter is not strictly required. It is just convenient if the luptiude variance at zero flux is comparable to its variance at a small signal-to-noise ratio. These two effects oppose each other. \citet{Lupton1999} minimize a total penalty made of the addition of those two effects modeled as costs. The optimal choice of $b$ with their penalty is $b =1.042\sigma$ where $\sigma^2$ is the variance of the flux. This assumes all objects have the same error.

In reality, measurement errors change as the observation conditions change. It could be possible to set different values of $b$ for different parts of the sky although it is unpractical. We can set $b$ for the whole footprint using a typical seeing quality and sky brightness for a given band. \citet{Lupton1999} show that even if the softening parameter is badly chosen, it does not result in catastrophic definition of luptitudes; we recover the expected behavior.

Our measurement errors in each band $x$, are computed using
\begin{equation}
\label{soft_b_sigma_lim}
\sigma_x = \frac{1}{n_{\sigma}}10^{\frac{22.5-l_x}{2.5}},
\end{equation}
where $l_x$ is the limiting magnitude of the survey in $x$ band  and $n_{\sigma}$ is the number of $\sigma$ at which the limiting magnitude is quoted. For the DES bands, the DES Y1 limiting magnitudes of \citet{Drlica-Wagner2018}, quoted at 10-$\sigma$, are used : $u=23.7$, $g=23.5$, $r=22.9$, $i=22.2$, $z=25$. DES Y3 has similar depth as Y1, but over the full survey area. For the VISTA bands, we use the VIDEO limiting magnitudes, which are quoted at 5-$\sigma$ : $Y=24.6$, $J=24.5$, $H=24.0$, $Ks=23.5$.

The limiting magnitudes used are conservative as the DES deep measurements are expected to be at a higher depth. Also, our simulations have true fluxes which have no errors. We test the sensitivity of our quoted uncertainties to the softening parameter by running our pheno-$z$ scheme with limiting magnitudes in all deep bands increased by one magnitude (thus decreasing the softening parameter). The result is presented in \autoref{tab:YJHK_comparisons} (entry `w/ decreased softening parameter'). There is no significant change in our metrics. Hence, we are insensitive to such a change of the softening parameter.

\section{Validation of feature and SOM size choice}
\label{app:val_features_size}
The choice of a lupticolor 128 by 128 Deep SOM coupled to a 32 by 32 Wide SOM must be validated empirically. To this end, we train a variety 
of 128 by 128 Deep SOM, trained using either colors or lupticolors, and a variety of 32 by 32 Wide SOM, trained on several different features including lupticolors, colors, lupticolors and a luptitude, colors and a magnitude. For this test the samples consist of galaxies randomly selected over the whole Y3 footprint to avoid any sample variance. Also, the number of galaxies in the redshift sample is sufficient to minimize the shot noise effect. The difference in mean redshift between the true distribution and the one estimated with our scheme is reported in \autoref{tab:features}. We also report the overlap, $\mathcal{O}$, between bins, i.e.\ the fraction of galaxies assigned to a bin which does not have the highest $\mathrm{d}n/\mathrm{d}z$ at their true redshift:
\begin{equation}
\mathcal{O} = \sum_{i=1}^{N_{\mathrm{bin}}}\int_{z\,:\,n^{i}(z)<\max_{j} n^{j}(z)} dz\;n^{i}(z)
\end{equation}
where $n^{i}(z)=p(z|i, \hat{s})N(i)$ is the unnormalized redshift distribution in bin $i$ and $N_{\mathrm{bin}}$ is the number of tomographic bins.
\begin{table*}
\caption{Calibration error, $\Delta \langle z \rangle$, and overlap between bins for different choices of features. \textit{Color} denotes a difference in magnitudes whereas \textit{lupticolor} denotes a difference in luptitudes. All those differences are with respect to the \textit{i} band. Magnitude and luptitude denote adding the \textit{i} band magnitude and luptitude, respectively. The Deep SOMs are 128 by 128 and the Wide SOMs are 32 by 32. All samples used in this test are randomly selected over the whole sky. There is no significant difference between the features used. We choose to use lupticolor for the Deep SOM and lupticolor + luptitude for the Wide SOM as it is a convenient way to deal with objects which are not measured in some bands (see \autoref{sec:choice_features}).}
\ra{1.2}
\centering
\begin{tabular}{ll*{5}{S[table-format=1.4, round-mode=places,round-precision=4, group-digits=false]}c}
\toprule
\multirow{2}{*}[-2pt]{\textbf{Deep SOM}}                                    & \multirow{2}{*}[-2pt]{\textbf{Wide SOM}}                                     & \multicolumn{5}{c}{\textbf{$\boldsymbol{\Delta \langle z \rangle}$ in bin}}                                                                                                    & \multirow{2}{*}[-2pt]{\textbf{Overlap}} \\   \cmidrule(lr){3-7}
  &  &  \multicolumn{1}{c}{\textit{1}} & \multicolumn{1}{c}{\textit{2}} & \multicolumn{1}{c}{\textit{3}} & \multicolumn{1}{c}{\textit{4}} & \multicolumn{1}{c}{\textit{5}} &      \\ \midrule
color                                         & lupticolor                                   & 0.00006                        & -0.00090                       & 0.00020                        & 0.00335                        & 0.01056                        & 0.33                           \\
lupticolor                                                        & lupticolor                                                    & -0.00110                       & -0.00220                       & -0.00110                       & 0.00263                        & 0.00799                        & 0.34                              \\ \rowcolor{Gray} lupticolor         & lupticolor + luptitude    &  -0.00292 & -0.00202 & -0.00042 & 0.00223 &  0.00724 & 0.37       \\
color                                                 & lupticolor + luptitude                                         & -0.00159                       & -0.00135                       & 0.00042                        & 0.00217                        & 0.01125                        & 0.37                              \\
color                                                     & color                                                        & 0.00050                        & -0.00107                       & 0.00013                        & 0.00374                        & 0.01052                        & 0.34                              \\
color                                                  & color + magnitude                                           & 0.00135                        & -0.00091                       & 0.00026                        & 0.00279                        & 0.01031                        & 0.38                              \\ \bottomrule                            
\end{tabular}
\label{tab:features}
\end{table*}
We find that the choice of features does not matter very much. The first 4 bins have a calibration error, for all features combination tested, of $\Delta \langle z \rangle < 0.005$ which is acceptable for our purpose. The last bin has a larger calibration error, reaching $\Delta \langle z \rangle > 0.01$, but it is the most suspect one in the simulations as it is constructed from a hard cutoff at $z=1.5$. We stick to the Deep lupticolor and Wide lupticolor-luptitude SOMs for the reasons mentioned in \autoref{sec:choice_features}.

\begin{figure}
\centering
\includegraphics[width=\linewidth]{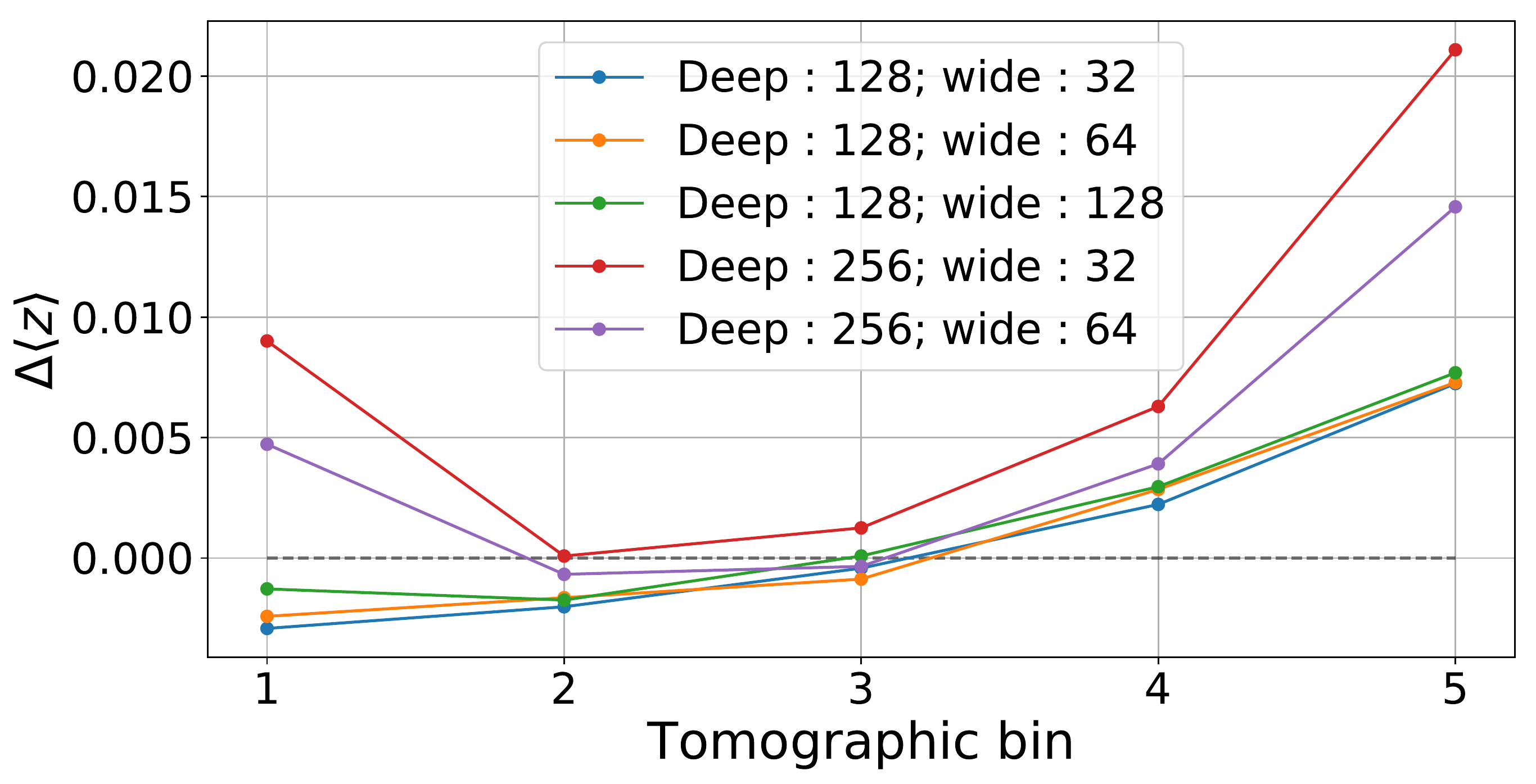}
\caption{Impact of the number of Deep and Wide SOM cells on the redshift distribution estimation. An unbiased method would give $\Delta\langle z\rangle=0$ for all bins (gray dashed line). The Deep SOM is trained on lupticolors and the Wide SOM on lupticolors and the luptitude in \textit{i} band. All SOMs are square and the size given in the legend is the number of cells on a side (e.g.\ 128 means a 128 by 128 SOM).   \label{fig:som_size}}
\end{figure}

To test the impact of the size of the SOMs on our scheme, we use a realistic redshift sample ($10^5$ galaxies) and different SOM sizes. The result, presented in \autoref{fig:som_size}, shows that increasing the size of the Deep SOM results in a larger calibration error whereas increasing the size of the Wide SOM does not result in any improvement. We therefore stick to the 128 by 128 Deep SOM and 32 by 32 Wide SOM.

\section{Calculation of Covariance and Inverse Covariance Matrices}
\label{app:covariance}

We present the analytic forms for calculating the covariance and inverse covariance matrices for two cases:
\begin{itemize}
    \item{differences of magnitudes or luptitudes with respect to a reference magnitude or luptitude, and including the reference magnitude or luptitude;}
   	\item{differences of magnitudes or luptitudes with respect to a reference magnitude or luptitude, \textit{not} including the reference magnitude or luptitude.}
\end{itemize}
For example, we might have the bands $g$, $r$, and $i$, and we might decide to use the $i$ band as the reference band. Let us call the errors in each band $\sigma_x$. We assume each band is independently measured. We define the four combinations of covariance terms between the reference band magnitude and the colors:
\begin{itemize}
    \item{ The covariance between the reference band magnitude and itself, \begin{equation}\Sigma_{i, i} = \sigma^2_i \ . \end{equation}}
    \item{ The covariance between the reference band magnitude and a color, \begin{equation}\Sigma_{i, g-i} = -\sigma^2_i \ . \end{equation}}
    \item{ The covariance between a color and itself, \begin{equation}\Sigma_{g-i, g-i} = \sigma^2_g + \sigma^2_i \ . \end{equation}}
    \item{ The covariance between one color and a second, \begin{equation}\Sigma_{g-i, r-i} = \sigma^2_i \ . \end{equation}}
\end{itemize}
If our input vector is $\vec{x} = (m_i, m_g-m_i, m_r-m_i)$, then its covariance matrix is
\begin{equation}
    \Sigma = \begin{bmatrix}
        \sigma^2_i & -\sigma^2_i & -\sigma^2_i  \\
        -\sigma^2_i & \sigma^2_i + \sigma^2_g & \sigma^2_i \\
        -\sigma^2_i & \sigma^2_i & \sigma^2_i + \sigma^2_r \\
    \end{bmatrix} \ .
\end{equation}
If our input vector is $\vec{x} = (m_g-m_i, m_r-m_i)$, then its covariance matrix is
\begin{equation}
    \Sigma = \begin{bmatrix}
    \sigma^2_i + \sigma^2_g & \sigma^2_i \\
    \sigma^2_i & \sigma^2_i + \sigma^2_r \\
    \end{bmatrix} \ .
\end{equation}

We are interested in the inverse covariance matrix for \autoref{assignment_som}. We could numerically invert the covariance matrices constructed from the above rules, but we find a significant speedup (about 60 times faster) from using analytic formulas for the inverse covariance. When our input vector is the reference magnitude and differences with respect to the reference band, for example $\vec{x} = (m_i, m_g-m_i, m_r-m_i)$, then the inverse covariance terms are as follows:
\begin{itemize}
    \item{ The inverse covariance between the reference band magnitude and itself, \begin{equation}\Sigma^{-1}_{i, i} = \sum_f \frac{1}{\sigma^2_f} \ , \end{equation} where the sum is over all flux passbands $f$.}
    \item{ The inverse covariance between the reference band magnitude and a color, \begin{equation}\Sigma^{-1}_{i, g-i} = \frac{1}{\sigma^2_g} \ . \end{equation}}
    \item{ The inverse covariance between a color and itself, \begin{equation}\Sigma^{-1}_{g-i, g-i} = \frac{1}{\sigma^2_g} \ . \end{equation}}
    \item{ The inverse covariance between one color and a second, \begin{equation}\Sigma^{-1}_{g-i, r-i} = 0 \ . \end{equation}}
\end{itemize}

When our input vector is only the difference with respect to the reference band, for example $\vec{x} = (m_g-m_i, m_r-m_i)$, then the inverse covariance terms are as follows:
\begin{itemize}
    \item{The inverse covariance between a color and itself, \begin{equation}\Sigma^{-1}_{g-i, g-i} = \frac{1}{\sigma^2_g} - \frac{1}{\sigma^4_g} \frac{1}{\sum_f \frac{1}{\sigma^2_f}} \ , \end{equation} where the sum is over all flux passbands $f$.}
    \item{The inverse covariance between one color and a second, \begin{equation}\Sigma^{-1}_{g-i, r-i} = -\frac{1}{\sigma^2_g \sigma^2_r} \frac{1}{\sum_f \frac{1}{\sigma^2_f}} \ , \end{equation} where the sum is over all flux passbands $f$.}
\end{itemize}
These terms are derived by considering the Sherman-Morrison formula:
\begin{equation}
	\label{eq:sherman-morrison}
	(C + u v^T)^{-1} = C^{-1} - \frac{C^{-1} u v^T C^{-1}}{1 + v^T C^{-1} u} \ ,
\end{equation}
where $\Sigma = C + u v^T$. In our case, $C$ is a diagonal matrix, $C_{j-i, k-i}$ = $\delta_{jk} \sigma^2_{j}$, and $u$ and $v$ are the same vector $[\sigma_i \ldots \sigma_i]$. We may suggestively write \autoref{eq:sherman-morrison} as a sum over the indices $m,n$:
\begin{equation}
(C + u u^T)^{-1}_{jk} = \frac{\delta_{jk}}{\sigma^2_j} - \frac{\sum_{mn} \frac{\delta_{jm}}{\sigma^2_j} \sigma^2_i \frac{\delta_{kn}}{\sigma^2_k}}{1 + \sum_{mn} \frac{\delta_{mn} \sigma^2_i}{\sigma^2_m}} \ .
\end{equation}
Carrying out the sums and simplifying, we find the above formulas for the inverse covariance matrix.

The variance in the measurement of a luptitude, $\sigma^2_{\mu_x}$, can be calculated from the variance in the measurement of its respective flux, $\sigma^2_{f_x}$:
\begin{equation}
	\sigma^2_{\mu_x} = \frac{a^2}{4b^2 + f_x^2} \sigma^2_{f_x},
\end{equation}
where $a=2.5 \log e$ and $b$ is a softening parameter \citep{Lupton1999}. Once this variance is found, the covariance and inverse covariance matrices may be found with the above formulas.


\section*{Affiliations}
$^{1}$ SLAC National Accelerator Laboratory, Menlo Park, CA 94025, USA\\
$^{2}$ Kavli Institute for Particle Astrophysics \& Cosmology, P. O. Box 2450, Stanford University, Stanford, CA 94305, USA\\
$^{3}$ \' Ecole Polytechnique F\' ed\' erale de Lausanne, Route Cantonale, 1015 Lausanne, Switzerland\\
$^{4}$ Institute of Science, Technology, and Policy, ETH Zurich, Universit\" atstrasse 41, 8092 Zurich, Switzerland\\
$^{5}$ Descartes Labs, Inc., 100 N Guadelupe St, Santa Fe, NM 87501, USA\\
$^{6}$ Department of Physics, Stanford University, 382 Via Pueblo Mall, Stanford, CA 94305, USA\\
$^{7}$ Institut d'Estudis Espacials de Catalunya (IEEC), 08034 Barcelona, Spain\\
$^{8}$ Institute of Space Sciences (ICE, CSIC),  Campus UAB, Carrer de Can Magrans, s/n, 08193 Barcelona, Spain\\
$^{9}$ Department of Physics and Astronomy, University of Pennsylvania, Philadelphia, PA 19104, USA\\
$^{10}$ Center for Cosmology and Astro-Particle Physics, The Ohio State University, Columbus, OH 43210, USA\\
$^{11}$ Infrared Processing and Analysis Center, Pasadena, CA 91125, USA\\
$^{12}$ Jet Propulsion Laboratory, California Institute of Technology, 4800 Oak Grove Dr., Pasadena, CA 91109, USA\\
$^{13}$ Instituci\'o Catalana de Recerca i Estudis Avan\c{c}ats, 08010 Barcelona, Spain\\
$^{14}$ Institut de F\'{\i}sica d'Altes Energies (IFAE), The Barcelona Institute of Science and Technology, Campus UAB, 08193 Bellaterra (Barcelona), Spain\\
$^{15}$ Department of Physics, Duke University Durham, NC 27708, USA\\
$^{16}$ Cerro Tololo Inter-American Observatory, National Optical Astronomy Observatory, Casilla 603, La Serena, Chile\\
$^{17}$ Fermi National Accelerator Laboratory, P. O. Box 500, Batavia, IL 60510, USA\\
$^{18}$ Institute of Cosmology and Gravitation, University of Portsmouth, Portsmouth, PO1 3FX, UK\\
$^{19}$ LSST, 933 North Cherry Avenue, Tucson, AZ 85721, USA\\
$^{20}$ Physics Department, 2320 Chamberlin Hall, University of Wisconsin-Madison, 1150 University Avenue Madison, WI  53706, USA\\
$^{21}$ Jodrell Bank Center for Astrophysics, School of Physics and Astronomy, University of Manchester, Oxford Road, Manchester, M13 9PL, UK\\
$^{22}$ Department of Physics \& Astronomy, University College London, Gower Street, London, WC1E 6BT, UK\\
$^{23}$ Centro de Investigaciones Energ\'eticas, Medioambientales y Tecnol\'ogicas (CIEMAT), Madrid, Spain\\
$^{24}$ Laborat\'orio Interinstitucional de e-Astronomia - LIneA, Rua Gal. Jos\'e Cristino 77, Rio de Janeiro, RJ - 20921-400, Brazil\\
$^{25}$ Department of Astronomy, University of Illinois at Urbana-Champaign, 1002 W. Green Street, Urbana, IL 61801, USA\\
$^{26}$ National Center for Supercomputing Applications, 1205 West Clark St., Urbana, IL 61801, USA\\
$^{27}$ Observat\'orio Nacional, Rua Gal. Jos\'e Cristino 77, Rio de Janeiro, RJ - 20921-400, Brazil\\
$^{28}$ Department of Physics, IIT Hyderabad, Kandi, Telangana 502285, India\\
$^{29}$ Kavli Institute for Cosmological Physics, University of Chicago, Chicago, IL 60637, USA\\
$^{30}$ Department of Astronomy/Steward Observatory, 933 North Cherry Avenue, Tucson, AZ 85721-0065, USA\\
$^{31}$ Department of Astronomy, University of Michigan, Ann Arbor, MI 48109, USA\\
$^{32}$ Department of Physics, University of Michigan, Ann Arbor, MI 48109, USA\\
$^{33}$ Instituto de Fisica Teorica UAM/CSIC, Universidad Autonoma de Madrid, 28049 Madrid, Spain\\
$^{34}$ Department of Physics, ETH Zurich, Wolfgang-Pauli-Strasse 16, 8093 Zurich, Switzerland\\
$^{35}$ Santa Cruz Institute for Particle Physics, Santa Cruz, CA 95064, USA\\
$^{36}$ Department of Physics, The Ohio State University, Columbus, OH 43210, USA\\
$^{37}$ Center for Astrophysics $|$ Harvard \& Smithsonian, 60 Garden Street | MS 42 | Cambridge, MA 02138, USA\\
$^{38}$ Australian Astronomical Optics, Macquarie University, North Ryde, NSW 2113, Australia\\
$^{39}$ Departamento de F\'isica Matem\'atica, Instituto de F\'isica, Universidade de S\~ao Paulo, CP 66318, S\~ao Paulo, SP, 05314-970, Brazil\\
$^{40}$ George P. and Cynthia Woods Mitchell Institute for Fundamental Physics and Astronomy, and Department of Physics and Astronomy, Texas A\&M University, College Station, TX 77843, USA\\
$^{41}$ Department of Astrophysical Sciences, Princeton University, Peyton Hall, Princeton, NJ 08544, USA\\
$^{42}$ School of Physics and Astronomy, University of Southampton,  Southampton, SO17 1BJ, UK\\
$^{43}$ Brandeis University, Physics Department, 415 South Street, Waltham, MA 02453, USA\\
$^{44}$ Instituto de F\'isica Gleb Wataghin, Universidade Estadual de Campinas, 13083-859, Campinas, SP, Brazil\\
$^{45}$ Computer Science and Mathematics Division, Oak Ridge National Laboratory, Oak Ridge, TN 37831, USA\\
$^{46}$ Argonne National Laboratory, 9700 South Cass Avenue, Lemont, IL 60439, USA\\

\bsp	
\label{lastpage}
\end{document}